\documentclass[a4paper,10pt]{JHEP3}
\usepackage[dvips]{graphicx}
\usepackage{cite}
\usepackage{amssymb}
\usepackage{amsmath}
\usepackage{latexsym}
\usepackage{xspace}
\usepackage{array}
\usepackage{url}
\usepackage{afterpage}

\usepackage{tabularx}
\newcolumntype{L}[1]{>{\raggedright\arraybackslash}p{#1}} 
\newcolumntype{C}[1]{>{\centering\arraybackslash}p{#1}}

\newcommand{\trule}{\rule[-1.5mm]{0mm}{6mm}}
\newcommand{\SqrtS}{\sqrt{S}}
\newcommand{\alphas}{\alpha_s}

\newcommand {\uL}{\tilde{u}_{L}}
\newcommand {\uR}{\tilde{u}_{R}}
\newcommand {\dL}{\tilde{d}_{L}}
\newcommand {\dR}{\tilde{d}_{R}}

\newcommand{\sq}{\ensuremath{\tilde{q}\xspace}}

\newcommand{\msq}{\ensuremath{m_{\sq}}}

\newcommand{\neu}{\ensuremath{\tilde{\chi}^{0}}}
\newcommand{\neuc}{\ensuremath{\tilde{\chi}}}
\newcommand{\neuone}{\ensuremath{\tilde{\chi}^{0}_1}}
\newcommand{\LSP}{\neuone}
\newcommand{\gluino}{\ensuremath{\tilde{g}}}
\newcommand{\gl}{\ensuremath{\tilde{g}}\xspace}
\newcommand{\mgl}{\ensuremath{m_{\gl}}}
\newcommand{\MeV}{{\rm Me\kern -1pt V}\xspace}
\newcommand{\GeV}{{\rm Ge\kern -1pt V}\xspace}
\newcommand{\TeV}{{\rm Te\kern -1pt V}\xspace}

\newcommand{\mhalf}{\ensuremath{m_{1/2}}\xspace}

\newcommand{\msusy}{\ensuremath{M_{\text{SUSY}}}\xspace}
\newcommand{\msbar}{\ensuremath{\overline{\text{MS}\xspace}}}

\newcommand{\be}{\begin{eqnarray*}}
\newcommand{\ee}{\end{eqnarray*}}
\newcommand{\bee}{\begin{eqnarray}}
\newcommand{\eee}{\end{eqnarray}}
\renewcommand{\eqref}[1]{eq.~(\ref{#1})}

\newcommand{\figref}[1]{figure~\ref{#1}}
\newcommand{\tabref}[1]{table~\ref{#1}}

\newcommand{\ord}{{\cal O}}
\newcommand{\fba}{\unskip\,{fb}\xspace}
\newcommand{\pba}{\unskip\,{pb}\xspace}

\usepackage{cancel}
\usepackage{dcolumn}
\newcolumntype{d}[0]{D{.}{.}{-1}}

\newcommand{\Slashed}[1]{\ensuremath{{#1}{\!}{\!}{\!}{\!}{\!}{\:}/}}

\newcommand{\eg}[0]{e.g.}

\newcommand{\ie}[0]{i.e.\xspace}

\newcommand{\softsusy}[0]{\texttt{SOFTSUSY}\xspace}
\newcommand{\sdecay}[0]{\texttt{SDECAY}\xspace}
\newcommand{\susyhit}[0]{\texttt{SUSY-HIT}\xspace}
\newcommand{\Prospino}[0]{\texttt{Prospino~2}\xspace}
\newcommand{\madgraph}[0]{\texttt{MadGraph~5}\xspace}
\newcommand{\FeynArts}[0]{\texttt{FeynArts}\xspace}

\newcommand{\FormCalc}[0]{\texttt{FormCalc}\xspace}

\newcommand{\LoopTools}[0]{\texttt{LoopTools}\xspace}

\newcommand{\fastjet}[0]{\texttt{FastJet~3.0.2}\xspace}

\newcommand{\si}{\ensuremath{\sigma}\xspace}

\newcommand{\pT}{\ensuremath{p^\mathrm{T}}\xspace}
\newcommand{\alphat}{\ensuremath{\alpha_T}\xspace}
\def\missingET{\ensuremath{\displaystyle{\not}E_T}\xspace}
\newcommand{\meff}{\ensuremath{m_{\text{eff}}}\xspace}

\newcommand{\Li}[1]{\text{Li}_2\left(#1\right)}

\title{NLO corrections to squark--squark production and decay at the LHC}

\author{W. Hollik, J. M. Lindert, D. Pagani\\
 Max-Planck-Institut f\"ur Physik, 
 F\"ohringer Ring 6, 
 D-80805 M\"unchen, Germany\\
 Email: \email{hollik@mpp.mpg.de}, \email{lindert@mpp.mpg.de}, \email{pagani@mpp.mpg.de}}

\abstract{
We present an analysis of the signature $2j+\missingET(+X)$ via squark--squark
production and direct decay into the lightest neutralino, $pp \to \tilde q \tilde q \to jj
\tilde\chi_1^0 \tilde\chi_1^0(+X)$,
in next-to-leading order QCD within the framework of the minimal supersymmetric standard model.
In our approximation the produced squarks are treated
on shell. Thus, the calculation of production and decay factorizes.
In this way, we provide a consistent, fully differential calculation of
NLO QCD factorizable corrections to the given processes.
Clustering final states into partonic jets, we investigate the experimental
inclusive signature $2j+\missingET$ for several benchmark scenarios.
We compare resulting differential distributions with leading-order
approximations rescaled by a flat K-factor and examine a possible impact for cut-and-count
searches for supersymmetry at the LHC.\\\\

}

\preprint{arXiv:1207.1071 \\ MPP-2012-106}

\begin{document}

\section{Introduction}
Supersymmetry (SUSY) \cite{Wess:1974tw} is one of the most appealing scenarios for
physics beyond the Standard Model (SM), which the ongoing experiments at 
the Large Hadron Collider (LHC) are searching for. 
SUSY, predicting new states with masses at the \TeV-scale
or below,  provides an elegant solution to the hierarchy problem, 
and gauge coupling unification can be achieved naturally. In addition, many
realizations of SUSY, particularly the Minimal Supersymmetric Standard Model (MSSM),
provide a viable dark matter candidate assuming the lightest SUSY particle (LSP) to be
stable due to R-parity conservation. Furthermore, the MSSM is in accordance
with the measured values of the 
muon anomalous magnetic moment and with electroweak precision observables,
and also with a light Higgs boson as indicated by the ATLAS and CMS experiments~\cite{:2012gk,:2012gu}. 

Within the MSSM and assuming conserved R-parity, SUSY particles (sparticles) are produced in pairs,
and related searches have been performed at LEP, the Tevatron, and the LHC using various
final-state signatures. Due to their color charge, squark and gluino production typically
gives the largest contribution to an inclusive SUSY cross section at a hadron collider 
like the LHC. Assuming the lightest neutralino $\neu_{1}$ to be the LSP, 
produced squarks and gluinos eventually decay into the neutralino which leaves the 
detectors unobserved. This results in the general experimental signature of jets+missing 
energy, which is one of the signatures that
has been searched for by the experiments at the LHC. Within the 
constrained MSSM (CMSSM) resulting limits can be used to exclude  squarks and 
gluinos with masses below $1-1.4~\TeV$ \cite{Chatrchyan:2011zy, Aad:2011ib}. Exact exclusion
 limits, however, depend on the detailed structure of the underlying model parameters 
and become much weaker, \eg, in parameter regions with compressed spectra where 
final-state jets do not pass the cuts applied in the experimental searches
\cite{LeCompte:2011fh,Asano:2011ik,Strubig:2012qd}. \\

Precise and reliable theoretical predictions for squark and gluino production are
necessary for several reasons: -- to set accurate exclusion limits, --
to possibly refine experimental search strategies in problematic parameter regions, and 
-- in case of discovery, to determine the parameters of the underlying model
\cite{Dreiner:2010gv}. 
The last point becomes more important as generic mass
bounds for squarks and gluinos are pushed to higher values and many of the
proposed sophisticated methods (see \eg \cite{Barr:2010zj} for a review) for parameter
determination might not be feasible with low signal statistics due to a rather heavy
spectrum. \\

Until now precision studies of sparticle production at the LHC have focussed on inclusive
cross sections, without taking into account phase-space cuts that have to be applied in any
experimental analysis. Although these inclusive quantities are of fundamental interest,
both for exclusion limits and for parameter determination, they are not directly
observable in high-energy collider experiments.  Furthermore, precise knowledge of
distributions of the decay products might help to determine the fundamental parameters
of the model \cite{Barr:2010zj,Miller:2005zp} or even permits the measurements of the
spin of the new particles \cite{Smillie:2005ar,MoortgatPick:2011ix} and thereby helps
to discriminate SUSY models from possible other extensions of the SM with
similar signatures \cite{Hallenbeck:2008hf,Hubisz:2008gg}. Thus, a fully differential
prediction including higher-orders in all relevant stages of the process (eventually 
matched to a NLO parton shower) is desirable. \\

In this paper we systematically study squark--squark production and the subsequent 
decay into the lightest neutralino at next-to-leading order (NLO) in QCD. Final state 
partons are clustered into jets and thus, we provide, for the first time at NLO, a fully differential 
description of the physical signature $2j+\missingET(+X)$ via on-shell squark-squark 
production and decay. In principle, our calculation does not depend on the hierarchy 
between the squarks and the gluino. However, in our numerical evaluation we only 
consider benchmark points where the mass $\msq$ of all light flavour squarks is smaller 
than the gluino mass $(\mgl > \msq)$; otherwise the decay of a squark into a gluino and 
a quark would be dominant. Investigating the squark-squark channel should be 
understood as a first step towards a fully differential prediction for all sparticle 
production channels at NLO. It is, however, also of practical importance, since
from recent searches at the LHC mass bounds 
for squarks and gluinos are generically pushed to higher values and here squark--squark 
production (initiated from valence-quarks) yields the dominant channel \cite{Mrenna:2011ek}.\\

First leading order (LO) cross section predictions for squark and gluino production
processes were already made many years ago
\cite{Kane:1982hw,Harrison:1982yi,Reya:1984yz,Dawson:1983fw,Baer:1985xz} and are
reviewed in \cite{Nakamura:2010zzi}. Also the calculation of NLO
corrections in perturbative QCD has been performed quite some time
ago \cite{Beenakker:1994an,Beenakker:1995fp,Beenakker:1996ch,Beenakker:1997ut}. These
corrections can be large ($10\% - 70\%$, depending on the process and the parameters)
and have to be included in any viable phenomenological study due to the otherwise
enormous scale uncertainties (including NLO corrections the scale uncertainty on
inclusive cross sections is typically reduced to an order of $20\% - 30\%$). Besides 
the scale uncertainty, PDF uncertainties dominate the error of theoretical predictions
of sparticle production processes. In a very recent publication 
\cite{Kramer:2012bx} general guidelines for the systematic treatment 
of these errors are presented.

Already in \cite{Beenakker:1996ch} differential distributions at NLO are presented for the
produced squarks and gluinos. Here, NLO correction factors (K-factors)
look rather flat in phase space. Therefore, in the experimental analyses, they are 
used as a global multiplicative factor to the LO cross section. However, a 
systematic study of the differential behaviour of these K-factors has never been
performed. Furthermore, in \cite{Beenakker:1996ch} and in the corresponding public computer 
code \Prospino~\cite{Beenakker:1996ed}, which can calculate LO cross sections and 
NLO K-factors efficiently, NLO corrections for squark--squark production are always 
summed over the various flavour and chirality combinations of the produced (light-flavour) 
squarks. Realistic physical observables do depend on the chiralities 
through the decay modes which are in general quite different. In this work we treat 
the individual squark chirality and flavour configurations independently. One goal of our
paper is to investigate the quality of these approximations: using flat K-factors and 
averaging on squark masses.

More recently also results beyond NLO in QCD were calculated,
based on resummation techniques
\cite{Langenfeld:2009eg,Kulesza:2008jb,Kulesza:2009kq,Beenakker:2009ha, Beenakker:2010nq,
Beneke:2010da,Beenakker:2011sf,Falgari:2012hx}. These corrections increase the inclusive
cross section by about $2\% - 10\%$ and further reduce the scale uncertainty. 
Moreover, electroweak contributions can also give sizeable corrections. 
At leading order
they were first calculated in \cite{Bornhauser:2007bf,Arhrib:2009sb} and at NLO in
\cite{Hollik:2007wf,Hollik:2008yi,Hollik:2008vm,Beccaria:2008mi,Mirabella:2009ap,
Germer:2010vn,Germer:2011an}. In detail, those corrections depend strongly
on the model parameters and on the flavour/chiralities of the  squarks. \\

A similar amount of work has been put into the calculation of higher order corrections
to decays of (coloured) sparticles, with focus mainly on 
the integrated decay widths and branching ratios. 
NLO QCD corrections to the 
decay of light squarks into neutralinos and charginos were first calculated in 
\cite{Hikasa:1995bw,Djouadi:1996wt} and to heavy squarks also in \cite{Djouadi:1996wt} 
and in \cite{Kraml:1996kz}. Corrections to the total decay width of light squarks are in
general moderate (below $10~\%$) and can change sign, depending on the involved mass
ratios. However, for very small mass splittings between the decaying squark and the
neutralinos/charginos these corrections increase significantly. 
Higher-order corrections to
the decay of top-squarks are in general sizeable, 
but they depend strongly on the mixing in the heavy squark sector. 
Related to this mixing also decays into weak gauge bosons or Higgs bosons can become
relevant \cite{Bartl:1994bu,Bartl:1998xk}, receiving large higher order corrections
\cite{Bartl:1997pb,Arhrib:1997nf,Bartl:1998xp}. Decays of a gluino into a light squark and
a quark at NLO QCD together with the decay of a light squark into a gluino and a quark
have been calculated in \cite{Beenakker:1996dw}. Corresponding decays involving 
stops were presented in \cite{Beenakker:1996de}. All these decays including their NLO QCD
corrections have been implemented in the public computer programs \sdecay
\cite{Muhlleitner:2003vg} and \susyhit \cite{Djouadi:2006bz}. 

Besides NLO QCD, also NLO electroweak corrections to squark decays into neutralinos and charginos
have been investigated in the literature \cite{Guasch:1998as,Guasch:2002ez} and can give
sizeable contributions. These corrections often compensate those from QCD on the level
of integrated decay widths, however, they depend strongly on the model parameters. 
Corresponding NLO electroweak corrections for third generation squark decays have been studied in
\cite{Arhrib:2004tj,Arhrib:2005ea,Li:2002ey,Weber:2007id}.\\

As already mentioned before, most of the discussed studies of higher-order corrections
focussed on inclusive observables or considered differential distributions in unphysical
final states, like unstable sparticles. Few studies were performed
investigating invariant mass distributions of SM particles emitted from cascade chains
including various higher-order corrections \cite{Drees:2006um,Horsky:2008yi}. Finally, in
\cite{Plehn:2005cq,Alwall:2008qv} production of sparticles was studied at tree
level matched to a parton shower including additional hard jets. In these works large 
deviations from the LO prediction with or without showering were found particularly in 
the high-\pT tail for scenarios with compressed spectra. In this paper we go beyond these 
estimates and provide a  fully differential description of production and decay of 
squark--squark pairs at NLO.\\

In the big picture of the complete calculation of NLO QCD corrections to $pp\to
2j+\missingET(+X)$, also $\sq\sq'^{*}$, $\tilde{g}\sq$ and $\tilde{g}\tilde{g}$
intermediate states can contribute to this signature. Already without systematically 
including decays of the squarks, the calculation of NLO corrections to on-shell 
production of such pairs of coloured sparticles carries problems of double counting. 
Parts of NLO corrections to one final state can be identified as LO of another final 
state where the decay is already included. The standard solution, used to avoid this
double counting problem, can not be straightforwardly extended to the 
calculation where off-shell effects are included. 
Moreover, complete NLO corrections to $pp\to q q' \neu_{1} \neu_{1}$ do not only include
factorizable contributions, \ie, contributions that can be classified as corrections to the 
production or to the decays, but also non-factorizable contributions, where such
a separation is not possible. In this paper we analyze the factorizable NLO corrections to 
squark--squark production and decay, which are expected to yield the dominant part of the
NLO contributions.   
Non-factorizable effects and off-shell contributions
will be analyzed in a forthcoming publication, providing a consistent conceptual approach
and evaluating their numerical effects.\\

The outline of this paper is as follows. In section \ref{sec:method} the method of 
combining consistently production and decay at NLO in the narrow-width-approximation
is described.
In the subsequent two sections the calculation of all required ingredients 
of this combination is explained, with respect to the squark production processes 
in section \ref{sec:prod}, and to the squark decays in \ref{sec:decay}.
In section \ref{sec:numerics} we present our
numerical results for representative benchmark points, and conclude 
with a summary in section \ref{sec:conclusion}.


\section{Method}
\label{sec:method}

We investigate the production of squark-squark pairs induced by proton-proton collisions,
with subsequent decays of the squarks into the lightest neutralinos. Since we are
interested in the experimental signature $2j+\missingET(+X)$, all contributions from
light-flavour squarks have to be included. Hence, the cross section is given by the sum over 
all independent flavour and chirality configurations,
\begin{align}\label{sumflavours}
d\sigma=\sum_{\sq_{ia}\sq_{jb}}\left[d\sigma(pp\rightarrow\sq_{ia}\sq_{jb}\rightarrow q_{i}
\neu_{1} q_{j} \neu_{1}(+X))+d\sigma(pp\rightarrow\sq^{*}_{ia}\sq^{*}_{jb}\rightarrow \bar{q}_{i}
\neu_{1} \bar{q}_{j} \neu_{1}(+X))\right] \, .
\end{align}
Indices $i,j$ denote the flavours of the (s)quarks and $a,b$ their chiralities.
At LO, the only partonic subprocesses that contribute to a given intermediate configuration 
$\sq_{ia}\sq_{jb}$ or $\sq^{*}_{ia}\sq^{*}_{jb}$ arise from quark and anti-quark pairs,
respectively, 
$q_{i}q_{j}\rightarrow\sq_{ia}\sq_{jb}\rightarrow q_{i} \neu_{1}q_{j} \neu_{1}$ and 
$\bar{q}_{i}\bar{q}_{j}\rightarrow\sq^*_{ia}\sq^*_{jb}\rightarrow \bar{q}_{i} \neu_{1}
\bar{q}_{j} \neu_{1}$.

For simplifying the notation, we will write
$qq'\rightarrow\sq\sq'\rightarrow q \neu_{1}q' \neu_{1}$ whenever
the specification of flavour and chiralities is not 
required~\footnote{In this notation $\sq=\sq'$ implies $q=q'$,
but not vice versa.}.
Moreover, we will perform the discussion without the
charge-conjugate subprocesses; they are, however, included in the final results.

In the considered class of processes, 
squarks appear as intermediate particles with mass $m_{\sq}$
and total decay width $\Gamma_{\sq}$. 
In the limit $\Gamma_{\sq}/m_{\sq} \to 0$ , 
the narrow width approximation (NWA),
their resonating contributions in the squared amplitude can 
be approximated by the replacement
\begin{align}\label{breit}
\frac{1}{(p^2-\msq^2)^{2}+ \msq^{2}\, \Gamma_{\sq}^{2}} \, \rightarrow \,
\frac{\pi}{\msq\,\Gamma_{\sq}}\,  \delta(p^2 - \msq^2) \, ,
\end{align}
for each squark with momentum $p$.

\smallskip
At LO in NWA for scalar particles, the phase-space integration
of the squared amplitude for the total cross section
of the $2\to4$ processes factorizes 
into a production and a decay part.
At the partonic level, the LO cross section gets the following form,
\begin{align}\label{bornnw}
 \hat{\si}_{\mbox{NWA}}^{(0)}(qq'\rightarrow\sq\sq'\to q \neu_{1} q' \neu_{1})\, = \,  
\hat{\sigma}^{(0)}_{qq'\rightarrow\sq\sq'}\cdot \text{BR}^{(0)}_{\sq \to q
\neu_{1}}\cdot \text{BR}^{(0)}_{\sq' \to q' \neu_{1}} \, ,
\end{align}
with the LO partonic production cross section
$\hat{\sigma}^{(0)}_{qq'\rightarrow\sq\sq'}$ 
and the LO  branching ratios $\rm BR^{(0)}$ for the squark decays into 
the lightest neutralino.
%
%
A direct generalization of \eqref{bornnw} yields the cross section in a
completely differential form, which can be written at the hadronic level 
as follows,
\begin{align}\label{borndiff}
d \si_{\mbox{NWA}}^{(0)}(pp\rightarrow\sq\sq'\to q \neu_{1} q' \neu_{1}) \, =\, 
d\sigma^{(0)}_{pp\rightarrow\sq\sq'} \;
\frac{1}{\Gamma^{(0)}_{\sq}} \,
 d\Gamma^{(0)}_{\sq \to q\neu_{1}} \;
\frac{1}{\Gamma^{(0)}_{\sq'}} \,
 d\Gamma^{(0)}_{\sq' \to q'\neu_{1} }\, .
\end{align}
Therein, $\Gamma^{(0)}_{\sq}$ and $\Gamma^{(0)}_{\sq'}$ denote the LO total
widths of the two squarks; $d\Gamma^{(0)}_{\sq \to q \neu_{1}}$ and
$d\Gamma^{(0)}_{\sq' \to q' \neu_{1}}$ 
are the respective differential decay distributions
boosted to the moving frames of $\sq$ and $\sq'$. 

\smallskip
The other basic ingredient, 
$d\sigma^{(0)}_{pp\rightarrow\sq\sq'}$, 
is the hadronic differential production
cross section, expressed in terms of the partonic cross section
$d\hat{\sigma}^{(0)}_{q q'\rightarrow\sq\sq'}$
as a convolution
\begin{align}
\label{diffhadr}
d\sigma^{(0)}_{pp\rightarrow\sq\sq'} \, = \,
\int_{\tau_{0}}^{1}d\tau \, \mathcal{L}_{qq'}(\tau)\, 
d\hat{\sigma}^{(0)}_{q q'\rightarrow\sq\sq'}(\tau)
\end{align}
with the  parton luminosity
\begin{align}
\label{luminosity}
\mathcal{L}_{qq'}(\tau)=\frac{1}{1+\delta_{qq'}}\, \int_{\tau}^{1} \frac{dx}{x}\, 
\left[ f_{q}(x,\mu_{F})\, f_{q'}\!\left(\frac{\tau}{x},\mu_{F}\right) \, 
 +\, (q\leftrightarrow q') \right] \, ,
\end{align} 
where
$f_{i}(x,\mu_{F})$ is the parton distribution
function (PDF) at the scale $\mu_F$ of the quark $i$ 
with momentum fraction $x$ inside the proton.
$\tau$ denotes the ratio between the squared center-of-mass 
energies of the partonic and hadronic processes,
$\tau=s/S$, and 
the kinematical production threshold corresponds to 
$\tau_{0}=(m_{\sq}+m_{\sq'})^2/S$.

\smallskip
The NWA cannot be extended to the complete set of NLO QCD corrections to
$pp\rightarrow\sq\sq'\rightarrow q \neu_{1}q' \neu_{1}$. Interactions between initial
and final state quarks, for example, do not allow to split the process into on-shell
production of squarks and subsequent decays.
The subset of factorizable corrections, however, can be obtained in NWA,
occurring as corrections to the production or to the decay processes, as illustrated in
\figref{fig:blobs}.
In the present article we focus on this class of corrections, which are expected to be
the dominant ones.
In an upcoming article we will investigate the non-factorizable corrections
and their numerical influence.

\smallskip
The differential cross section for on-shell production
of squark--squark pairs and subsequent decays including the NLO factorizable corrections
can be written as a formal expansion in $\alphas$,
\begin{align}
\label{masterformel}
d\sigma^{(0+1)}_{\mbox{NWA}}(pp\rightarrow\sq\sq' \rightarrow q\neu_{1}q'\neu_{1}(+&X))
\,=\, \frac{1}{\Gamma^{(0)}_{\sq}\,\Gamma^{(0)}_{\sq'}} \,
\Big[d\sigma^{(0)}_{pp\rightarrow\sq\sq' }\, 
     d\Gamma^{(0)}_{\sq \to q\neu_{1}} \,d\Gamma^{(0)}_{\sq' \to q'\neu_{1}}\,
     \Big(1-\frac{\Gamma^{(1)}_{\sq}}{\Gamma^{(0)}_{\sq}}-
            \frac{\Gamma^{(1)}_{\sq'}}{\Gamma^{(0)}_{\sq'}}\Big) \nonumber\\[0.2cm]
+&\, d\sigma^{(0)}_{pp\rightarrow\sq\sq' }\, 
     d\Gamma^{(1)}_{\sq \to q\neu_{1}}\, d\Gamma^{(0)}_{\sq' \to q' \neu_{1}}
\,+\,d\sigma^{(0)}_{pp\rightarrow\sq\sq' }\, d\Gamma^{(0)}_{\sq \to q\neu_{1}}\,
     d\Gamma^{(1)}_{\sq' \to q' \neu_{1}} \nonumber\\[0.2cm]
+&\, d{\sigma}^{(1)}_{pp\rightarrow\sq\sq'(X)}\, d\Gamma^{(0)}_{\sq \to q\neu_{1}}\,
     d\Gamma^{(0)}_{\sq' \to q' \neu_{1}}\Big]  \, ,
\end{align}
with the NLO contributions to cross section and widths 
$d\sigma^{(1)},\Gamma^{(1)}, \dots$ in obvious notation.
The LO term in the first line of \eqref{masterformel}
gets a global correction factor from the 
NLO contribution to the total widths; the second and third line involve 
the NLO corrections to the decay distributions and the production cross 
section, respectively~\footnote{An analogous treatment has been used, \eg, for the calculation
of NLO corrections of top pair production and
decay~\cite{Melnikov:2009dn,Campbell:2012uf}.}. \\

%
\FIGURE{
\includegraphics[width=0.48\textwidth]{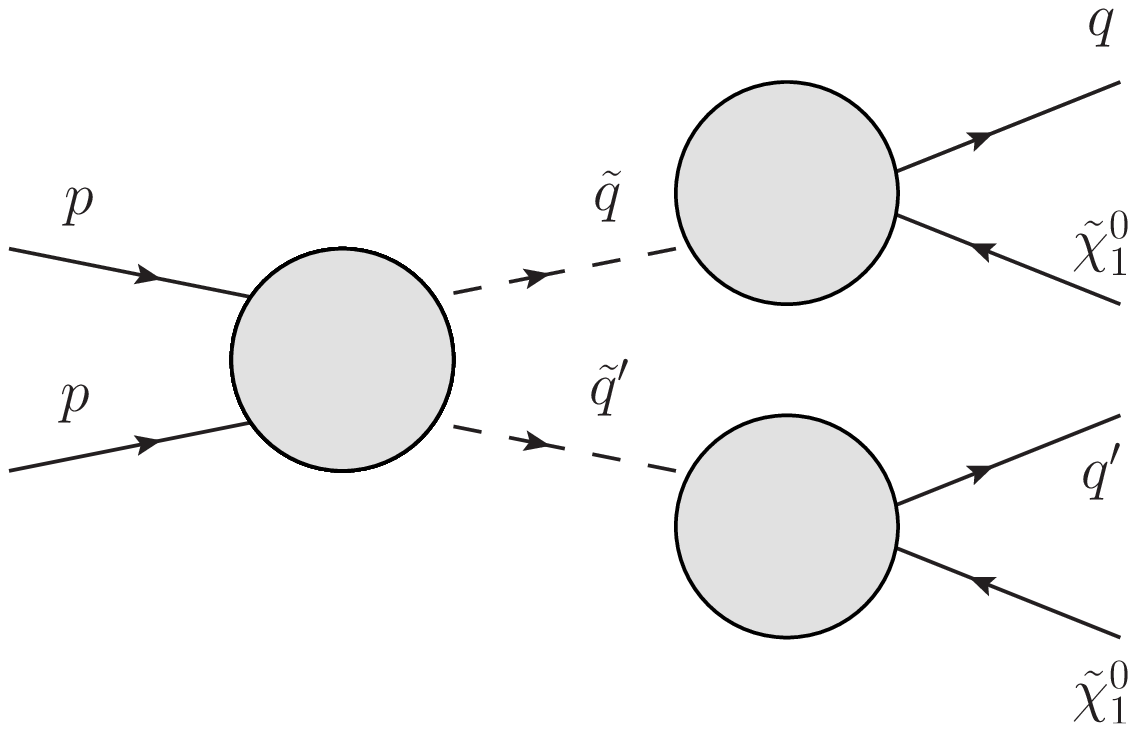}
\caption{General structure of the factorizable NLO QCD corrections to the signature $2j+\missingET(+X)$ via squark--squark
production and direct decay into the lightest neutralino.
\label{fig:blobs}
 }}
In order to evaluate the terms contained in \eqref{masterformel}, we produce, for all
different combinations of light flavours and chiralities, weighted events for squark-squark 
production and squark decays.
Production events for $pp \to \sq\sq'(X)$ are generated in the laboratory frame. 
Decay events for $\sq\to q \neu_{1}(g)$ 
and $\sq'\to q' \neu_{1}(g)$ are generated in the respective squark rest frame.
Finally, $pp \to qq'\neu_{1}\neu_{1}(+X)$ events are obtained by boosting the decay events 
from the squark rest frames, defined by the production events, into the laboratory frame. 
The weights of the $pp \to qq'\neu_{1}\neu_{1}(+X)$  events are obtained combining
the different LO and NLO weights of production and decay 
according to \eqref{masterformel}. 
Phenomenological results derived by these combinations are presented in 
section \ref{sec:numerics}. 
The treatment of the various entries in \eqref{masterformel}
is described in the following sections \ref{sec:prod} and \ref{sec:decay}.


\section{Squark--squark production}
\label{sec:prod}

\subsection{LO squark--squark production}
\label{sec:prodLO}

Amplitudes and cross sections for squark production depend on the flavours
(indices $i, j$) and on the chiralities (indices $a,b$) of the squarks.  We
consider light-flavour squarks only, treating quarks as massless.\\

\FIGURE{
\centering
\hspace{2cm}
\includegraphics{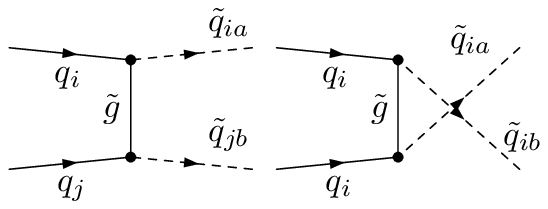}
\hspace{2cm}
\caption{Tree level Feynman diagrams for squark-squark production.}
\label{fig:feynman_tree}
}

If the two produced squarks are of the same flavour, 
the contributing Feynman diagrams correspond to $t$- and $u$-channel 
gluino exchange (\figref{fig:feynman_tree}).
For squarks of the same chirality, the partonic cross section reads as follows, 
\begin{align}
\label{LOa}
\frac{d\hat{\sigma}^{(0)}_{q_{i}q_{i}\rightarrow\sq_{ia}\sq_{ia}}}{dt}  =
\frac{\pi\alphas^{2}}{9s^{2}} \,m_{\gl}^{2}s\,
\left(\frac{1}{(t-m_{\gl}^{2})^{2}}+\frac{1}{(u-m_{\gl}^{
2})^{2}}+\frac{2/3}{(u-m_{\gl}^{2})(t-m_{\gl}^{2})}\right)\, ,
\end{align}
where $s,t$ and $u$ are the usual Mandelstam variables for $2\to 2$ processes.
For different chiralities, $m_{q}=0$ implies 
vanishing interference between the $t$- and $u$-channel diagrams,
yielding
\begin{align}
\label{LOb}
\frac{d\hat{\sigma}^{(0)}_{q_{i}q_{i}\rightarrow\sq_{ia}\sq_{ib}}}{dt} =
\frac{2\pi\alphas^{2}}{9s^{2}} \,
\left( 
 \frac{-st-(t-m^{2}_{\sq_{ia}})(t-m^{2}_{\sq_{ib}})} {(t-m_{\gl}^{2})^{2}}
 +\frac{-su-(u-m^{2}_{\sq_{ia}} ) (u-m^{2}_{\sq_{ib}}) }   {(u-m_{\gl}^{2})^{2}} 
\right)  \, .
\end{align}
If the two squarks are of different flavours, there is no $u$-channel exchange diagram;
the partonic cross section for equal chiralities is hence given by
\begin{align}
\label{LOc}
\frac{d\hat{\sigma}^{(0)}_{q_{i}q_{j}\rightarrow\sq_{ia}\sq_{ja}}}{dt} =
\frac{2\pi\alphas^{2}}{9s^{2}} \,
\frac{m_{\gl}^{2}s}{(t-m_{\gl}^{2})^{2}} \, ,
\end{align}
and for different chiralities by
\begin{align}
\label{LOd}
\frac{d\hat{\sigma}_{q_{i}q_{j}\rightarrow\sq_{ia}\sq_{jb}}^{(0)}}{dt} =
\frac{2\pi\alphas^{2}}{9s^{2}} \,
\frac{(-t+m^{2}_{\sq_{ia}})(t-m^{2}_{\sq_{ib}})-st}{(t-m_{\gl}^{2}
)^{2}}\,.
\end{align}
Besides the dominating QCD contributions, 
there are also tree-level electroweak production
channels \cite{Bornhauser:2007bf,Germer:2010vn} 
with chargino and neutralino exchange,
which can interfere with the QCD amplitude providing 
a contribution to the cross-section of $\ord(\alpha \alphas)$.
In principle these terms can be numerically of similar importance as the NLO QCD
$\ord(\alphas^3)$ corrections we are investigating. 
For the present study, the electroweak contributions are neglected.

\subsection{NLO squark--squark production}
\label{sec:prodNLO}
The NLO QCD corrections to squark--squark production have been known for many years
\cite{Beenakker:1996ch} and an efficient public code (\Prospino) is available for the
calculation of total cross sections at NLO. However, in order to study systematically the
$2j+\missingET(+X)$ signature emerging from production of squark--squark pairs and subsequent
decays into the lightest neutralino, also the complete differential cross section is
necessary. To this purpose, we perform an independent (re)calculation 
of the NLO QCD corrections,
where we treat the masses for $\sq_L$, $\sq_R$ and all chirality and flavour
configurations independently. In \cite{Beenakker:1996ch} different squark chiralities are
treated as mass degenerate and NLO contribution are always summed over all chirality and
flavour combinations.

NLO calculations involve, in intermediate steps, infrared and collinear divergences. 
Since our calculation does not involve any diagrams with non-Abelian vertices, infrared 
singularities can be regularized by a gluon mass ($\lambda$). Collinear singularities, 
in analogy, can be regularized by a quark mass ($m_q$), that is kept at zero
everywhere else in
the calculation. The cancellation of these two kinds of singularities is 
obtained by summing the virtual loop contributions and
the real gluon bremsstrahlung part, with subsequent mass factorization
in combination with the choice of the parton densities.

The complete NLO corrections to the differential cross section 
can be written symbolically in the following way,
\begin{align}
\label{NLO-general}
d\sigma^{(1)}_{pp\to\sq\sq'(X)} \, =\, 
d\sigma^{\text{virtual+soft}}_{pp\to\sq\sq'(g)} 
+d\sigma^{\text{coll}}_{pp\to\sq\sq'(g)}       
+ d\sigma^{\text{hard}}_{pp\to\sq\sq' g}  
+d\sigma^{\text{real-quark}}_{pp\to\sq\sq'\bar{q}^{(\prime)}}\, .
\end{align}
With $d\sigma^{\text{virtual+soft}}_{pp\to\sq\sq'(g)}$ we denote the summed contributions from 
the renormalized virtual corrections and soft gluon emission; 
$d\sigma^{\text{coll}}_{pp\to\sq\sq'(g)}$ corresponds to initial state
collinear gluon radiation including the proper subtraction term for the collinear divergences;  
$d\sigma^{\text{hard}}_{pp\to\sq\sq' g}$ denotes the remaining hard gluon emission
outside the soft and collinear phase space regions.
$d\sigma^{\text{real-quark}}_{pp\to\sq\sq'\bar{q}^{(\prime)}}$ is the contribution from real quark emission
from additional quark--gluon initial states contributing at NLO.

Technically, the calculation of the loop corrections and real radiation contributions
is performed separately for every flavour and chirality combination,
$q_{i}q_{j}\to\sq_{ia}\sq_{jb}$,
with the help of
\FeynArts \cite{Hahn:2000kx} and \FormCalc \cite{Hahn:1998yk,Hahn:2001rv}.
Appendix \ref{app:diagrams} shows a collection of the contributing Feynman diagrams. 
Loop integrals are numerically evaluated with \LoopTools \cite{Hahn:1998yk}.

\subsubsection{Virtual corrections and real gluon radiation}
In the term $d\sigma^{\text{virtual+soft}}_{pp\to\sq\sq'(g)}$ 
the virtual and soft contributions are added at the parton level, according to
\begin{eqnarray}
\label{virt-soft}
d\sigma^{\text{virtual+soft}}_{pp\rightarrow\sq\sq'(g)} &=&
\int_{\tau_{0}}^{1}d\tau \;  \mathcal{L}_{qq'}(\tau)\, 
d\hat{\sigma}^{\text{virtual+soft}}_{qq'\to\sq\sq'(g)}(\tau)\, , \nonumber  \\
d\hat{\sigma}^{\text{virtual+soft}}_{qq'\to\sq\sq'(g)}(\tau) &=&
d\hat{\sigma}^{\text{virtual}}_{qq'\to\sq\sq'}
+
d\hat{\sigma}^{\text{soft}}_{qq'\to\sq\sq'(g)}
\, . 
\end{eqnarray} 
The fictitious gluon mass  $\lambda$ for infrared regularization cancels in the sum
of $d\hat{\sigma}^{\text{virtual}}_{qq'\to\sq\sq'}$ and $d\hat{\sigma}^{\text{soft}}_{qq'\to\sq\sq'(g)}$.
\\

At NLO, UV finiteness requires renormalization by 
inclusion of appropriate counterterms, which can be found explicitly
in~\cite{Hollik:2008yi}.
All mass and field renormalization constants are 
determined according to the on-shell scheme.
The renormalization of the QCD coupling constant 
($\delta g_{s}=g_{s}\, \delta Z_{g_s}$) has to be done in accordance 
with the scheme for $\alpha_{s}$ in the PDFs, the $\msbar$ scheme with five flavours;
this corresponds to the renormalization constant~\cite{Beenakker:1996ch}
\begin{align}
\label{dzgmsbar}
\delta Z_{g_{s}} = 
-\frac{\alphas}{4 \pi}\Big[\Delta\frac{\beta_{0}}{2}
+\frac{1}{3}\log\frac{m_{t}^{2}}{\mu_R^{2}}
+\log\frac{m_{\tilde{g}}^{2}}{\mu_R^{2}}
+\frac{1}{12}\sum_{\tilde{q}}\log\frac{m_{\tilde{q}}^{2}}{\mu_R^{2}}\Big]\, ,
\end{align}
with the UV divergence $\Delta = 2/\epsilon -\gamma_E + \log(4\pi)$
and the renormalization scale $\mu_R$.
$\beta_{0}=3$ is the leading term of the $\beta$ function for the QCD coupling in the MSSM.
We choose to use dimensional regularization
for the calculation. This breaks the supersymmetric Slavnov-Taylor
identity that relates 
the $qqg$ vertex function and the $q\sq\gluino$ vertex function
at one-loop order.
However, this identity can be restored (see~\cite{Beenakker:1996ch,Hollik:2001cz}) 
by an extra finite shift of the $\hat{g}_{s}$ coupling in the $q\sq\gluino$ vertex with respect
to $g_s$ in the $qqg$ vertex, 
\begin{align}
\hat{g}_{s}=g_{s}(1+\delta Z_{{\hat g_{s}}})\,,\qquad\delta Z_{\hat{g}_{s}} =
\delta Z_{g_{s}}+\frac{\alphas}{3\pi}\, .
\end{align} 

The second term $d\hat{\sigma}^{\text{soft}}_{qq'\to\sq\sq'(g)}$ 
in \eqref{virt-soft} contains the contributions from real gluon emission
integrated over the soft-gluon phase space with $E_{g}<\Delta E$.
It is similar to the case  of soft-photon emission \cite{tHooft:1978xw,Denner:1991kt},
yielding a multiplicative correction factor to the LO cross section.
In the case of gluons, however, the
color structures are different for emission from $t$ and $u$ channel diagrams and hence
the various bremsstrahlung integrals enter the cross section 
with different weights. Accordingly, we decompose the partonic LO
cross section for $qq'\to\sq\sq'$ in the following way in obvious notation, 
\begin{align}
\label{ttuuut}
d\hat{\sigma}^{(0)}_{qq'\to\sq\sq'} =
d\hat{\sigma}_{\sq\sq'}^{(tt)}+d\hat{\sigma}_{\sq\sq'}^{(ut)}+d\hat{\sigma}_{\sq\sq'}^{(uu)} = 
\left[C_{\sq\sq'}^{(tt)}+C_{\sq\sq'}^{(ut)}+C_{\sq\sq'}^{(uu)}\right]
     d\hat{\sigma}^{(0)}_{qq'\to\sq\sq'}\, ,
\end{align} 
where the coefficients $C^{(tt,ut,uu)}_{\sq\sq'}$ for the individual channels 
can be easily read off
from the LO cross sections in eqs.~(\ref{LOa})--(\ref{LOd}).
Defining $\epsilon_{i} =1$ for incoming 
and $\epsilon_{i} = -1$ for outgoing particles,
the soft gluon contribution at partonic level can be written as follows,
using the label assignment
$\{q,q',\tilde{q},\tilde{q}'\} \leftrightarrow \{1,2,3,4\}$, 
\begin{align}
\label{softintegrals}
d\hat\sigma^{\text{soft}}_{qq'\to\sq\sq'(g)} =
-\frac{\alphas}{2\pi} \Big\{ 
  \sum_{i,j=1;i\le j}^4 \epsilon_i \epsilon_j\, \mathcal{I}_{ij}
  \Big\} \, 
  d\hat{\sigma}^{(0)}_{qq'\to\sq\sq'}\, .
\end{align} 
The $\mathcal{I}_{ij}$ involve the bremsstrahlung integrals
and the weight factors $C^{(tt,ut,uu)}_{\sq\sq'}$. Explicit 
expressions are listed in \eqref{Iij} of Appendix~\ref{app:soft-collinear}.
\\

$d\sigma^{\text{virtual+soft}}_{pp\rightarrow\sq\sq'(g)}$ 
still depends on the quark mass ($m_q$) owing to the 
initial-state collinear singularities. This dependence cancels
by adding the real collinear radiation term
$d\sigma^{\text{coll}}_{pp\rightarrow\sq\sq' (g)}$ 
resulting from gluon emission into the hard collinear region
and mass factorization for the PDFs
via adding a proper subtraction term, 
\begin{align}
\label{Coll}
d\sigma^{\text{coll}}_{pp\to\sq\sq'(g)} = \,
d\sigma^{\text{coll-cone}}_{ pp\to\sq\sq' (g)}\,
+ \, d\sigma^{\text{sub-pdf}}_{pp\to\sq\sq'} \, .
\end{align}
The collinear gluon emission into a narrow cone around the emitting particle
yields the following contribution that corresponds to
the results of~\cite{Baur:1998kt} with the replacement 
$\alpha Q_{q}^2\to(4/3)\alphas$, 
\begin{align}
\label{collcross}
d\sigma^{\text{coll-cone}}_{pp\to\sq\sq' (g)} \, = \,
\int_{\tau_{0}}^{1}d\tau\int_{\tau}^{1} \frac{dx}{x}
\int_{x}^{1-\delta_{s}} \frac{dz}{z} \;
\mathcal{L}^{\text{coll}}_{qq'}(\tau,x,z) \;
d\hat{\sigma}^{\text{coll-cone}}_{qq'\to\sq\sq'(g)}(\tau,z)\, .
\end{align}
The luminosity 
$\mathcal{L}^{\text{coll}}_{qq'}$ and the partonic cross section
$d\hat{\sigma}^{\text{coll-cone}}$ can be found   
in eqs.~(\ref{lumcoll}) and (\ref{partoniccollinear})
of Appendix~\ref{app:soft-collinear}.

\smallskip
The subtraction term for phase-space slicing, in accordance with the $\msbar$ scheme,
can be written in the following way, 
\begin{align}
\label{subpdf-cross}
d\sigma^{\text{sub-pdf}}_{pp\to\sq\sq'}  = &
-2 \int_{\tau_{0}}^{1}d\tau\int_{\tau}^{1}\frac{
dx}{x}\int_{x}^{1-\delta_{s}}\frac{dz}{z} \;
\mathcal{L}^{\rm coll}_{qq'}(\tau,x,z) \;
d\hat{\sigma}^{\rm sub1}_{qq'\to\sq\sq'}(\tau,z) \nonumber \\
& -2 \int_{\tau_{0}}^{1}d\tau \; 
\mathcal{L}_{qq'}(\tau) \; d\hat{\sigma}^{\rm sub2}_{qq'\to\sq\sq'}(\tau)\, ,
\end{align}
where $d\hat{\sigma}^{\rm sub1}_{pp\to\sq\sq'}$ and 
$d\hat{\sigma}^{\rm sub2}_{pp\to\sq\sq'}$
are defined in eqs.~(\ref{subpdf1}) and (\ref{subpdf2}) 
of Appendix~\ref{app:soft-collinear}.

\smallskip
Finally, we have to add $d\sigma^{\text{hard}}_{pp\to\sq\sq' (g)}$ from 
the residual hard gluon emission outside the collinear region,
which cancels the dependence on the slicing parameters $\delta_s$ for the soft
and $\delta_\theta$ for the collinear region.

\subsubsection{Real quark radiation}
\label{sec:quarkrad}
The $\ord(\alphas)$ corrections to $pp\to\sq_{ia}\sq_{jb}(X)$ get also a contribution
from the gluon-initiated subprocesses $q_{i}g\to\sq_{ia}\sq_{jb}\bar{q}_{j}$ and
$q_{j}g\to\sq_{ia}\sq_{jb}\bar{q}_{i}$, which 
also have to be included for a consistent treatment of NLO PDFs. 
Diagrams for these two subprocesses can be divided into resonant
(\figref{fig:radq-t}(a) and \ref{fig:radq-u}(a)) and non-resonant
(\figref{fig:radq-t}(b) and \ref{fig:radq-u}(b)) diagrams, where in 
the resonant diagrams the intermediate gluino can be on-shell.
This resonant production channel via on-shell gluinos corresponds
basically to LO production of a squark--gluino pair (with
subsequent gluino decay). Such contributions are generally classified as squark--gluino
production and have to be removed here
\footnote{The same type of problem is discussed, \eg, in \cite{Frixione:2008yi} 
in the context of NLO corrections 
to single top quark production, where quark radiation creates
configurations corresponding to top quark pair production.}.
In a general context, combining production and decay for all colored SUSY particles ($\sq,\sq^{*}$ and $\tilde{g}$), also off-shell configurations from resonant diagrams appear. In this context, the off-shell contributions from resonant diagrams in \figref{fig:radq-t}(a) and \ref{fig:radq-u}(a) can, in the same way, be classified as production and decay of squark--gluino pairs.
The most important difference in calculations with and without including decays is the role of the colored supersymmetric particles. In one case they belong to the final state, in the other one they are intermediate states. Thus, due to the quark radiation at NLO, a separation of squark--gluino and squark--squark channel contributions to $pp\rightarrow \text{2j(+X)}+\missingET$ is only an intermediate organisational instrument. It is important to remember that our NLO calculation of the production of squark--squark pairs is meant as a necessary ingredient of \eqref{masterformel}. Our primary goal is the consistency of the calculation at NLO of the full process with decays included and not only of the production of squarks.\\

In the following, we address the structure of the various terms
contributing to the real quark radiation and describe 
two different approaches to perform the
subtraction of the contributions corresponding to squark--gluino production 
(with subsequent gluino decay).\\

In the case of different flavours $i\neq j$, there are two parton
processes which provide NLO  
differential cross sections for real quark emission, given by
\begin{align}\label{xsec-qradij} 
d\hat{\sigma}_{q_{i}g\to\sq_{ia}\sq_{jb}\bar{q}_{j}} & \sim
 d\Pi_{(2\rightarrow3)}\left[\, \overline{
\left|\mathcal{M}_{\text{nonres},i}^{\phantom{*}}\right|^2}+2\text{Re}\overline{\left(
\mathcal{M}_{\text{nonres} , i }^{\phantom{*}} \mathcal { M }_{\text{res},i}^{*}\right)}+\overline{
\left|\mathcal{M}_{\text{res},i}^{\phantom{*}}\right|^2} \, \right] \,,\\
d\hat{\sigma}_{q_{j}g\to\sq_{ia}\sq_{jb}\bar{q}_{i}} & \sim
 d\Pi_{(2\rightarrow3)}\left[\, \overline{
\left|\mathcal{M}_{\text{nonres},j}^{\phantom{*}}\right|^2}+2\text{Re}\overline{\left(
\mathcal{M}_{\text{nonres}, j }^{\phantom{*}} \mathcal { M }_{\text{res},j}^{*}\right)}+\overline{
\left|\mathcal{M}_{\text{res},j}^{\phantom{*}}\right|^2}\, \right] \,,
\nonumber
\end{align}  
where {\it overline} represents the usual summing and
averaging of external helicities and colours and $d\Pi_{(2\rightarrow3)}$ 
is the usual phase-space element for three particles in the final state.
$\mathcal{M}_{\text{res},i}$ and $\mathcal{M}_{\text{nonres},i}$
correspond to the diagrams of \figref{fig:radq-t}(a) and
\figref{fig:radq-t}(b),
respectively, and 
$\mathcal{M}_{\text{res},j}$ and $\mathcal{M}_{\text{nonres},j}$ to
those of 
\figref{fig:radq-u}(a) and \figref{fig:radq-u}(b).

For the case of equal flavours $i=j$, we have
\begin{align}\label{xsec-qrad}
d\hat{\sigma}_{q_{i}g\to\sq_{ia}\sq_{ib}\bar{q}_{i}}\sim 
 d\Pi_{(2\rightarrow3)} \left[\, \overline{
\left|\mathcal{M}_{\text{nonres}}^{\phantom{*}}\right|^2}+2\text{Re}\overline{\left(
\mathcal{M}_{\text{nonres}}^{\phantom{*}} \mathcal { M }_{\text{res}}^{*}\right)}+\overline{
\left|\mathcal{M}_{\text{res}}^{\phantom{*}}\right|^2}\, \right]\, ,
\end{align}
with  $\mathcal{M}_{\text{res}}$ from the diagrams of \figref{fig:radq-t}(a) and
\figref{fig:radq-u}(a), which we will call in this case $\mathcal{M}_{\text{res,1}}$ and $\mathcal{M}_{\text{res,2}}$;
 $\mathcal{M}_{\text{nonres}}$ is the part from the  diagrams of \figref{fig:radq-t}(b)
  and \figref{fig:radq-u}(b).
The term  $2\text{Re}\overline{\left(
\mathcal{M}_{\text{res,1}} \mathcal { M }_{\text{res,2}}^{*}\right)}$
appears only for equal chiralities ($a=b$) and flavors of the squarks. 
We describe the subtractions for the case with equal flavor; 
analogous arguments apply to the different-flavor case.\\
\FIGURE[t]{
\centering
\includegraphics{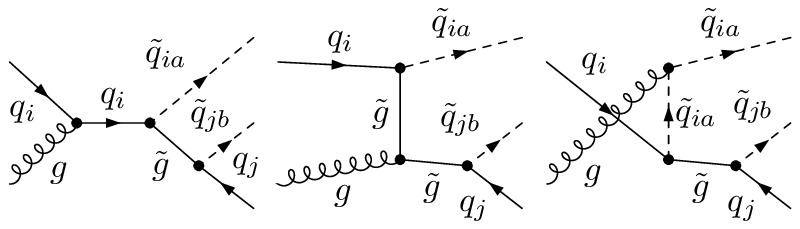}
\hspace*{1.0cm}
\includegraphics{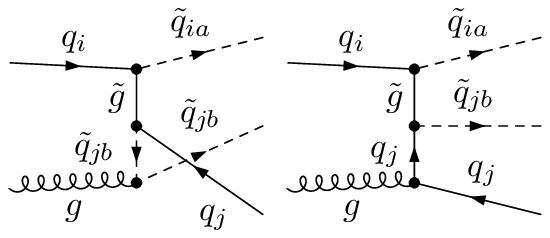}
\small (a) \hspace*{7.3cm} (b) 
\caption{Resonant (a) and non-resonant (b) diagrams contributing to $q_{i}g\to\sq_{ia}\sq_{jb}\bar{q}_{j}$.}
\label{fig:radq-t}
}

\FIGURE[t]{
\centering
\includegraphics{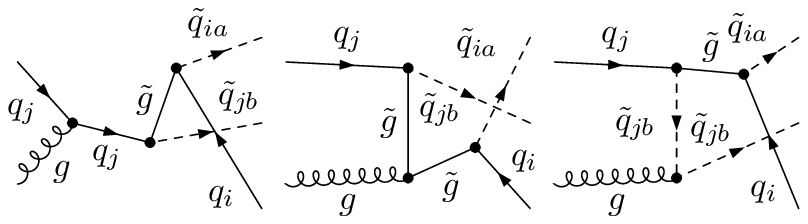}
\hspace*{1.0cm}
\includegraphics{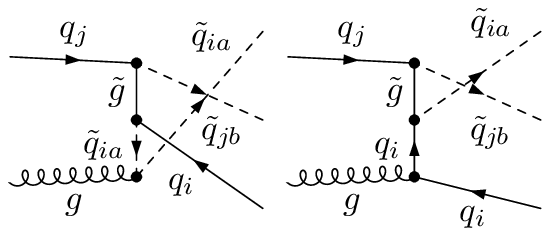}
\small (a) \hspace*{7.3cm} (b) 
\caption{Resonant (a) and non-resonant (b) diagrams contributing to $q_{j}g\to\sq_{ia}\sq_{jb}\bar{q}_{i}$.}
\label{fig:radq-u}
}

We refer to two strategies, used already in the literature
\footnote{The DR and DS schemes defined here are almost equal to the
  approaches extensively studied in \cite{Frixione:2008yi}.},
to avoid the double-counting of terms contained also in the calculation of the squark--gluino channel contribution to $pp\rightarrow \text{2j(+X)}+\missingET$:
\begin{itemize}
\item DS: Diagram Subtraction,
\item DR: Diagram Removal.
\end{itemize}

In the DS scheme the contribution from the LO on-shell production of 
a squark--gluino pair with the gluino decaying into a squark is removed: 
\begin{align}\label{xsec-qrad-ds}
d\hat{\sigma}_{q_{i}g\to\sq_{ia}\sq_{ib}\bar{q}_{i}}^{DS}
~\sim ~&\left[\, \overline{
\left|\mathcal{M}_{\text{nonres}}^{\phantom{*}}\right|^2}+2\text{Re}\overline{\left(
\mathcal{M}_{\text{nonres}}^{\phantom{*}} \mathcal { M }_{\text{res}}^{*}\right)}+\overline{
\left|\mathcal{M}_{\text{res}}^{\phantom{*}}\right|^2}\, \right]~ d\Pi_{(2\rightarrow3)} \nonumber \\
&-\left[\, \overline{
\left|\mathcal{M}_{\text{res,1}}^{\phantom{*}}\right|^2}+\overline{
\left|\mathcal{M}_{\text{res,2}}^{\phantom{*}}\right|^2}\, \right]~d\Pi_{(2\rightarrow 2) \times (1\rightarrow 2)} \ .
\end{align}
In \eqref{xsec-qrad-ds}, $d\Pi_{(2\rightarrow 2) \times (1\rightarrow  2)}$ 
is the phase-space with three particles in the final state
applying consistently the on-shell condition
$(p_{\sq}+p_{q})^2=m_{\tilde{g}}^{2}$ for the two different resonant cases. 
Eq.~(\ref{xsec-qrad-ds}) is conceptually equal to the DS scheme
explained in \cite{Frixione:2008yi} and the ``Prospino scheme'' in \cite{Beenakker:1996ch,Binoth:2011xi};
in practice there is a small difference with respect to our approach, which
is explained in more detail in appendix \ref{app:DS}. We subtract at global level exactly what we would obtain from LO
on-shell production of a squark--gluino pairs with the gluino decaying
into a squark. This is done by producing two different sets of events
corresponding to the two lines of \eqref{xsec-qrad-ds}, respectively.
In \cite{Beenakker:1996ch,Frixione:2008yi,Binoth:2011xi} a local
subtraction of the on-shell contribution involving a mapping or
reshuffling of momenta from the general $d\Pi_{2\rightarrow 3}$
phase-space into an equivalent on-shell configuration is
performed. These two implementations of the DS scheme give slightly
different results even in the limit $\Gamma_{\tilde{g}}\rightarrow 0$. 
The threshold conditions $p_{\tilde{g}}^2>m_{\sq}^2$ and 
$\sqrt{s}>\left(\sqrt{p^{2}_{\tilde{g}}}+m_{\sq}\right)$ 
in the local subtraction, together with the convolution of the PDFs and the precise
on-shell mapping, produce small differences from numerical results of the global subtraction. The DS scheme, both in the local approach discussed in \cite{Frixione:2008yi} and in the global approach, defined in \eqref{xsec-qrad-ds}, is gauge invariant in the limit $\Gamma_{\tilde{g}}\rightarrow 0$. The decay width of the gluino is used as a numerical regulator and not as a physical parameter. \\

In an extreme approach, the quark radiation calculation could even be
completely excluded from the NLO corrections in the squark--squark channel. 
Then, all diagrams, resonant and non-resonant, 
constituting a gauge invariant subset,
have to be included in the squark--gluino production and decay channel
(in this way, we would alter the organisational separation of squark/gluino channels). 
Since the term $\overline{\left|\mathcal{M}_{\text{nonres}}\right|^2}$
contains initial state collinear singularities, also the subtraction
term of the PDFs has to be excluded and computed within the squark--gluino channel.
It is worth to mention, that even if we want to include in all production and decay channels only on-shell configurations for the resonant intermediate supersymmetric particles (as performed here for the squark--squark channel), quark radiation in the NLO corrections introduces unavoidably off-shell contributions. 
\\ 

The DR scheme represents, in a certain sense, 
an intermediate step between the DS scheme and a complete removal as explained above. 
Here, one  removes, from a diagrammatic  perspective, the minimal
set of contributions 
in the squared amplitude that contain a resonant gluino.
In our calculation this results in
\begin{align}\label{xsec-qrad-dr}
d\hat{\sigma}_{q_{i}g\to\sq_{ia}\sq_{ib}\bar{q}_{i}}^{DR}&\sim&
d\Pi_{(2\rightarrow3)}\left[\, \overline{
\left|\mathcal{M}_{\text{nonres}}^{\phantom{*}}\right|^2}+2\text{Re}\overline{\left(
\mathcal{M}_{\text{nonres}}^{\phantom{*}} \mathcal { M }_{\text{res}}^{*}\right)}+\delta_{ab}~2\text{Re}\overline{\left(
\mathcal{M}_{\text{res,1}}^{\phantom{*}} \mathcal { M }_{\text{res,2}}^{*}\right)}\, \right] .
\end{align}
In the different flavor cases the third term in \eqref{xsec-qrad-dr} does not appear. Comparing \eqref{xsec-qrad-dr} with \eqref{xsec-qrad}, it is clear that the removed terms are $\overline{
\left|\mathcal{M}_{\text{res,1}}\right|^2}$ and $\overline{
\left|\mathcal{M}_{\text{res,2}}\right|^2}$. 
In the definition of DR given in \cite{Frixione:2008yi} also the interference term  $2\text{Re}\overline{\left(
\mathcal{M}_{\text{nonres}}^{\phantom{*}} \mathcal { M }_{\text{res}}^{*}\right)}$ is removed 
(with a study of  the impact of the inclusion of this contribution),
whereas we keep this interference term. 
Although the DR scheme  formally violates gauge invariance, a consistent description is
achieved when the procedure presented here is combined 
with off-shell contributions.
It should not be forgotten that the narrow-width approximation, both in the DR and the DS scheme
is not an exact description in any case; as an approximation it has a natural uncertainty arising 
from missing off-shell contributions and non-factorizable NLO corrections. 
These will be studied in a forthcoming paper.
In our numerical results we basically employ the DR scheme, however, we
compare it with results in the DS scheme, 
both for inclusive K-factors and for differential distributions. \\

Finally, for the practical calculation 
of the real quark radiation contributions, in both schemes one has to perform the
phase space integration over the final state quark.
The squared non-resonant terms
lead, as mentioned before, to initial state collinear singularities.  
Again, these singular  terms have to be subtracted
since they are factorized and absorbed into the PDFs.
Like in the case of gluon radiation, we divide the emission of a quark into a collinear
and a non-collinear region (since no IR singularities occur,  a
separation into soft and hard quark emission is not required),
\bee\label{qradcollenon}
d\sigma^{\text{real-quark}}_{pp\rightarrow\sq_{ia}\sq_{jb}\bar{q}^{(\prime)}} =
\sum_{k=i,j}\frac{1}{1+
\delta_{i,j}}\left[d\sigma^{\text{coll-quark}}_{pp\to\sq_{ia}\sq_{jb}\bar{q}_{k}}
+d\sigma^{\text{noncoll-quark}}_{ pp\to\sq_{ia}\sq_{jb}\bar{q}_{k}}\right]\, .
\eee
The non-collinear contribution 
\begin{align}\label{qrad-split}
d\sigma^{\text{noncoll-quark}}_{pp\to\sq_{ia}\sq_{jb}\bar{q}_{j/i}}=\int_{\tau_{0}}^{1}
d\tau \, \mathcal{L}^{\text{noncoll-quark}}_{i/j}(\tau) \;
d\hat{\sigma}_{q_{i/j}g\rightarrow\sq_{ia}\sq_{jb}\bar{q}_{j/i}}(\tau)\, ,
\end{align}
contains $\mathcal{L}^{\rm noncoll-quark}_{i}(\tau)$ as given in \eqref{lumnoncollqrad}. 
The collinear emission together with the subtraction terms for the PDFs instead can be written
as follows,
\begin{align}\label{qrad-hadr}
d\sigma^{\text{coll-quark}}_{pp\to\sq_{ia}\sq_{jb}\bar{q}_{j/i}}=
 \int_ { \tau_{0}}^{1}d\tau\int_{\tau}^{1}
\frac{dx}{x}\int_{x}^{1}\frac{dz}{z}\,
\mathcal{L}^{\text{coll-quark}}_{i/j}(\tau,x,z) \;
d\hat{\sigma}^{\text{coll-quark}}_{q_{i/j}g\rightarrow\sq_{ia}\sq_{jb}\bar{q}_{j/i}}(\tau,z) \nonumber\, ,\\
\end{align}
with $\mathcal{L}^{\text{coll-quark}}_{i}(\tau,x,z)$ and
$d\hat{\sigma}^{\text{coll-quark}}_{q_{i/j}g\rightarrow\sq_{ia}\sq_{jb}\bar{q}_{j/i}}(\tau,z)$ 
defined in \eqref{lumcollqrad} and \eqref{qrad-part}
of Appendix \ref{app:soft-collinear}.\\



\section{Squark decay}
\label{sec:decay}

\subsection{Squark decay at LO}
\label{sec:decayLO}
The LO decay width for a squark decaying into a neutralino and a quark, 
$\sq_{ia}\rightarrow q_{i}\neu_{j}$, 
depends on the flavour and chirality of the squark.  
For $m_{q}=0$ the width can be written as follows,
\begin{align}
\label{LOdwn}
\Gamma^{(0)}_{\sq_{ia}\rightarrow q_{i} {\neu_{j}}} =
\frac{\alpha}{4}m_{\sq_{ia}}
\Big(1-\frac{{m^2_{\neu_{j}}}}{{m^2_{\sq_{ia}}}}\Big) f_{a}^{2}\, .
\end{align}
The coupling constants $f_{a}$ can be expressed in terms of the isospin $I^{q}_{3L}$ 
and the charge $e_{q}$ of the quark, together with
the neutralino mixing matrix $(N_{jk})$ including the electroweak
mixing angle through
$s_{W}=\sin\theta_W$ and $c_{W}=\cos\theta_W$ , 
\begin{align}
f_{L}=&\sqrt{2}\Big[e_{q}N'_{j1}+(I^{q}_{3L}-e_{q}s_{W}^{2}
)\frac{1}{c_{W}s_{W}}N'_{j2}\Big]\, ,\\
f_{R}=&-\sqrt{2}\Big[e_{q}N'_{j1}-e_{q}\frac{s_{W}}{c_{W}}N'_{j2}\Big]\, ,\\
N'_{j1}=&c_{W}N_{j1}+s_{W}N_{j2},\qquad N'_{j2}=-s_{W}N_{j1}+c_{W}N_{j2}\, .
\end{align}
For a scalar particle decaying in its rest frame 
there is no preferred direction, and hence the  
differential decay distribution is isotropic. 
For squark decays into neutralino and quark, 
the decay distribution is thus simply given by  
\begin{align}
\label{diff.decay.born}
d\:\!\Gamma^{(0)}_{\sq\rightarrow q\neu_{j}}
\, = \,  \frac{1}{4\pi}\, \Gamma^{(0)}_{\sq\rightarrow q\neu_{j}} \;
d\:\!\!\cos\;\!\!\theta\; d\phi
\end{align}
with polar angle $\theta$ and azimuth $\phi$ referring to 
the quark momentum.

\subsection{NLO squark decay distribution}
\label{sec:decayNLO}
The differential decay width for $\sq \to q\neu_{j}$  
at NLO is obtained in analogy to the steps in section \ref{sec:prodNLO}
by adding the virtual loop corrections
and the real gluon bremsstrahlung contribution
from the soft, collinear, and hard non-collinear phase space regions,
yielding the full NLO contribution in the form  
\bee\label{NLO-general-decay}
d\:\!\Gamma^{(1)}_{\sq\to q\neu_{j} }=
d\:\!\Gamma^{\text{virtual}}_{\sq\to q\neu_{j}}+d\:\!\Gamma^{\text{soft}}_{\sq\to q\neu_{j}(g)}
+d\:\!\Gamma^{\text{coll}}_{\sq\to q\neu_{j}(g)}+d\:\!\Gamma^{\text{hard}}_{\sq\to q\neu_{j}g}\, .
\eee
The virtual corrections  $d\Gamma^{\text{virtual}}_{\sq\to q\neu_{j}}$ 
for $m_{q}=0$ correspond to the two vertex loop diagrams 
in \figref{fig:decayNLO}(a) and the vertex counter term
(indicated by the cross in \figref{fig:decayNLO}(a)),   
which consists of the wave-function renormalization constants 
of the external quark and squark line.
As for the production amplitudes, the renormalization constants are 
determined in the on-shell renormalization scheme. 
 Details on the vertex counter term
can be found in~\cite{Hollik:2008yi}, 
and the explicit analytical expression is given 
in \eqref{Fg} of Appendix \ref{app:decay-an}.

\FIGURE[t]{
\centering
\includegraphics{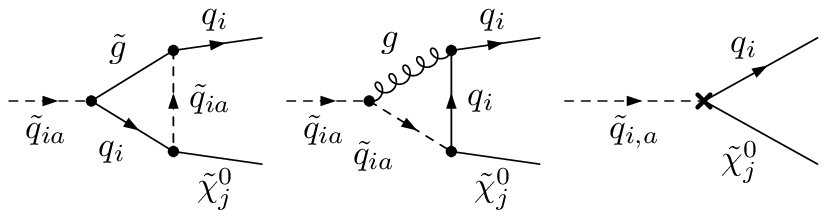} 
\hspace{1.1cm}
\includegraphics{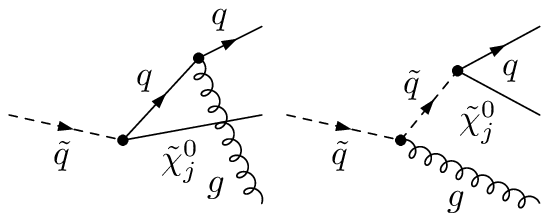}
\small (a) \hspace*{7.8cm} (b) 
\caption{Loop and counterterm diagrams (a) and gluon radiation diagrams (b) for squark decays.}
\label{fig:decayNLO}
}

\smallskip
The term $d\:\!\Gamma^{\text{soft}}_{\sq\to q \neu_{j}(g)}$ can be calculated by the same
strategy as for $d\hat{\sigma}^{\text{soft}}_{qq'\to\sq\sq' (g)}$, yielding
\begin{align}
\label{dGammasoft}
  d\Gamma^{\rm soft}_{\sq \to q\neu_{j}(g)}  \,= \,
-\frac{2\alphas}{3\pi} \, 
 \Big\{  \sum_{i,j=1;i<j}^2 \epsilon_i \epsilon_j\, \mathcal{I}_{ij} \Big\} \, 
 d\:\! \Gamma^{(0)}_{\sq \to q\neu_{j}}  \, ,
\end{align}
where the $\mathcal{I}_{ij}$ are defined in \eqref{Iij2}.

Collinear divergences now emerge from the final state. Making again use of the
results of~\cite{Baur:1998kt}, the  collinear  emission of gluons 
with  energy larger than $\Delta E$ into a cone with opening angle $\Delta\theta$
yields the contribution
\begin{align}
d\:\!\Gamma^{\rm coll}_{\sq\to q \neu_{j}(g)} \, = & \,
d\:\!\Gamma^{(0)}_{\sq\to q \neu_{j}}   \\ \nonumber
& \cdot \frac{2\alpha_s}{3 \pi} \, 
  \left[ \frac{9}{2} - \frac{2}{3} \pi^2 -
    \frac{3}{2} \,\log\left(\frac{2E_{q,\text{max}}^2\,\delta_{\theta}}{m_{q}^2}\right)
   +2 \log\left(\delta_{s}\right) 
  \left(1-\log\left(\frac{2 E_{q,\text{max}}^2\,\delta_{\theta}}{m_{q}^2}\right)\right)\right]\, ,
\end{align}
where  $\delta_{s}=2\Delta E/m_{\sq}$,
$\delta_{\theta}=1-\cos\left(\Delta\theta\right)\simeq\Delta\theta^2/2$,  and
$E_{q,\text{max}}=\frac{m_{\sq}^{2}-m_{\neu_{j}}^2}{2m_{\neu_{j}}}$, the maximum energy
available for the quark in the squark rest frame.
Gluons with $\theta<\Delta \theta$ are recombined with the emitter
quark into a quark with momentum $p_{\text{recomb}}=p_{q}+p_{g}$. In general, differential
distributions in the quark momenta 
are dependent on the slicing parameter $\Delta \theta$.
However, this dependence will disappear once a jets algorithm 
that is much more inclusive in the recombination of quarks and
gluons, is applied (see section \ref{sec:numerics}).

The contribution $d\Gamma^{\text{hard}}_{\sq\to q \neu_{j}}$ from real emission of hard
gluons ($E_{g}>\Delta E$) at large angles ($\theta>\Delta \theta $) is evaluated
by numerical integration of the squared matrix elements, obtained from the diagrams in 
\figref{fig:decayNLO}(b).

\subsection{Total decay width}
\label{sec:decayTOT}

The total squark decay width $\Gamma^{(0)}_{\sq}$ at LO is obtained by summing the partial decay widths
of the 6 different possible decay channels into neutralinos and charginos (assuming always $\mgl > \msq$). 
The partial decay widths into neutralinos are given directly  by \eqref{LOdwn}. 
For charginos, the partial decay widths $\Gamma^{(0)}_{\sq \to q'\tilde{\chi}^{\pm}_{j}}$ are also described 
by the formula~(\ref{LOdwn}),  with the specification $f_{R}=0 $ and  
\begin{align}
\label{LOchar}
f_{L}=\frac{V_{j1}}{s_{W}} \quad \text{for} \quad \sq=\tilde{u}, \tilde{c}, \qquad 
f_{L}=\frac{U_{j1}}{s_{W}} \quad \text{for} \quad \sq=\tilde{d}, \tilde{s},
\end{align}
for the coupling constants. Here, $U,V$ are the mixing matrices in the chargino sector.

\smallskip
For the total decay width at NLO, one has  to calculate the NLO 
QCD corrections for each channel, which can be done analytically 
performing the full phase space integration over the three-particle
final state with the radiated gluon and adding the loop contributions.
The six partial decay widths contributing to $\Gamma^{(0+1)}_{\sq}$ at NLO
can be expressed in terms of their respective LO result and a
NLO form factor $F^{QCD}$, \eqref{Fr}, which depends only on the mass ratios
of the involved massive particles,
%
%
\begin{eqnarray}
\label{eq:FQCD}
& & \Gamma^{(0+1)}_{\sq \to q \neu_{j}/ q'\tilde{\chi}^{\pm}_{j}}   \,=\,
\Gamma^{(0)}_{\sq \to q \neu_{j}/ q'\tilde{\chi}^{\pm}_{j}} \, \left[1+\frac{4}{3}\frac{\alphas}{\pi} \,
F^{QCD}\left(\frac{m_{\neu_{j}/\tilde{\chi}^{\pm}_{j}}}{m_{\sq}}, \frac{m_{\sq}}{m_{\tilde{g}}}\right)\right] \, .
\end{eqnarray}
The result for $F^{QCD}$ in\cite{Djouadi:1996wt} is confirmed by our independent  
derivation. Details can be found in Appendix~\ref{app:decay-an}.
In the threshold limit
$\frac{m_{\neu_{j}/\tilde{\chi}^{\pm}_{j}}}{m_{\sq}} \to 1$ the
correction in~\eqref{Fr} becomes very large and the fixed-order calculation is not reliable
(relative corrections remain finite only after proper resummation,
 see for example~\cite{Horsky:2008yi}).
However, for all parameter points considered in the numerical evaluation in 
section \ref{sec:numerics} the corrections are still sufficiently small and 
threshold problems are negligible.


\section{Phenomenological evaluation}
\label{sec:numerics}
Let us now turn to the numerical evaluation of the process under consideration.
In the following, we first specify in \ref{sec:inputs} the input
parameters and benchmark scenarios we consider in our numerical evaluation. Then, in 
\ref{sec:observables} a brief introduction to all considered distributions and observables 
is given. Finally, in \ref{sec:results}, we present our numerical results in three different 
ways. Firstly, we compare our results for inclusive production of squark--squark pairs with
\Prospino and investigate inclusive K-factors for different chirality and flavour 
combinations. Secondly, we present several differential distributions for various 
benchmark scenarios and center-of-mass energies. Thirdly, we investigate the impact 
of the considered higher-order corrections on total event rates and thus on cut-and-count 
searches for supersymmetry at the LHC.

\subsection{Input parameters }
\label{sec:inputs}
%
Standard Model input parameters are chosen according to \cite{Nakamura:2010zzi},
\begin{align}
M_Z & = 91.1876~{\GeV}\,,& M_W & = 80.399~\GeV\,,& G_F &=1.16637\cdot 10^{-5}~\GeV\,,
\nonumber\\
m_b^{\msbar}(m_b) & = 4.2~\GeV\,,&
m_t & = 173.3~{\GeV}\,,&  m_{\tau} & = 1.777~{\GeV}\,.
\label{eq_SMinputs}
\end{align}
We use the PDF sets CTEQ6.6~\cite{Nadolsky:2008zw} 
interfaced via the LHAPDF package~\cite{Whalley:2005nh} 
both for LO and NLO contributions. The strong coupling constant
$\alphas^{\msbar}(\mu_R)$ is also taken from this set of PDFs. Factorization scale
$\mu_F$ and renormalization scale $\mu_R$ are, if not stated otherwise, set to a common
value, $\mu=\mu_F=\mu_R=\overline{\msq}$, with $\overline{\msq}$ being the average mass
of all light-flavour squarks of a given benchmark point.\\
For SUSY parameters, we refer to three different benchmark scenarios. First, we
investigate the well studied CMSSM parameter point SPS1a \cite{Allanach:2002nj}.
Although being practically excluded by recent searches at the 
LHC~\cite{Chatrchyan:2011zy,Aad:2011ib,Dolan:2011ie}, 
this point still serves as a valuable benchmark to
compare with numerous numerical results available in the literature.
Second, we study the benchmark point CMSSM10.1.5 introduced in \cite{AbdusSalam:2011fc}.
Due to its larger \mhalf parameter, compared to SPS1a, squark and gluino masses
are considerably larger, resulting in a generally reduced production cross section at the
LHC. Not excluded yet, this parameter point can be tested in the near future. The
overall spectrum, though shifted to larger masses, is very similar to the one of SPS1a.
Third, we consider a phenomenological benchmark point defined at the scale $Q=1~\TeV$.
We follow the definitions of \cite{AbdusSalam:2011fc}, where such a point sits on a line called 
p19MSSM1. It can be parametrized by essentially one parameter, the 
gaugino mass parameter $M_1$. A unified parameter for the gluino and the 
light-generations sfermion soft masses $M_3=m_{\tilde f^{\text{1st/2nd gen}}_{L/R}}$ is fixed
to $M_3=m_{\tilde f^{\text{1st/2nd gen}}_{L/R}}=1.2~M_1$ on this line and we choose 
$M_1=300~\GeV$ for our benchmark scenario p19MSSM1A. All other masses and 
parameters as well as the soft masses for the first two generation right-handed 
sleptons are at a higher scale and irrelevant for our analysis.
This benchmark point is chosen to study a particular parameter region with rather light
squarks and gluinos which is difficult to exclude experimentally. Due to a small mass splitting between the $\LSP$ and the light squarks (and gluino) resulting jets tend to be very soft and thus escape the experimental analyses. Particularly in such parameter regions precise theoretical prediction of the resulting SUSY signal including
higher orders on the level of distributions seems to be necessary for a conclusive study.  \\
%
%

\TABULAR[ht]{c||c|c|c|c|c}{
  \hline\trule
\bf{benchmarkpoint}   & $\boldsymbol{m_0}$
  & $\boldsymbol{m_{1/2}}$
  & $\boldsymbol{A_0}$
  & $\boldsymbol{\tan\beta}$
  & $\boldsymbol{{\rm sign}(\mu)}$\\
  \hline
  SPS1a & 100~\GeV & 250~\GeV & $-$100~\GeV & 10 & $+$\\
  \hline
  10.1.5 & 175~\GeV & 700~\GeV & 0 & 10 & $+$\\
  \hline
}{High energy input parameters for the two considered CMSSM scenarios. 
\label{tab:cmssmBenchmarks}
}
\TABULAR[ht]{c||c|c|c|c|c|c}{
  \hline\trule
\bf{benchmarkpoint}   & $\boldsymbol{M_1}$
  & $\boldsymbol{M_2}$
  & $\boldsymbol{M_3}$
  & $\boldsymbol{A_i}$
  & $\boldsymbol{\tan\beta}$
  & $\boldsymbol{{\rm sign}(\mu)}$\\
  \hline
  p19MSSM1A & 300~\GeV & 2500~\GeV & 360~\GeV & 0 & 10 & $+$\\
  \hline
}{Low energy input parameters for the p19MSSM1A scenario. The first two
generation sfermion soft-masses (apart from the right handed sleptons) equal the gluino
mass $m_{\tilde f^{\text{1st/2nd gen}}_{L/R}}=M_3$. All other parameters are at a higher
scale $m_{\tilde f^{\text{3rd gen}}_{L/R}} = m_{\tilde e_R, \tilde\mu_R} = \mu = M_A =
M_2$.
\label{tab:p19Benchmark}
}
\TABULAR[ht]{c||c|c|c|c|c|c}{
  \hline\trule
\bf{benchmarkpoint}   & $\boldsymbol{\uL}$
  & $\boldsymbol{\uR}$
  & $\boldsymbol{\dL}$
  & $\boldsymbol{\dR}$
  & $\boldsymbol{\gluino}$
  & $\boldsymbol{\neu_1}$
\\
  \hline
  SPS1a & 563.6 & 546.7 & 569.0 &  546.6 & 608.5 & 97.0\\
   \hline
  10.1.5 & 1437.7 & 1382.3 &  1439.7 &  1376.9 & 1568.6 & 291.3 \\
  \hline
  p19MSSM1A & 339.6 &  394.8 & 348.3 & 392.7 & 414.7 &  299.1   \\
  \hline
}{On-shell masses of the squarks, the gluino, and the lightest neutralino within the
different SUSY scenarios considered. All masses are given in \GeV.
\label{tab:lowmasses}  
  }

\TABULAR[ht]{cc||c|c|c|c|c}{
\hline\trule
\bf{benchmarkpoint}   &
  & $\boldsymbol{\uL}$
  & $\boldsymbol{\uR}$
  & $\boldsymbol{\dL}$
  & $\boldsymbol{\dR}$
  & $\boldsymbol{\gluino}$
\\
  \hline
  SPS1a&$\Gamma^{(0)}$ & $5.361$ & $1.148$ & $5.253$ & $0.287$ & $6.849$\\
  &$\Gamma^{(0+1)}$ & $5.357$ & $1.131$ & $5.255$ & $0.283$ & \\
  \hline
  10.1.5&$\Gamma^{(0)}$& $12.47$ &$ 2.854 $& $12.46$ & $0.710$ & $10.04$ \\
  &$\Gamma^{(0+1)}$& $12.31$ & $2.821$ & $12.30$ & $0.702$ &  \\  
  \hline
  p19MSSM1A&$\Gamma^{(0)}$ &$2.414\cdot 10^{-3}$& $0.1625$ & $ 3.411\cdot 10^{-3}$ &
$3.917\cdot 10^{-2}$& $ 3.441$\\
  &$\Gamma^{(0+1)}$ &$2.497\cdot 10^{-3}$ & $0.1621$ & $3.503\cdot 10^{-3}$ & $3.912\cdot
10^{-2}$& \\
\hline
}{Leading order $\Gamma^{(0)}$ and next-to-leading order $\Gamma^{(0+1)}$ total widths of light squarks 
and gluino for the considered SUSY scenarios. All widths are given in \GeV.
\label{tab:width}
  }

Parameters of the CMSSM benchmark scenarios are defined universally at the GUT scale 
and are shown in \tabref{tab:cmssmBenchmarks}. They act as boundary conditions for the
renormalization group running of the soft-breaking parameters down to the SUSY scale
\msusy. This running is performed with the program \softsusy \cite{Allanach:2001kg} which
also calculates physical on-shell parameters for all SUSY mass eigenstates. We
use the resulting on-shell parameters directly as input for our calculation.
Low scale soft input parameters for the p19MSSM1A benchmark scenario are given in
\tabref{tab:p19Benchmark}. The physical spectrum is equivalently calculated with \softsusy.
For all considered benchmark scenarios we summarize relevant low energy physical masses in
\tabref{tab:lowmasses}. Due to non-vanishing Yukawa corrections implemented in \softsusy
the physical on-shell masses for second-generation squarks are slightly different from
their first-generation counterparts. To simplify our numerical evaluation we set
all second-generation squark masses to their first-generation counterparts. We checked
that in the results this adjustment is numerically negligible. However, the general setup of
our calculation is independent of this choice. For all considered benchmark scenarios the
gluino is heavier than all light flavour squarks $\mgl > \msq$. Thus, these squarks
decay only into electroweak gauginos and quarks. For SPS1a and 10.1.5, right-handed squarks
dominantly decay directly into the lightest neutralino \LSP\ (due to its bino nature). In
contrary, left-handed squarks decay dominantly into heavier (wino-like) neutralinos and
charginos, which subsequently decay via cascades into a \LSP, and only a small
fraction decays directly into a \LSP. In this paper we only investigate the direct decay
of any light flavour squark into the lightest neutralino. For point p19MSSM1A all
neutralinos and charginos, but the lightest one, are heavier than any light-flavour
squark. Thus, only the direct decay is allowed and all channels contribute equally to the
signature under consideration.

In \tabref{tab:width} we list all needed total decay widths of the squarks at LO and
NLO, calculated as explained in section \ref{sec:decay}. NLO corrections in
the total decay widths are of the order of a few percent for all three benchmark
scenarios~\footnote{
Numerically we observed a disagreement with the partial
decay widths at NLO for p19MSSM1A obtained from \sdecay,
despite the fact that the NLO contributions in \sdecay
are based on the analytical results
calculated in \cite{Djouadi:1996wt} and in Appendix \ref{app:decay-an}.
After corresponding with the authors, this problem was solved by
correcting a typo in \sdecay. We thank M. M\"uhlleitner for
helpful discussions.
}
. 
The total decay width of the gluino is calculated with \sdecay at LO and also
listed in \tabref{tab:width}.
In the calculation presented here this width is not used explicitly. Instead, we numerically employ
the limit $\Gamma_{\gluino} \to 0$. However we checked that, using the physical widths, all 
the results showed in the following present negligible differences.

Besides physical quantities, in our calculation phase-space slicing and regulator parameters 
enter as inputs in the calculation of virtual and real NLO contributions, as explained in 
sections~\ref{sec:prod} and~\ref{sec:decay}. 
In the results shown in this article we set $\delta_s = 2\cdot10^{-4}$,
$\delta_{\theta} = 10^{-4}$ and $m_q = 10^{-1.5}~\GeV$, both for production and decay. 
Numerically we checked carefully that varying their values our results remain unchanged on the level of individual
distributions once jets are recombined using a clustering algorithm, as explained below. We made sure that this holds
for all terms of \eqref{masterformel} individually.\\

\pagebreak
\subsection{Observables and kinematical cuts}
\label{sec:observables}
The physical signature we have in mind when calculating on-shell squark production
and decay into the lightest neutralino is $2j+\missingET(+X)$. Numerical results of
our calculation are presented in this spirit. In order to arrive at an experimentally well defined
two-jet-signature we always employ the anti-$k_{T}$\cite{Cacciari:2005hq} jet-clustering
algorithm implemented in \fastjet \cite{Cacciari:2011ma}. Thus, we provide a realistic
prediction on the level of partonic jets~\footnote{With the term partonic jets we mean 
that the jet-clustering-algorithm has been applied to events as produced from our calculation. 
No QCD showering or hadronization is included in the simulation.}.
In general we use a jet radius of
$R=0.4$, as in the SUSY searches performed by the ATLAS collaboration \cite{Aad:2011ib}.
CMS instead uses a radius of $R=0.5$ \cite{Chatrchyan:2011zy}. We employ $R=0.5$ in the
distributions and signatures used by CMS (\ie particularly the \alphat distribution as
described below). Although  we did not perform a systematic study, our results seem to be
independent of this choice. After performing the jet clustering we sort the partonic jets by
their transverse momentum \pT, and in the following analysis we keep only jets with
\begin{align}
\pT_{j_{1/2}} &> 20~\GeV , \hspace{.5cm} \vert \eta_j \vert < 2.8 , \label{basecuts}\\
\pT_{j_i} &> 50~\GeV,  \hspace{.5cm}  \vert \eta_j \vert < 3.0~(\text{for CMS observables})  \label{basecutsCMS} \, .
\end{align}
Cuts of \eqref{basecuts} are used everywhere but in the observables used specifically by
CMS ($\alpha_T$, as defined below), where cuts of \eqref{basecutsCMS} are applied.

\smallskip
Before showing results for the experimental signature $2j+\missingET(+X)$,
we compare in section \ref{sec:inclsuivesigma} values for NLO total cross
sections of squark-squark production, without decay included, with results obtained from \Prospino.
In section \ref{sec:differential} we investigate the effect of NLO corrections, for different benchmark points, on the following differential
distributions:
\begin{itemize}
 \item the transverse momentum of the two hardest jets $\pT_{1/2}$,
 \item the pseudorapidity of the two hardest jets $\eta_{1/2}$,
 \item the missing transverse energy \missingET, 
 \item the effective mass $\meff = \sum\limits_{i=1,2} \pT_{i} + \missingET$,
 \item the scalar sum of the \pT of all jets (visible after cuts of \eqref{basecutsCMS}),  $H_T = \sum\limits_{i=1,2(,3)}
\pT_i$,
 \item the invariant mass of the two hardest jets $m_{\text{inv}}(jj)$,
 \item the cosine of the angle between the two hardest jets $\cos\Theta_{jj}$, which
depends on the spin of the produced particles and therefore might help to distinguish SUSY
from other BSM models \cite{Hallenbeck:2008hf},
 \item $\cos{\hat\Theta} =\text{tanh}\left( \frac{\Delta\eta_{jj} }{2} \right)$,
$\Delta\eta_{jj}=\eta_1 - \eta_2$, introduced in \cite{MoortgatPick:2011ix} as a possible
observable for early spin determination at the LHC,
 \item the \alphat variable, first defined in \cite{Randall:2008rw}, where for hard real
radiation events with three jets and $\pT_{3}>50~\GeV$, these jets are reclustered into 
two pseudojets by minimizing the difference of the respective $H_T$ of the two pseudojets, as 
explained in \cite{Khachatryan:2011tk,Allanach:2011ut}. Furthermore, in all \alphat distributions we 
require $H_T > 350~\GeV$ as in \cite{Khachatryan:2011tk}. 
\end{itemize}
Searches for sparticle production performed by ATLAS are based on $\pT$, $\missingET$ and
$\meff$ cuts; CMS instead uses $\alphat$ to reduce SM backgrounds. In section
\ref{sec:rates} we examine NLO corrections in the resulting event rates after cuts.
Explicitly we employ the following cuts used by ATLAS,
\begin{align}
\label{atlascuts}
\pT_{j_1} > 130~\GeV,~  \pT_{j_2} > 40~\GeV,&~ \vert \eta_{j_{1/2}}\vert < 2.8 ,~
\Delta\phi(j_{1/2},\vec\missingET) > 0.4, \\
\meff > 1~\TeV,~& \missingET / \meff > 0.3, \nonumber  
\end{align}
in their two-jet analysis \cite{Aad:2011ib}. Here, $\Delta\phi(j_{1/2},\vec\missingET)$ denotes the angular 
separation between the two hardest jets and the direction of missing energy.
Instead the CMS signal region \cite{Khachatryan:2011tk} is defined as
\begin{align}
\label{cmscuts}
\pT_{j_{1/2}} > 100~\GeV,~  \vert \eta_{j_{1}}\vert < 2.5,~ \vert \eta_{j_{2}}\vert < 3.0, \\
H_T > 350~\GeV,~ \displaystyle{\not}H_T / \missingET < 1.25 ,~ \alphat > 0.55, \nonumber
\end{align}
where $\displaystyle{\not}H_T$ is calculated from $\vec H_T$.

\subsection{Results}
\label{sec:results}


\subsubsection{Inclusive cross sections}
\label{sec:inclsuivesigma}

In \tabref{tab:inclusive} we list inclusive LO cross sections and corresponding NLO
K-factors for the three benchmark scenarios SPS1a, 10.1.5, p19MSSM1A, varying the LHC center of mass
energy $\SqrtS=7, 8, 14~\TeV$. K-factors, here and in the following, are always defined as ratios
between NLO and LO predictions, where both, as stated above, are calculated using the same
NLO PDFs and associated $\alphas$. On the one hand, we list inclusive cross sections, 
$\sigma^{(0)}_{pp\to \sq \sq'}$, and K-factors for just the production of squark-squark 
pairs summed over all flavour and chirality combinations. These K-factors are calculated
in the DR scheme defined in section \ref{sec:quarkrad}. If not stated otherwise this scheme is
used in the following. On the other hand, we list cross section predictions 
for combined production and decay at LO $\sigma^{(0)}_{2j+\Slashed{E}_T(+X)}$ and the 
corresponding NLO K-factors $K_{2j+\Slashed{E}_T(+X)}$ calculated using \eqref{masterformel}, 
again summed over all flavour and chirality combinations. Here, cuts defined in 
\eqref{basecuts} are applied and cross sections are strongly decreased due to the 
branching into the lightest neutralino. All K-factors of the combined process are 
bigger than the corresponding K-factors of inclusive production. For point 10.1.5 
(and thus rather heavy squarks) these increments are less than $5~\%$. For the 
other two benchmark points (and thus smaller squark masses) increments in the 
K-factors can be of the order of $10~\%$ and increase with higher center of mass 
energies. In general, however, K-factors decrease with higher center of mass energies 
and increase with higher masses, both, for inclusive production and combined
production and decay.\\
 
\TABULAR[ht]{c|c||r|c||c|c}{
\hline
   \bf{benchmark}  
& $\boldsymbol{\sqrt{S}}$~[TeV] &
$\boldsymbol{\sigma^{(0)}_{pp\to \sq \sq'}}$ & $\boldsymbol{K^{DR}_{pp\to \sq
\sq'}}$  &
$\boldsymbol{\sigma^{(0)}_{2j+\Slashed{E}_T(+X)}}$  &
$\boldsymbol{K_{2j+\Slashed{E}_T(+X)}}$
\\
\hline
\hline
	&$7$	& $1.02$\pba	& $1.37$& $0.37$\pba & $1.41$	\\
SPS1a	&$8$	& $1.49$\pba	& $1.35$& $0.53$\pba & $1.40$ \\
	&$14$	& $5.31$\pba	& $1.28$& $1.74$\pba & $1.36$ \\
\hline	
	&$7$	& $0.90$\fba	& $1.57$& $0.45$\fba & $1.61$ \\
10.1.5	&$8$	& $2.62$\fba	& $1.52$& $1.24$\fba & $1.56$\\	
	&$14$	& $50.04$\fba	& $1.40$& $20.41$\fba & $1.44$	\\
\hline
	&$7$	& $7.90$\pba	& $1.40$& $6.31$\pba & $1.50$ \\
p19MSSM1A&$8$	& $10.48$\pba	& $1.39$& $8.35$\pba & $1.50$ \\
	&$14$	& $29.01$\pba	& $1.34$& $22.60$\pba & $1.47$\\
\hline
}{LO cross sections $\sigma^{(0)}_{pp\to \sq \sq'}$ and NLO
K-factors for inclusive squark-squark production, $K^{DR}_{pp\to
\sq \sq'}$, LO cross sections of inclusive combined squark-squark production 
and decay $\sigma^{(0)}_{2j+\Slashed{E}_T(+X)}$ and corresponding 
K-factor $K_{2j+\Slashed{E}_T(+X)}$, for the three benchmark 
scenarios SPS1a, 10.1.5, p19MSSM1A and center of mass energies $\SqrtS=7,8,14~\TeV$. In
the combined predictions cuts of \eqref{basecuts} are applied. \label{tab:inclusive} }

In \tabref{tab:1015channels} we compare the inclusive production with combined production
and decay for benchmark point 10.1.5 and a center of mass energy of $\SqrtS=14~\TeV$.
Here, we list results for individual flavour and chirality combinations. In general, our
calculation is set up to treat all $36 + c.c$  possible flavour and chirality
combinations independently. However, for simplicity and to save computing time we always
sum combinations with identical masses and matrix elements into 16 channels, both, for 
production and in the combination. This categorization follows the four 
possibilities discussed in section \ref{sec:prodLO}. For example, the channel $\tilde u_L\tilde u_L$ also
includes $\tilde c_L\tilde c_L$ (and as everywhere else in this paper,  the charge
conjugated processes). Similarly, the channel $\tilde u_L\tilde d_L$ also includes $\tilde
c_L\tilde s_L$, $\tilde u_L\tilde s_L$ and $\tilde c_L\tilde d_L$; and the channel $\tilde
u_L\tilde c_R$ also includes $\tilde u_R\tilde c_L$.
K-factors, both, in just the production and in the combined result, vary by up to $15~\%$
between different channels. Thus, an independent treatment seems adequate, as in general,
squarks of different chiralities and thus different channels have very different decays and 
kinematic distributions. This can easily be seen from the very different order of magnitude of
the various values of $\sigma^{(0)}_{2j+\Slashed{E}_T(+X)}$ in \tabref{tab:1015channels}. As already 
seen in \tabref{tab:inclusive}, K-factors increase comparing inclusive production and the 
combined result (where the cuts given in \eqref{basecuts} are applied). \\

\TABULAR[ht]{c||c|c|c||c|c|c}{
\hline
  \bf{channel}  & $\boldsymbol{\sigma^{(0)}_{pp\to \sq \sq'}}$ &
$\boldsymbol{\sigma^{(0+1)}_{pp\to \sq \sq'}}$ & $\boldsymbol{K^{DR}_{pp\to \sq \sq'}}$
			                         &
$\boldsymbol{\sigma^{(0)}_{2j+\Slashed{E}_T(+X)}}$  &
$\boldsymbol{\sigma^{(0+1)}_{2j+\Slashed{E}_T(+X)}}$  &
$\boldsymbol{K_{2j+\Slashed{E}_T(+X)}}$ \\
& [fb]&[fb]&&[fb]&[fb]&
\\
\hline
\hline
$\tilde u_L\tilde u_L$	&$7.08$		&$9.44$		&$1.33$	& $1.22\cdot 10^{-3}$   & $1.68\cdot 10^{-3}$   & $1.38$ \\
$\tilde u_R\tilde u_R$	&$8.64 $		&$11.5 $		&$1.33$	& $8.25$ & $11.36$     		& $1.38$ \\
$\tilde d_L\tilde d_L$	&$1.07 $		&$1.44 $ 		&$1.36$	& $2.82\cdot 10^{-4}$  & $3.96\cdot 10^{-4}$  & $1.40$ \\
$\tilde d_R\tilde d_R$	&$1.39 $		&$1.88 $		&$1.35$	& $1.33$ & $1.84$   			& $1.39$ \\
$\tilde u_L\tilde u_R$	&$6.00 $		&$8.49 $		&$1.42$	& $7.78\cdot 10^{-2}$   & $11.33\cdot 10^{-2}$  & $1.45$ \\
$\tilde d_L\tilde d_R$    &$8.20\cdot 10^{-1}$    &$1.19  $      	&$1.45$  & $1.32\cdot 10^{-2}$  & $1.96\cdot 10^{-5}$   & $1.49$ \\
$\tilde u_L\tilde d_L$	&$8.25 $		&$11.9 $		&$1.44$	& $1.76\cdot 10^{-3}$   & $2.62\cdot 10^{-3}$   & $1.49$ \\
$\tilde u_R\tilde d_R$	&$ 10.5 $		&$15.1 $		&$1.44$	& $10.00$ & $14.92$     		& $1.49$  \\
$\tilde u_L\tilde c_L$	&$ 3.28\cdot 10^{-1}$	&$4.33\cdot 10^{-1}$	&$1.32$	& $5.65\cdot 10^{-5}$   & $7.73\cdot 10^{-5}$   & $1.37$ \\
$\tilde u_R\tilde c_R$	&$ 4.29\cdot 10^{-1}$	&$5.74\cdot 10^{-1}$	&$1.34$	& $4.09\cdot 10^{-1}$   & $5.68\cdot 10^{-1}$   & $1.39$ \\
$\tilde d_L\tilde s_L$	&$ 1.95\cdot 10^{-1}$	&$2.75\cdot 10^{-1}$	&$1.41$	& $5.16\cdot 10^{-5}$   & $7.5097\cdot 10^{-5}$ & $1.46$ \\
$\tilde d_R\tilde s_R$	&$ 2.71\cdot 10^{-1}$	&$3.87\cdot 10^{-1}$ 	&$1.42$	& $2.59\cdot 10^{-1}$   & $3.82$    		& $1.48$ \\
$\tilde u_L\tilde d_R$	&$ 2.44 $		&$3.50 $		&$1.44$	& $3.16\cdot 10^{-2}$   & $4.67\cdot 10^{-2}$   & $1.48$ \\
$\tilde u_R\tilde d_L$	&$ 2.40 $		&$3.46 $		&$1.44$	& $3.87\cdot 10^{-2}$   & $5.70\cdot 10^{-2}$   & $1.48$ \\
$\tilde u_L\tilde c_R$	&$ 1.69\cdot 10^{-1}$   &$2.39\cdot 10^{-1}$	&$1.41$	& $2.19\cdot 10^{-3}$   & $3.18\cdot 10^{-3}$   & $1.46$ \\
$\tilde d_L\tilde s_R$	&$ 9.51\cdot 10^{-2}$	&$1.39\cdot 10^{-1}$	&$1.46$	& $1.52\cdot 10^{-3}$   & $2.29\cdot 10^{-3}$	& $1.50$ \\
\hline
sum	& $50.04$		&$69.86$		&$1.40$ & $20.41$ & $29.32$		& $ 1.44$ \\
\hline 
}{For the benchmark point 10.1.5 and a center of mass energy of $\SqrtS=14~\TeV$
inclusive production cross sections at LO $\sigma^{(0)}_{pp\to \sq \sq'}$ and NLO
$\sigma^{(0+1)}_{pp\to \sq \sq'}$ together with the corresponding K-factors $K_{pp\to \sq
\sq'}$ are listed for all different flavour and chirality channels (as explained in
the text). Also listed for all channels are LO $\sigma^{(0)}_{2j+\Slashed{E}_T(+X)}$ and
NLO $\sigma^{(0+1)}_{2j+\Slashed{E}_T(+X)}$ predictions of combined production and decay
and the corresponding K-factor $K_{2j+\Slashed{E}_T(+X)}$, where the cuts of
\eqref{basecuts} are applied. All cross sections are given in fb.
\label{tab:1015channels}}

In \tabref{tab:compare} we compare the different schemes defined in \ref{sec:quarkrad}. 
In order to also consistently compare with \Prospino results, we set, just here, 
the mass of all squarks to the average mass $\overline{\msq}$ for all bechmark points.
In \tabref{tab:compare} we show LO cross-sections and NLO K-factors from our calculation in the DR scheme, 
$K^{DR}_{pp\to\sq \sq'}$, and in the DS scheme, $K^{DS}_{pp\to\sq \sq'}$. We also 
list K-factors obtained from \Prospino, $K^{\text{Prospino}}_{pp\to \sq \sq'}$, where we 
adjusted \Prospino to use the same set of PDFs and definition of the strong coupling 
$\alpha_s(\mu_R)$ as in our calculation. The use of an average mass results in a small shift
in the LO cross section and also in the NLO K-factor $K^{DR}_{pp\to\sq \sq'}$ between \tabref{tab:inclusive}
and \tabref{tab:compare}.
Numerical differences between K-factors in the DR scheme and the DS scheme
are of the order of a few percent for SPS1a and p19MSSM1A and negligible for 10.1.5, as, for
a heavier spectrum the gluon contribution in the PDFs is suppressed. Differences between 
$K^{DS}_{pp\to\sq \sq'}$ and $K^{\text{Prospino}}_{pp\to \sq \sq'}$ originate in the different
on-shell subtraction approach. We checked numerically, excluding real quark radiation 
altogether, that inclusive NLO corrections from our calculation and results from 
\Prospino are in perfect agreement.\\

\TABULAR[ht]{c|c||r|c|c|c}{
\hline
   \bf{benchmark}  
& $\boldsymbol{\sqrt{S}}$~[TeV] &
$\boldsymbol{\sigma^{(0)}_{pp\to \sq \sq'}}$ & $\boldsymbol{K^{DR}_{pp\to \sq
\sq'}}$ & $\boldsymbol{K^{DS}_{pp\to \sq \sq'}}$ & $\boldsymbol{K^{\text{Prospino}}_{pp\to \sq \sq'}}$ 
\\
\hline
\hline
	&$7$	& $1.01$\pba & $1.37$ &$1.39$  & $1.41$ \\
SPS1a	&$8$	& $1.48$\pba & $1.35$ 	& $1.38$ & $1.40$ \\
	&$14$	& $5.31$\pba & $1.28$ & $1.34$ & $1.38$ \\
\hline	
	&$7$	& $0.89$\fba & $1.58$ & $1.58$ & $1.59$ \\
10.1.5	&$8$	& $2.59$\fba & $1.53$ & $1.53$  & $1.54$  \\	
	&$14$	& $49.87$\fba& $1.39$ &  $1.40$ & $1.41$  \\
\hline
	&$7$	& $7.65$\pba & $1.39$ & $1.41$ & $1.37$ \\
p19MSSM1A&$8$	& $10.17$\pba & $1.37$ & $1.41$ &$1.37$ \\
	&$14$	& $28.34$\pba & $1.31$ & $1.39$ &$1.38$ \\
\hline
}{LO cross sections $\sigma^{(0)}_{pp\to \sq \sq'}$ and NLO
K-factors for inclusive squark-squark production from our computation in the DR scheme, $K^{DR}_{pp\to
\sq \sq'}$, in the DS scheme $K^{DS}_{pp\to\sq \sq'}$ and also from \Prospino, $K^{\text{Prospino}}_{pp\to \sq \sq'}$. 
All squark masses taken to the average squark mass $\overline{\msq}$. \label{tab:compare} }

\subsubsection{Differential Distributions}
\label{sec:differential}

Now we turn to the investigation of differential distributions in various observables.\\

First, we compare the differential scale dependence between our LO and NLO
prediction. We do this by varying at the same time renormalization and factorization scale
between $\mu/2 < \mu < 2 \mu$, with $\mu=\overline{\msq}$. For SPS1a and an
energy of $\SqrtS=14~\TeV$ resulting LO and NLO bands are shown in blue and red in
\figref{fig:bandplotssps1a} for differential distributions in $\pT_1, \pT_2, \eta_1,
\eta_2, \missingET$ and $H_T$. Clearly, in all considered distributions the scale
dependence and thus the theoretical uncertainty is greatly reduced by our NLO calculation.
At the same time, one should also note that large parts of the NLO bands are outside the
LO bands. Still, for example in the \pT distributions, in the high-pT tail the NLO bands
move entirely inside the LO bands.\\

\FIGURE{
\includegraphics[width=.49\textwidth]{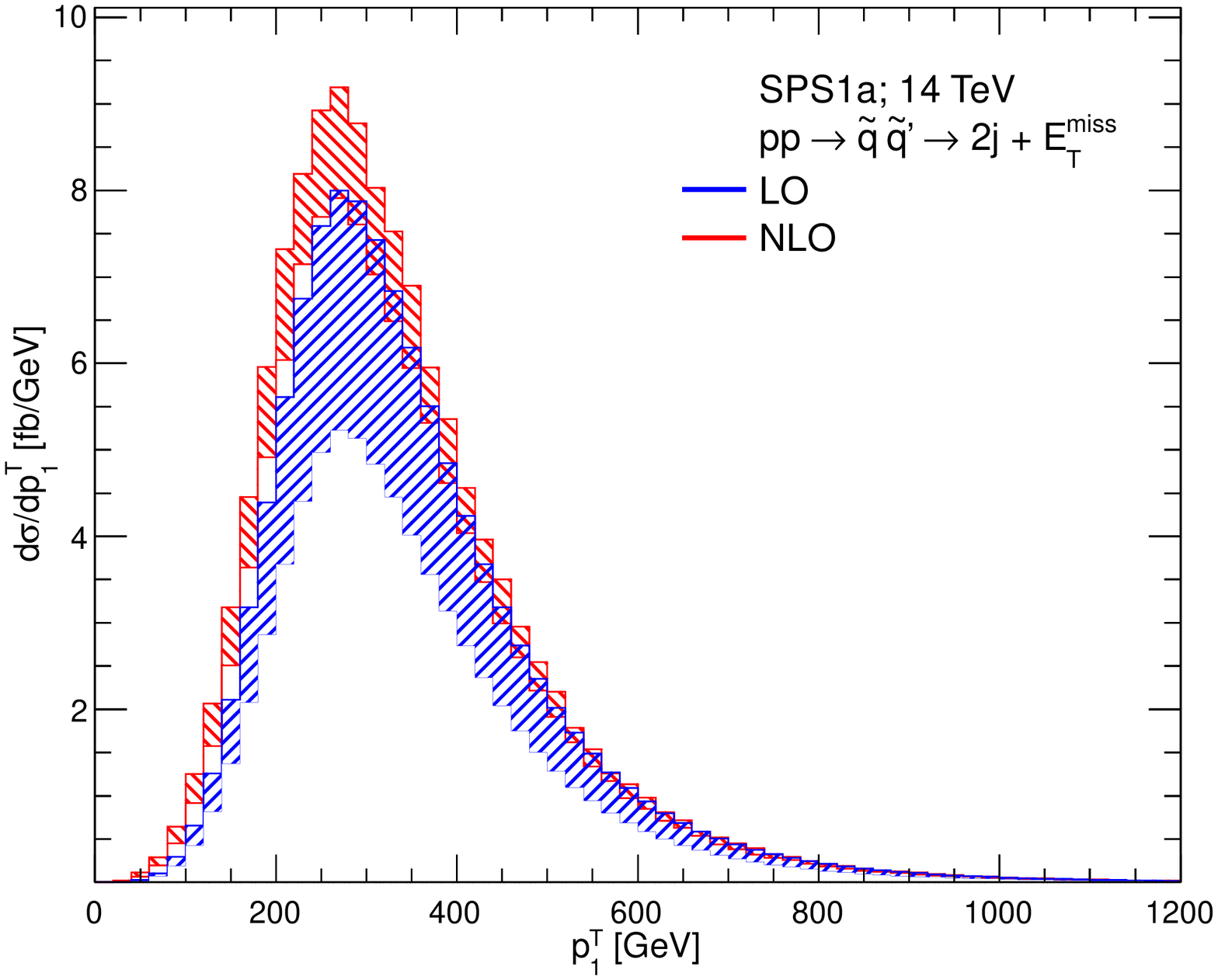}
\includegraphics[width=.49\textwidth]{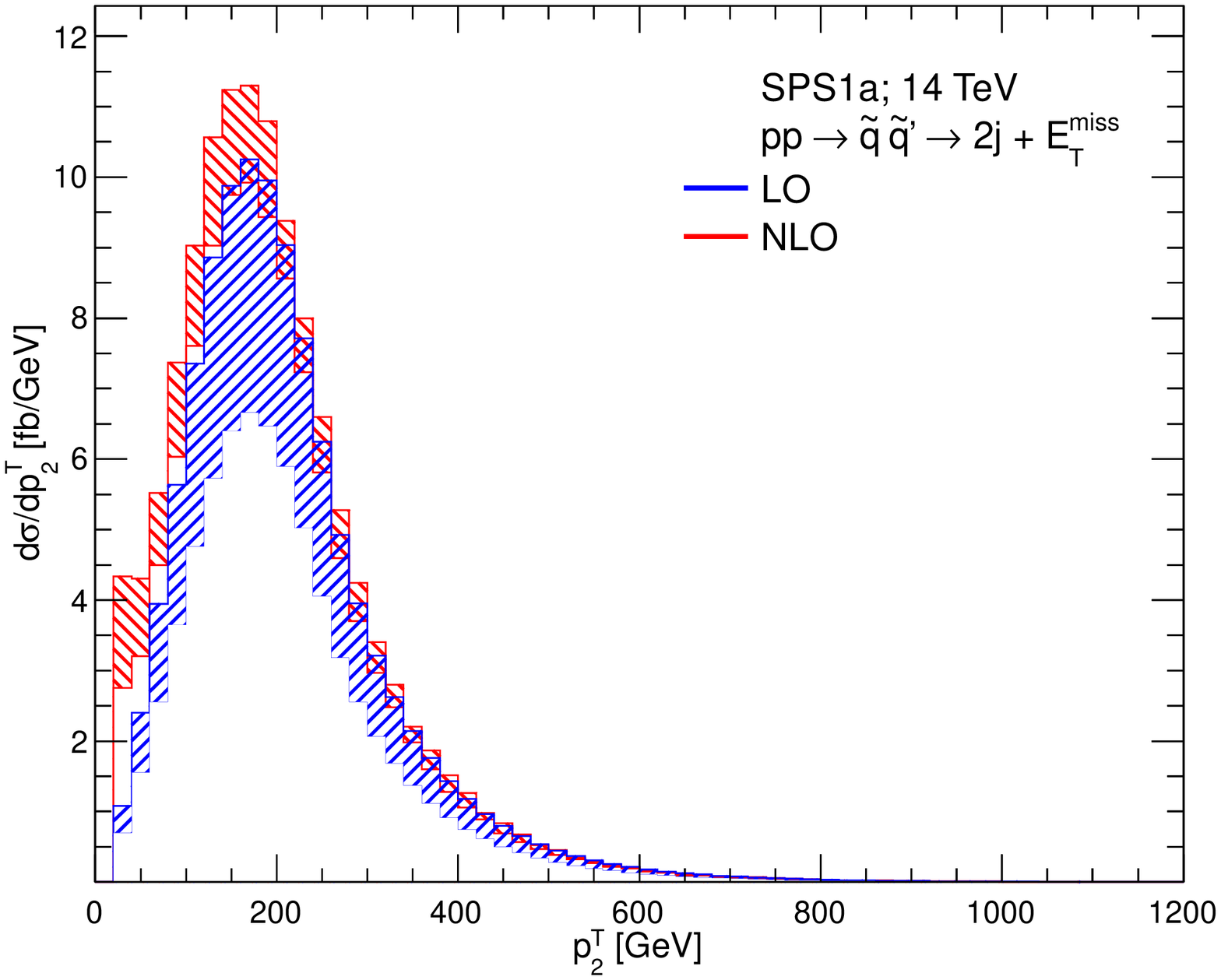}\\
\includegraphics[width=.49\textwidth]{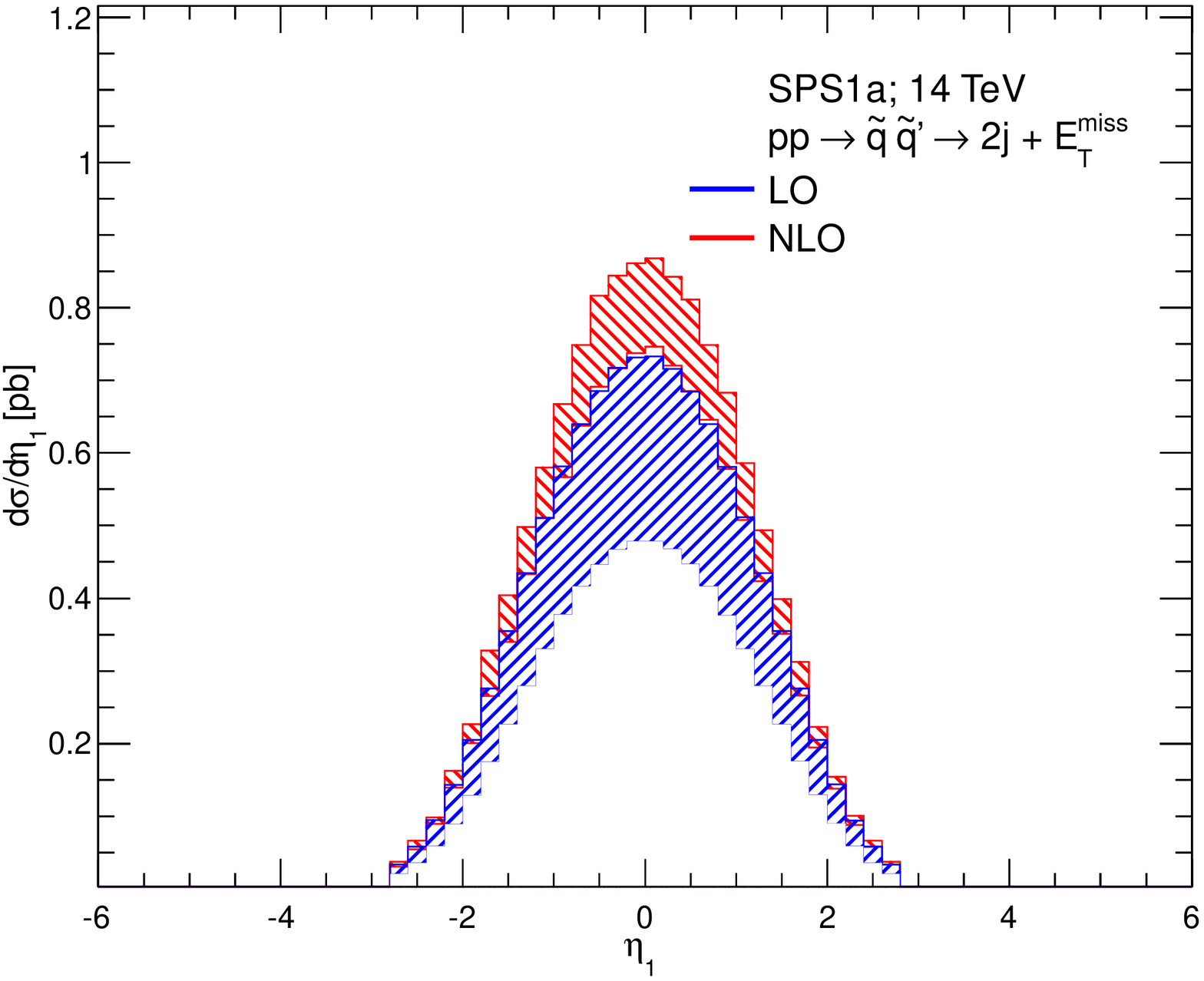}
\includegraphics[width=.49\textwidth]{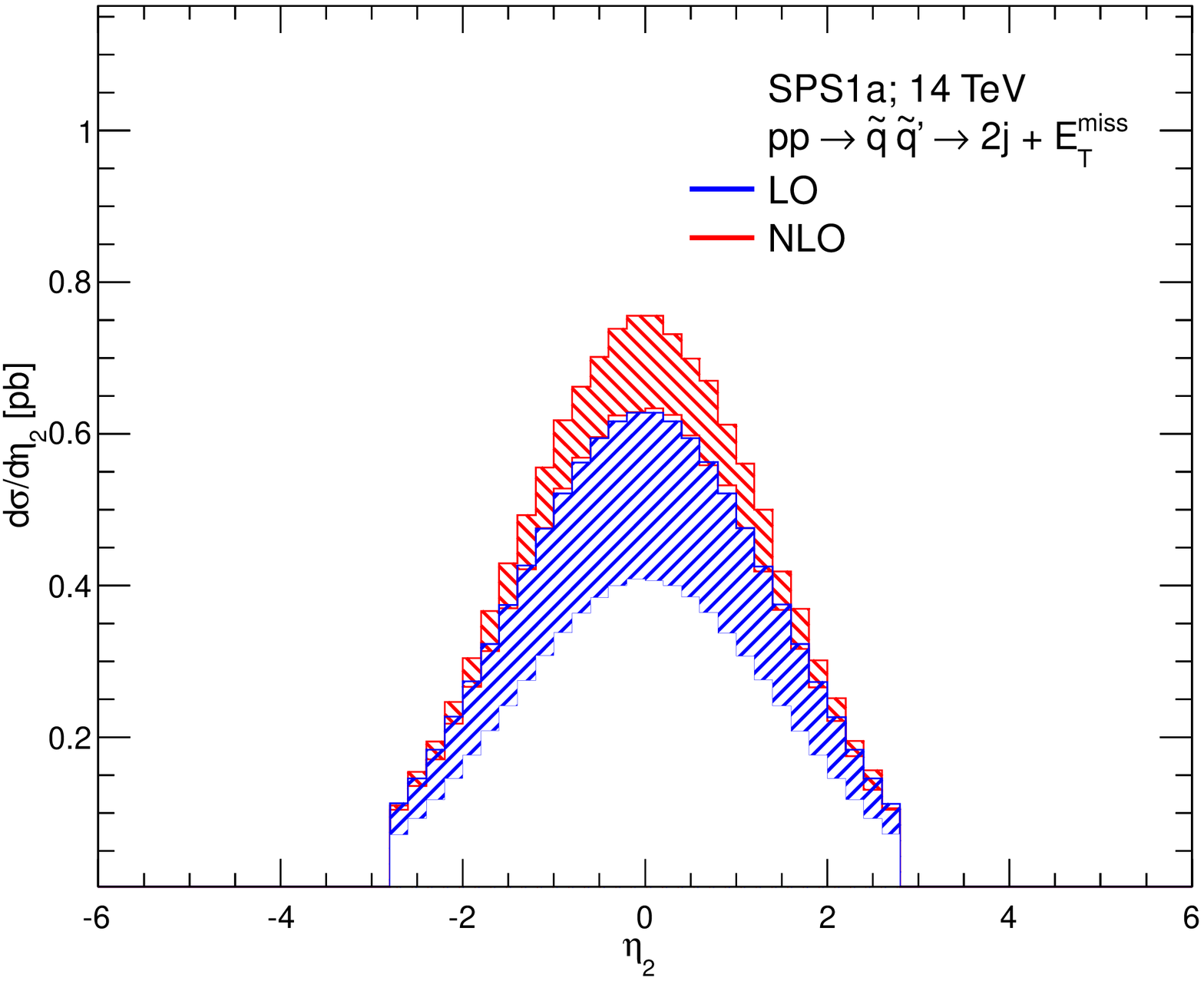}\\
\includegraphics[width=.49\textwidth]{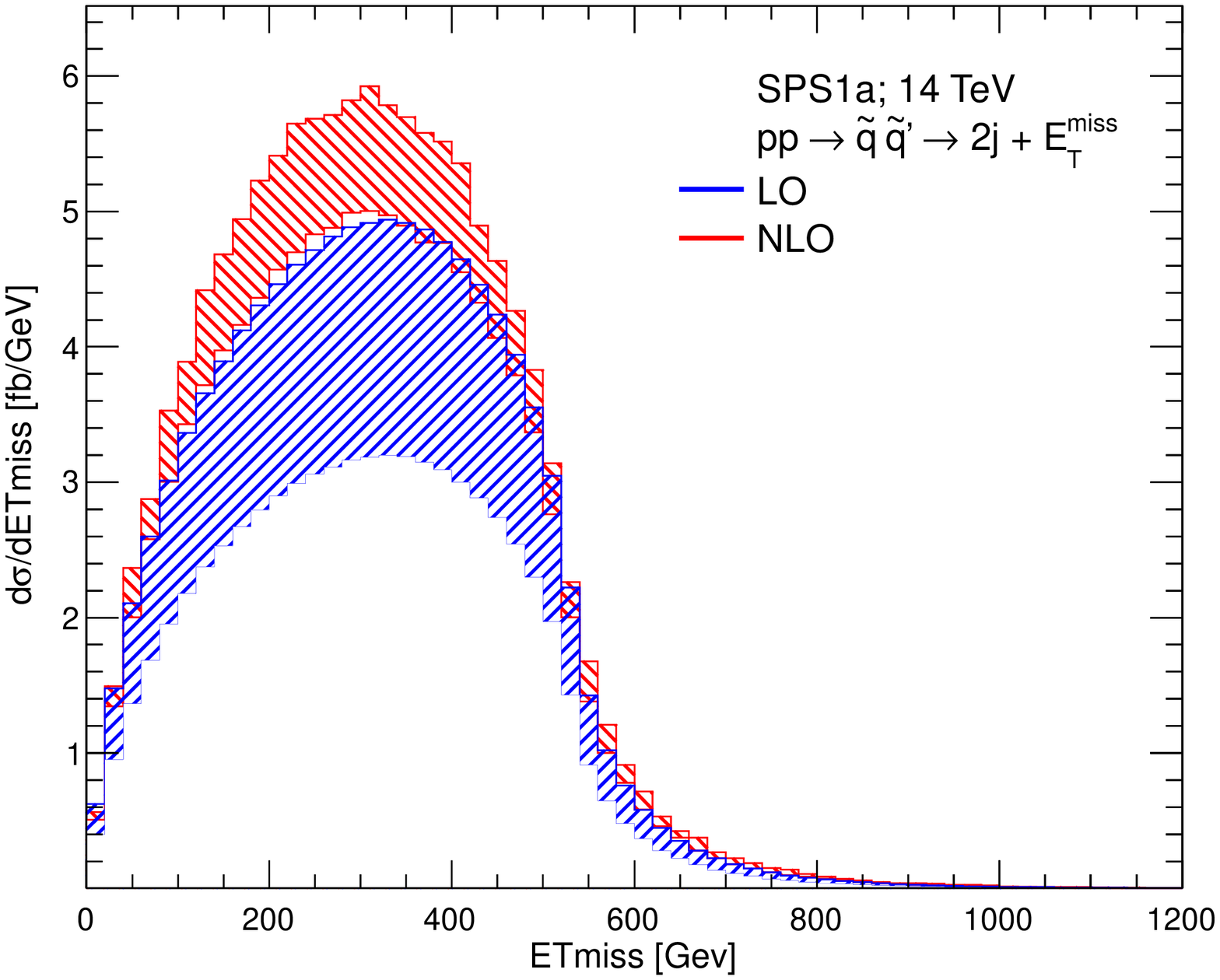}
\includegraphics[width=.49\textwidth]{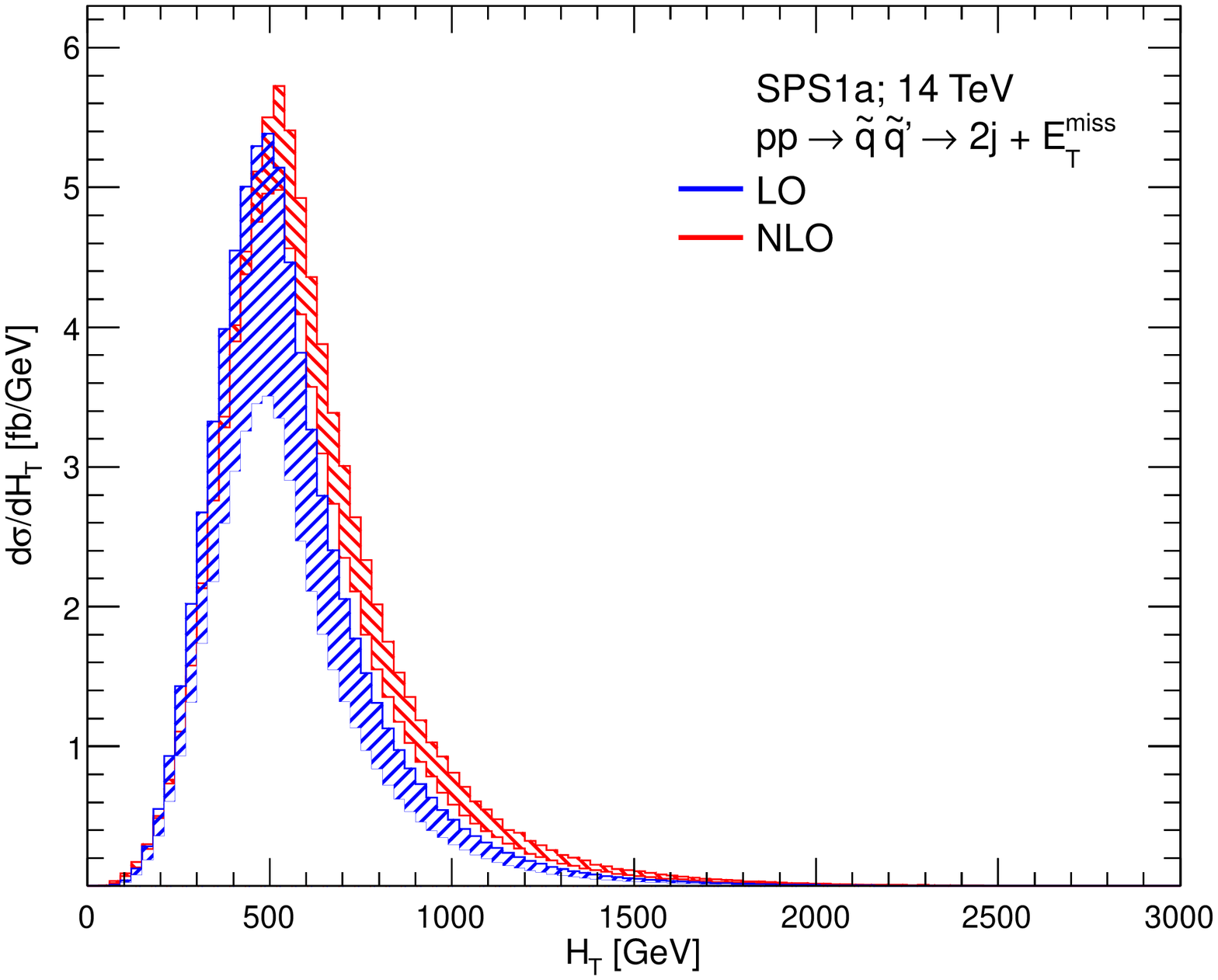}
\caption{Differential distributions in  $\pT_1, \pT_2$ (in fb/GeV), $\eta_1,
\eta_2$ (in pb/GeV), $\missingET$ and $H_T$ (in fb/GeV) for SPS1a and
$\SqrtS=14~\TeV$, where for, both, LO (blue) and NLO (red) the common renormalization and
factorization scale is varied between $\mu/2$ and $2 \mu$, with
$\mu=\overline{\msq}$.
\label{fig:bandplotssps1a}
 }}

Second, in \figref{fig:bandplotsschemes} we illustrate the difference between the schemes introduced
in section \ref{sec:quarkrad} for the benchmark point SPS1a and a center of mass 
energy $\SqrtS=14~\TeV$. In \figref{fig:bandplotsschemes} we show distributions in $\missingET$ and $H_T$. These are the distributions where we observe the largest deviations between the DS and DR schemes.
The upper part of these plots show the same band plots as already displayed at the bottom
of \ref{fig:bandplotssps1a}, however in a log scale. In the lower part we show, for the DR scheme, 
the ratio of the NLO results at $\mu=2\overline{\msq}$ and $\mu=\overline{\msq}/2$ over 
the LO results at $\mu=\overline{\msq}$. We also display the ratio between the NLO result 
in the DS scheme and the LO result, both at the central $\mu=\overline{\msq}$. In these two 
distributions the difference between the two schemes increases in the tail of the distributions. 
However the DS scheme remains within the theoretical uncertainty of the DR scheme. As explained
the chosen distributions $\missingET$ and $H_T$ show the largest differences we observe for the
benchmark points and energies considered.\\

\FIGURE{
\includegraphics[width=.48\textwidth]{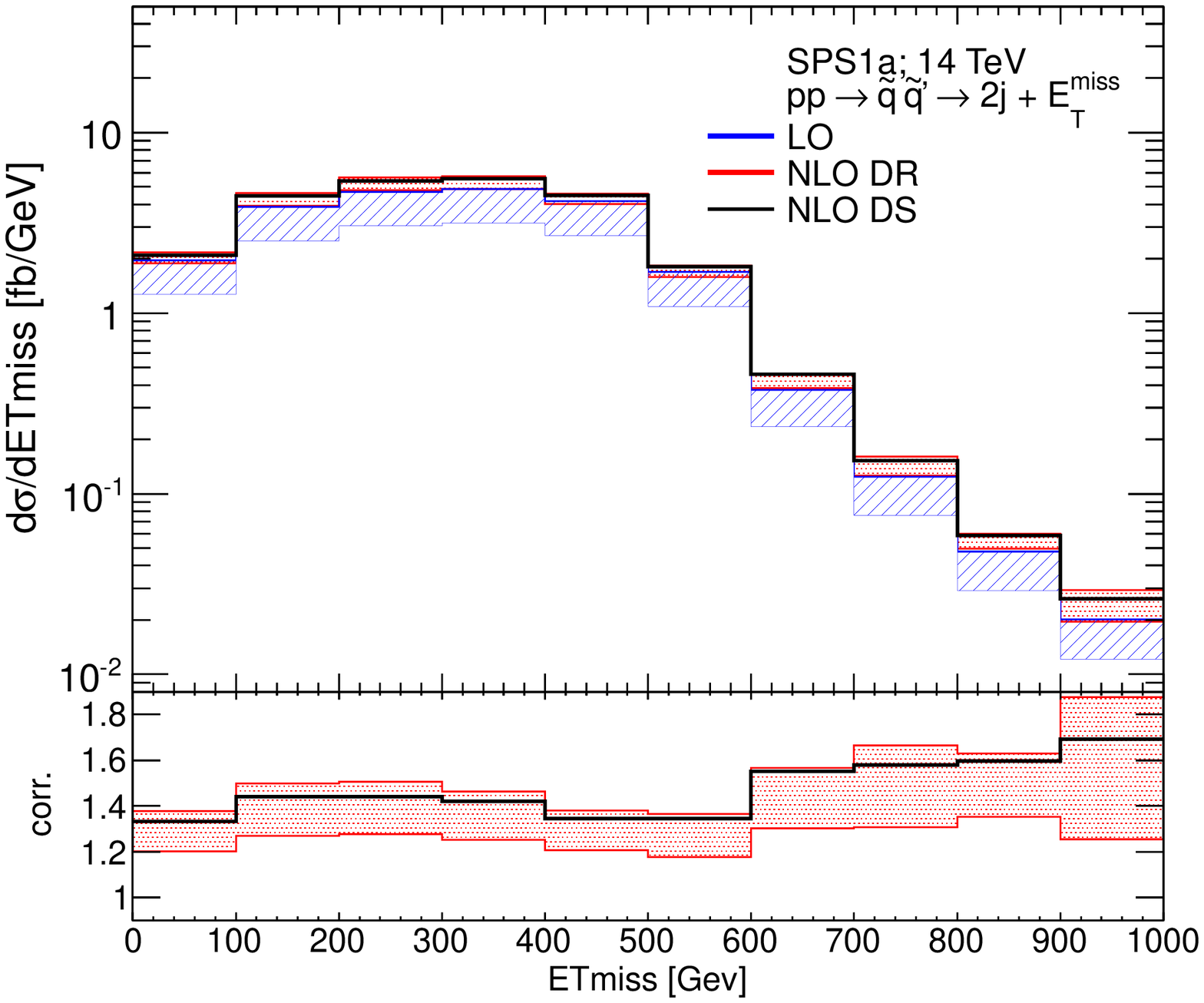}
\includegraphics[width=.48\textwidth]{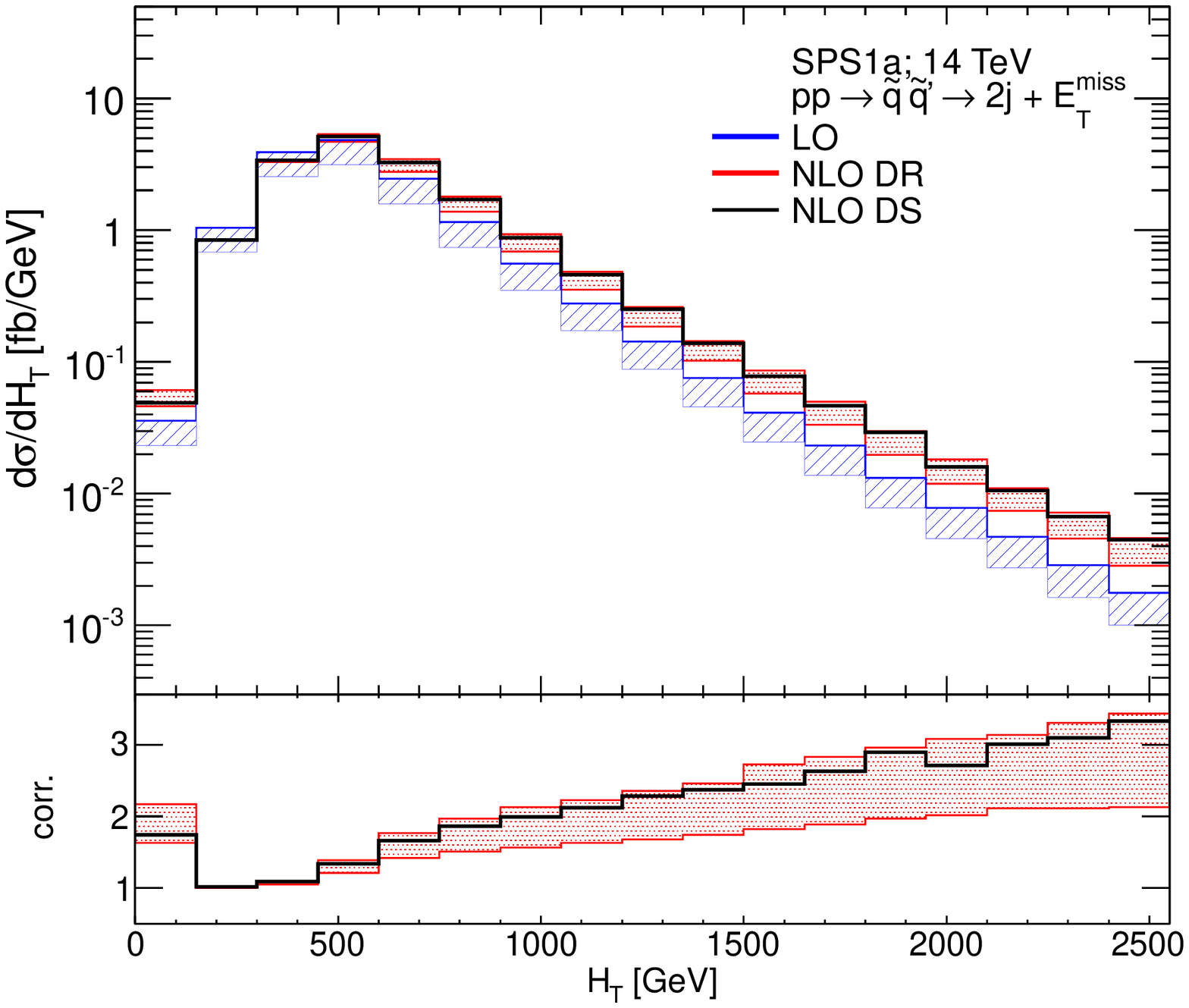}
\caption{Differential distributions in  $\missingET$ and $H_T$ (in fb/GeV) for SPS1a and
$\SqrtS=14~\TeV$. In the upper part the common renormalization and
factorization scale is varied between $\mu/2$ and $2 \mu$, with $\mu=\overline{\msq}$ for LO (blue) and $\text{NLO}^{DR}$ (red). In the
lower part we show in red the ratio of the $\text{NLO}^{DR}$ uncertainty band and the LO result (at scale $\mu=\overline{\msq}$). We also show in black the ratio of the NLO result in the DS scheme and the central LO result.
\label{fig:bandplotsschemes}
 }}

Third, we investigate the change in the shape of distributions relevant for searches for
supersymmetry at the LHC when going from LO to NLO. Here, we present distributions for
a center of mass energy $\SqrtS=14~\TeV$. Lower center of mass energies show qualitatively
the same behaviour. For benchmark point SPS1a plots are shown in \figref{fig:plotssps1a},
for 10.1.5 in \figref{fig:plots1015} and for p19MSSM1A in \figref{fig:plotsp19}. We present
distributions in $\pT_1$, $\pT_2$, $\meff$, $\missingET$ (all in fb/\GeV), where the ATLAS
jet choice $R=0.4$ and cuts of \eqref{basecuts} are applied. Also distributions in $H_{T}$
(in fb/\GeV) and in $\alpha_T$ (in pb) are displayed, where the CMS jet choice $R=0.5$
and corresponding cuts of \eqref{basecuts} are applied. In the $\alpha_T$ distribution,
events are reclustered into two pseudojets and a cut of $H_T>350~\GeV$ is applied. In the
upper 
part of any plot we show each distribution at LO in black, NLO in red and in blue the
LO prediction rescaled by the ratio, $K^{\text{NLO}}$, between the integrated NLO and 
LO result. In the lower part of any plot we show the NLO divided by the rescaled
$\text{LO}\cdot K^{\text{NLO}}$ distribution. In this way we present corrections 
purely in the shape and not in the normalization of the distributions.
For SPS1a and 10.1.5 corrections are qualitatively very similar and rather flat for 
$\pT_1$, $\pT_2$ and $\missingET$, as expected from \cite{Beenakker:1996ch}. Corrections in the (inclusive) $H_T$ distribution grow for larger $H_T$ and can be sizeable. This can be explained from
the high-\pT behaviour of the contribution from hard real gluon radiation to this observable.
Corrections to the shape of the $\alpha_T$ observable change sign at the physical boundary \cite{Randall:2008rw}
$\alpha_T=0.5$ and fall off continuously in the signal region $\alpha_T>0.55$.\\

Looking at the distributions of p19MSSM1A in \figref{fig:plotsp19} a completely different 
behaviour of the NLO corrections cannot be missed. The tail of the $\pT_1$, $\pT_2$, 
$\meff$ and $\missingET$ distributions completely departs from the LO predictions. This 
can be understood from the following considerations. Due to the small mass splitting between 
squarks and the \LSP\ for benchmark point p19MSSM1A, jets from squark decays tend to be soft. 
Now, the \pT of an  additional jet (which can not be distinguished from the decay jets) from 
hard gluon radiation in the production can easily be of the same order as the ones from squark 
decays and result in the given distortions. Such a behaviour for compressed spectra was already partly 
discussed in \cite{Alwall:2008qv}, where sparticle production and decay including additional hard jets matched to a parton shower 
was investigated. We verified our findings by comparing LO predictions plus real hard gluon radiation 
in the production stage with a corresponding calculation performed with \madgraph \cite{Alwall:2011uj}.
The tail of the considered distributions can adequately only be described by additional gluon radiation,
which should thus be seen as the LO prediction for these phase-space regions.  Still, only our
full NLO calculation allows a consistent treatment of the entire distributions. Turning to the $\alpha_T$
distribution, clearly shapes of LO and NLO prediction are different and, here, we refrain from  
showing explicitly corrections in the shape or a rescaled LO prediction.\\

Next to the distributions shown in figures \ref{fig:plotssps1a}, \ref{fig:plots1015} and 
\ref{fig:plotsp19}, we also investigated pseudorapidity distributions of the two hardest jets $\eta_{1/2}$.
Here, in the relevant region $\vert\eta_{1/2} \vert < 3.0$ corrections in the shapes are 
always smaller than about $5~\%$ for all benchmark scenarios and energies. \\

In \figref{fig:angular} we turn our attention towards angular distributions between the
two hardest jets. On the left we show distributions in the invariant mass of the two
hardest jets $m_{\text{inv}}(jj)$, on the right distributions in the cosine of the angle between the two
hardest jets $\cos\Theta_{jj}$ are presented. Again results are shown for all three benchmark points and a
center of mass energy $\SqrtS=14~\TeV$. Corrections in these distributions can be quite large.
In general, in the full NLO results one observes an increase for small angles between
the two hardest jets (up to $20~\%$ in the $\cos\Theta_{jj}$ distributions). In the 
high-invariant-mass tail for SPS1a and 10.1.5 corrections are negative and grow 
to $40~\%$ in the considered invariant mass range. Such corrections could potentially 
be absorbed into a dynamical renormalization/factorization scale definition, \eg, $\mu=H_T$; 
in-detail investigation is left to future work. In the invariant mass distribution of p19MSSM1A we observe
the same deviation of the NLO result from the LO shape as already 
discussed above. \\

Finally, in \figref{fig:thetaprime} we investigate NLO corrections to the
$\cos{\hat\Theta}$ distribution for the benchmark points SPS1a (top left), 10.1.5 (top right) 
and p19MSSM1A (bottom) at a center of mass energy of $\SqrtS=14~\TeV$. Corrections up to
$15~\%$ are observable. Still, the general shape and thus the potential for extraction of
spin information about the intermediate squarks seems to be robust under higher order
corrections.

\FIGURE{
\includegraphics[width=.49\textwidth]{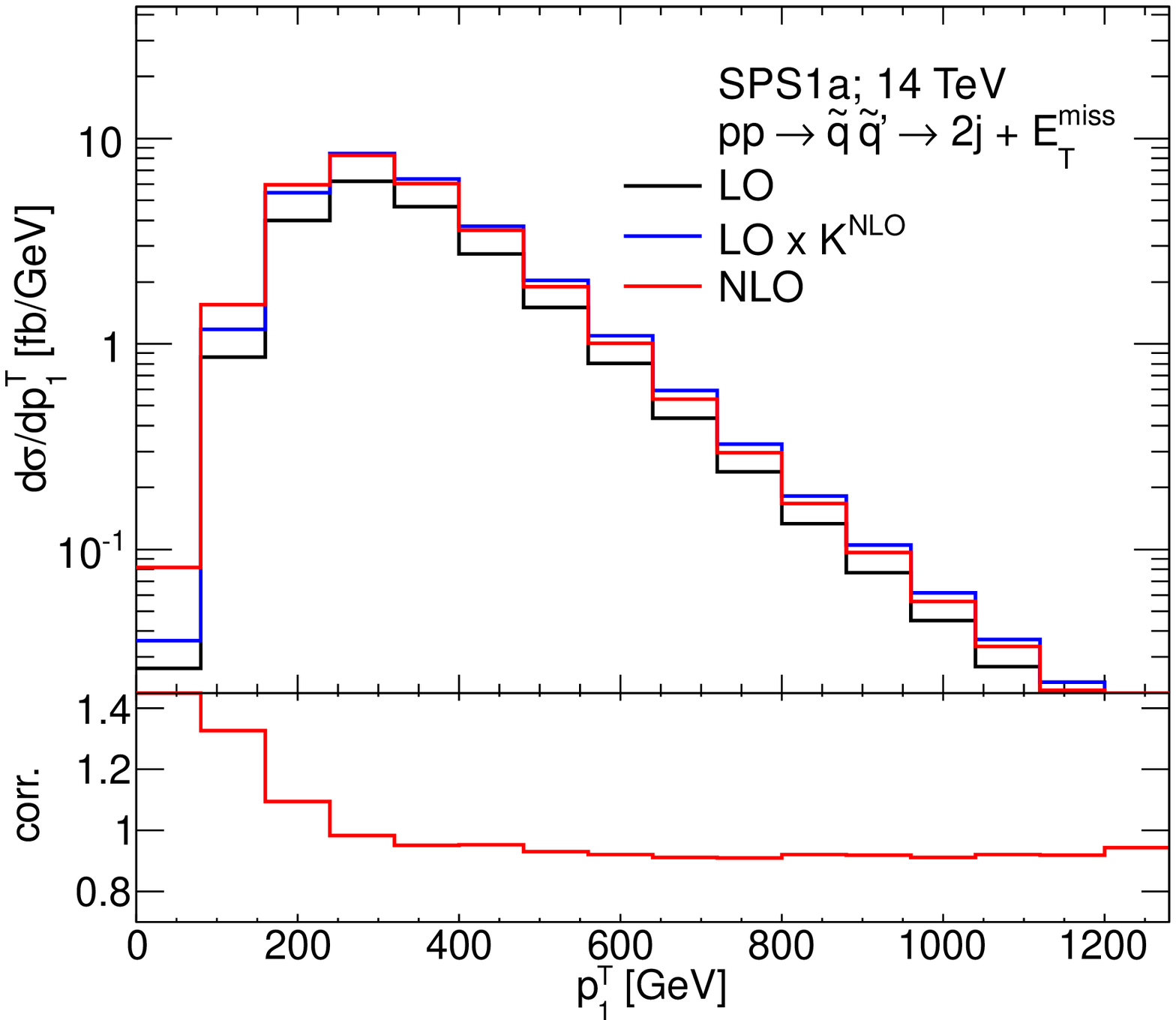}
\includegraphics[width=.49\textwidth]{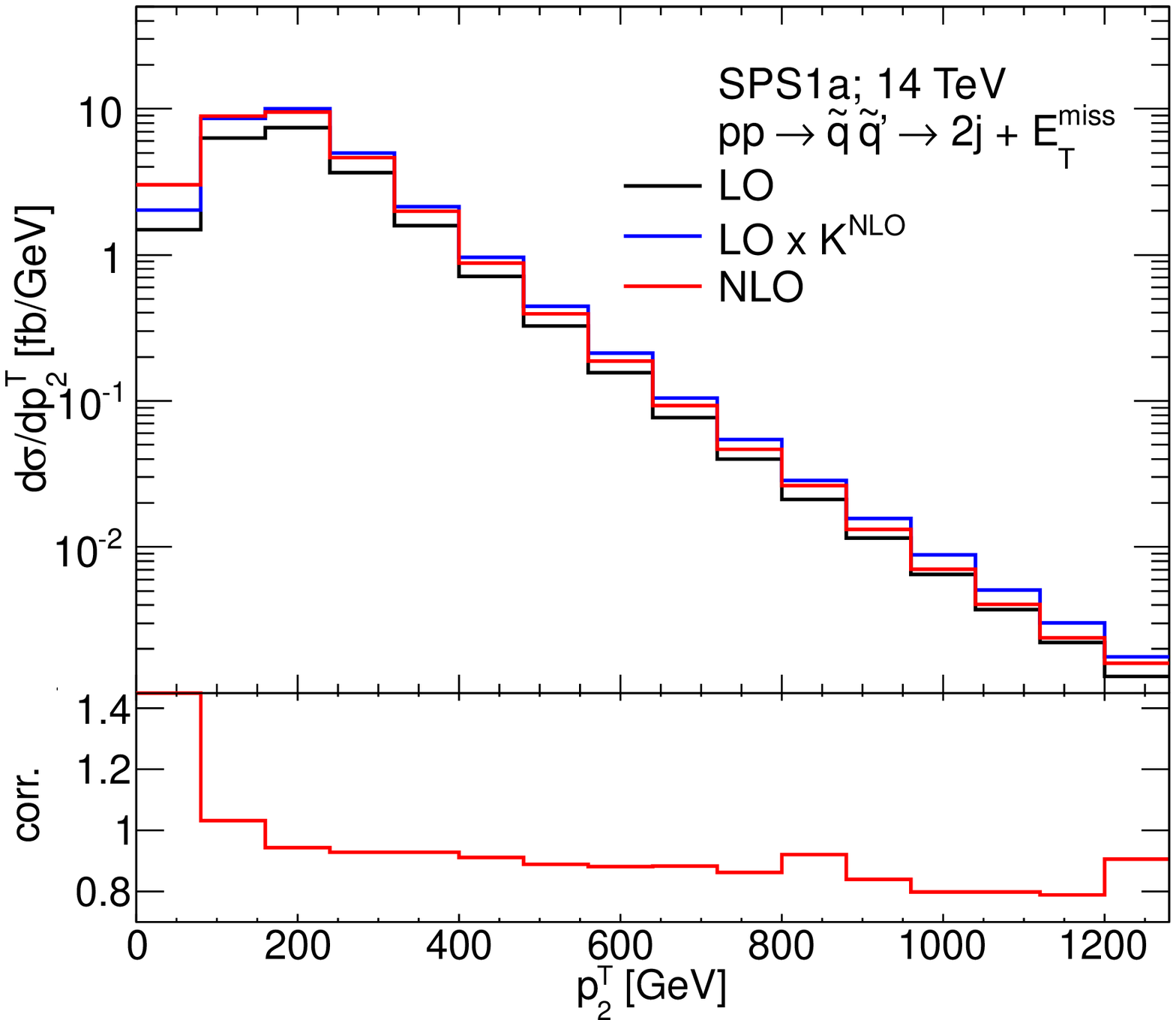}\\
\includegraphics[width=.49\textwidth]{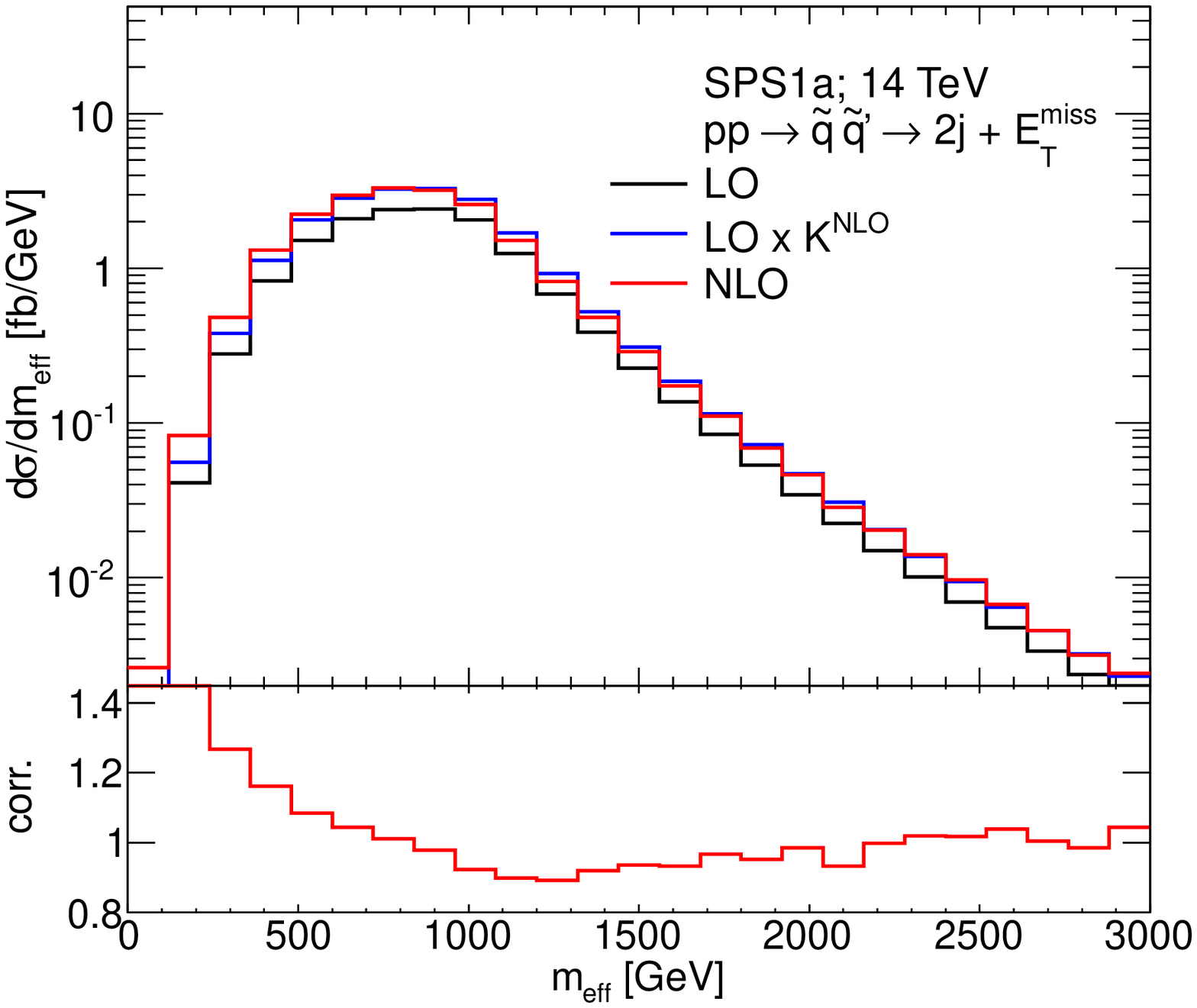}
\includegraphics[width=.49\textwidth]{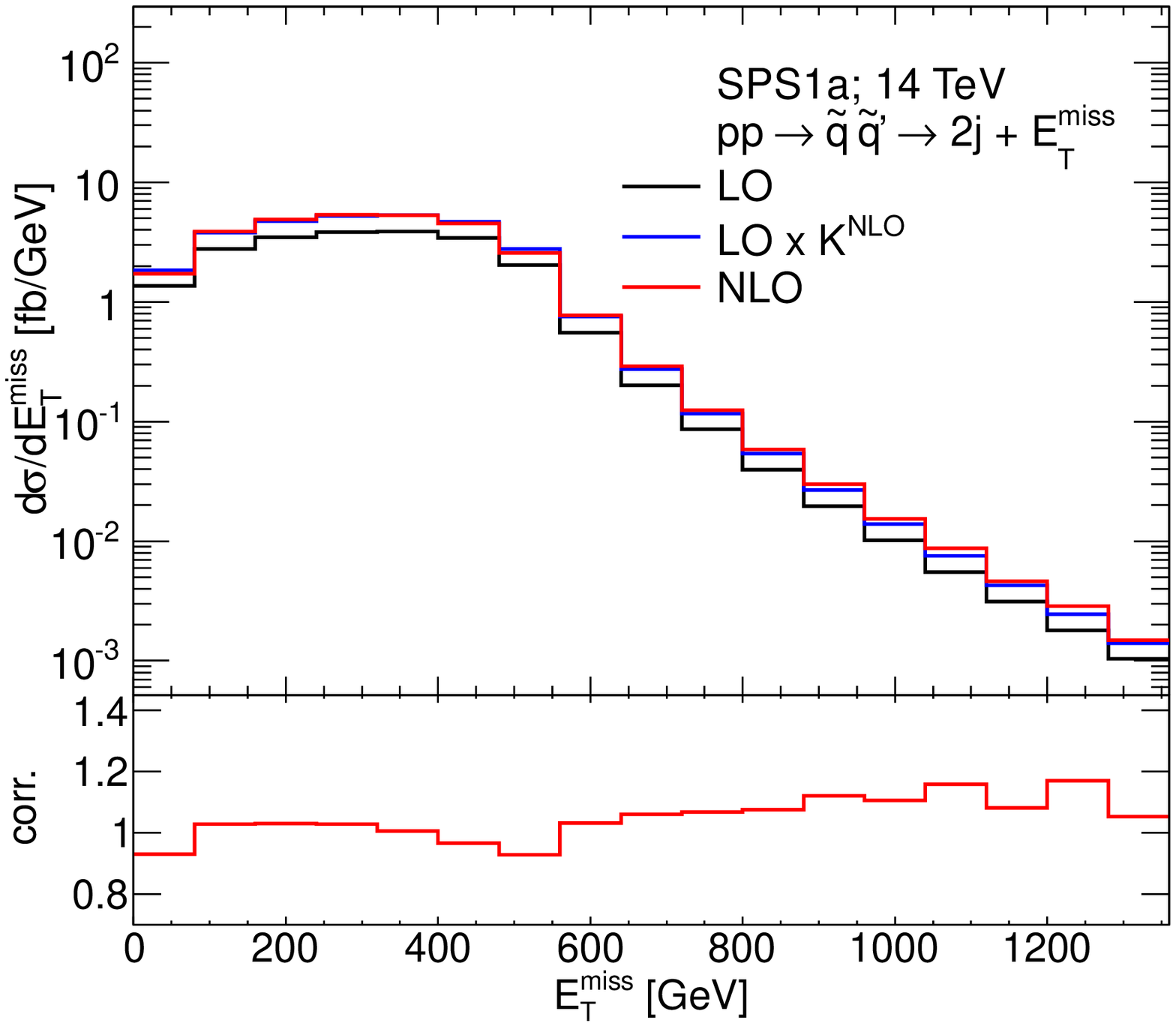}\\
\includegraphics[width=.49\textwidth]{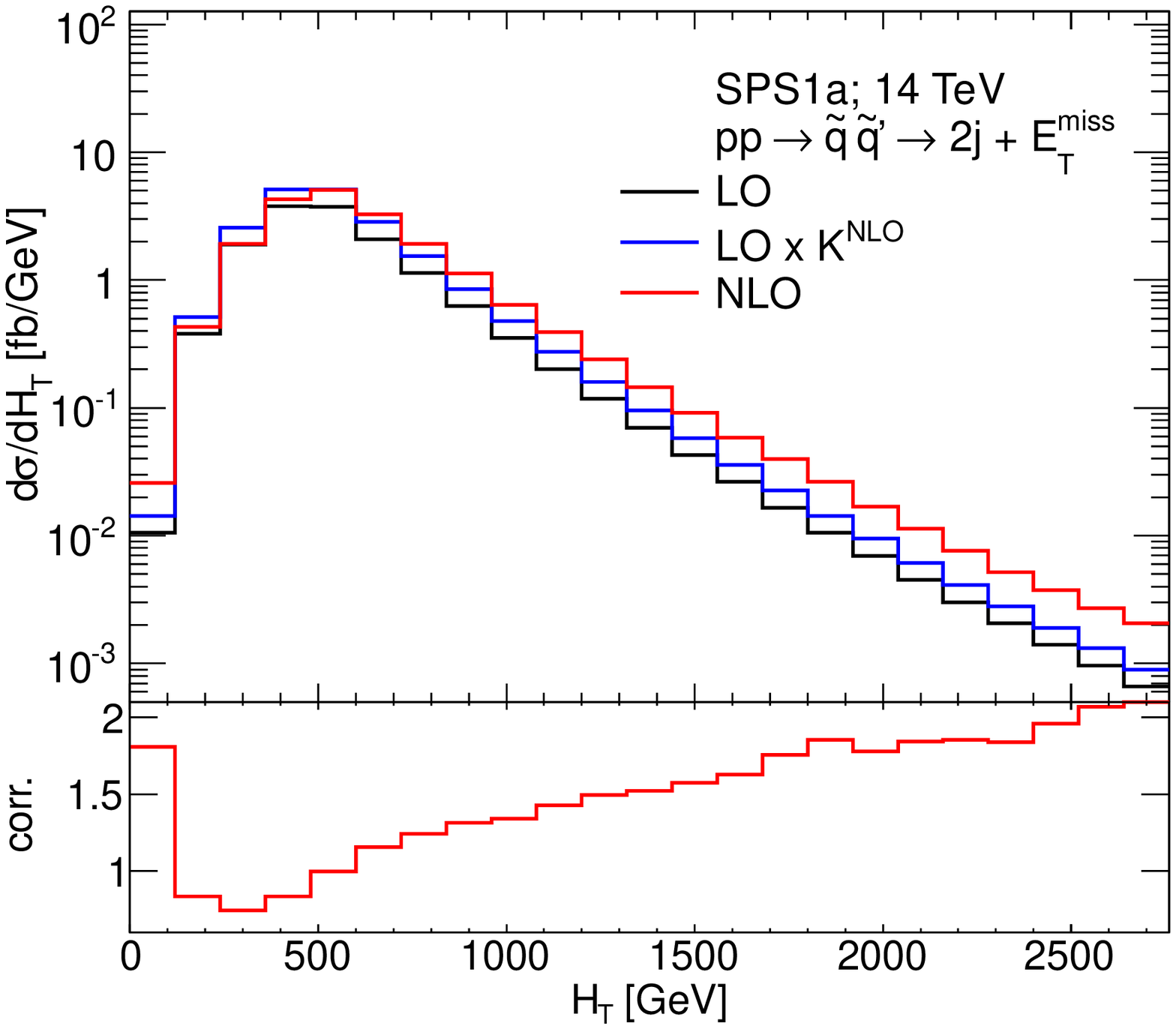}
\includegraphics[width=.49\textwidth]{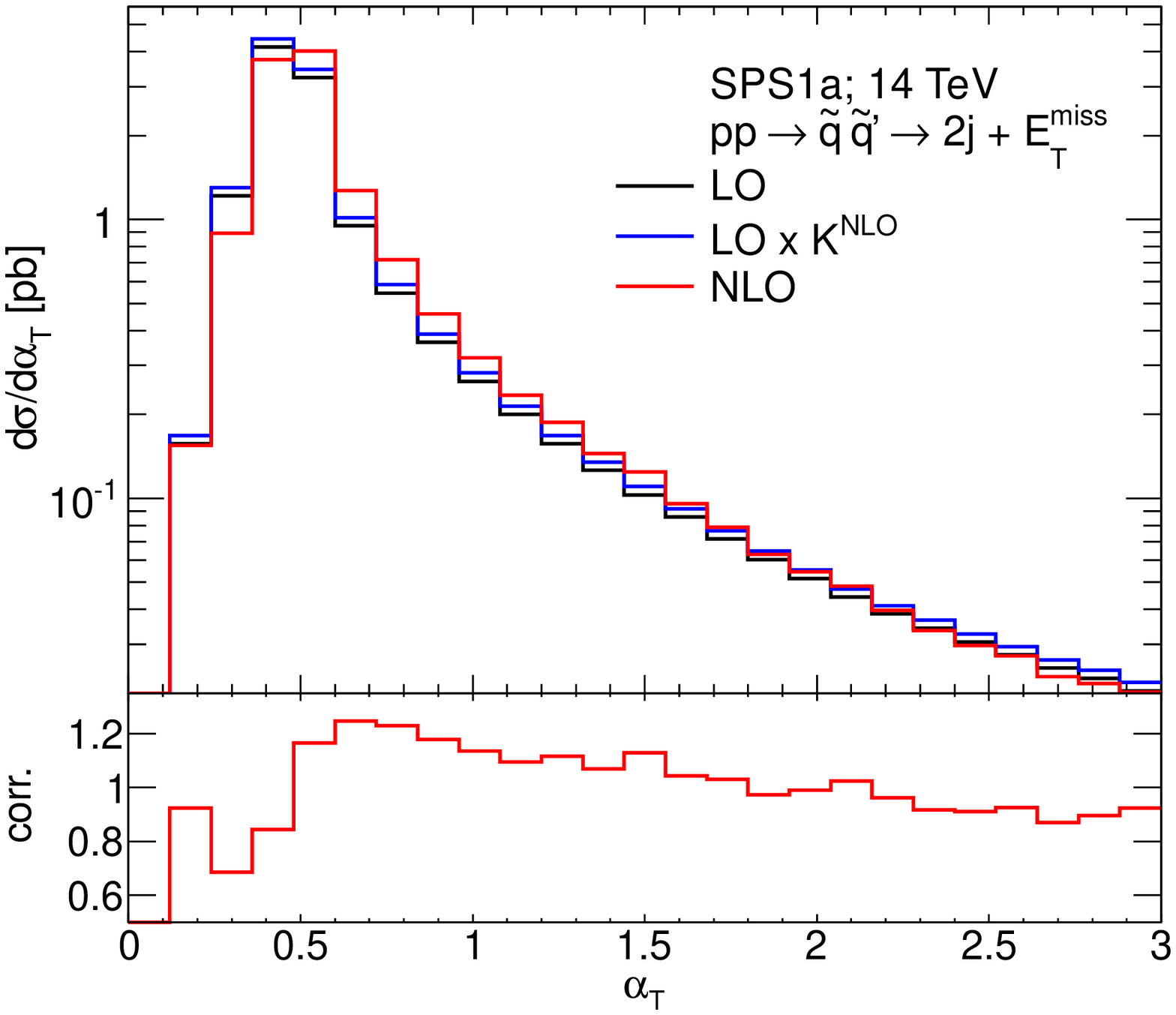}
\caption{Differential distributions of benchmark point SPS1a at a center of mass energy
$\SqrtS=14$. In the upper part of the plots we show in black LO, in red NLO  
and in blue LO distributions rescaled by the ratio $K^{\text{NLO}}$
between the integrated NLO and LO results. In the lower part of the plots NLO
corrections in the shapes are shown, defined as the full NLO divided by the rescaled
$\text{LO}\cdot K^{\text{NLO}}$ distribution.  From top left to bottom right we show
differential distributions in $\pT_1$, $\pT_2$, $\meff$, $\missingET$, $H_{T}$ (all in 
fb/\GeV)  and in $\alpha_T$ (in pb).}
\label{fig:plotssps1a}
}

\FIGURE{
\includegraphics[width=.49\textwidth]{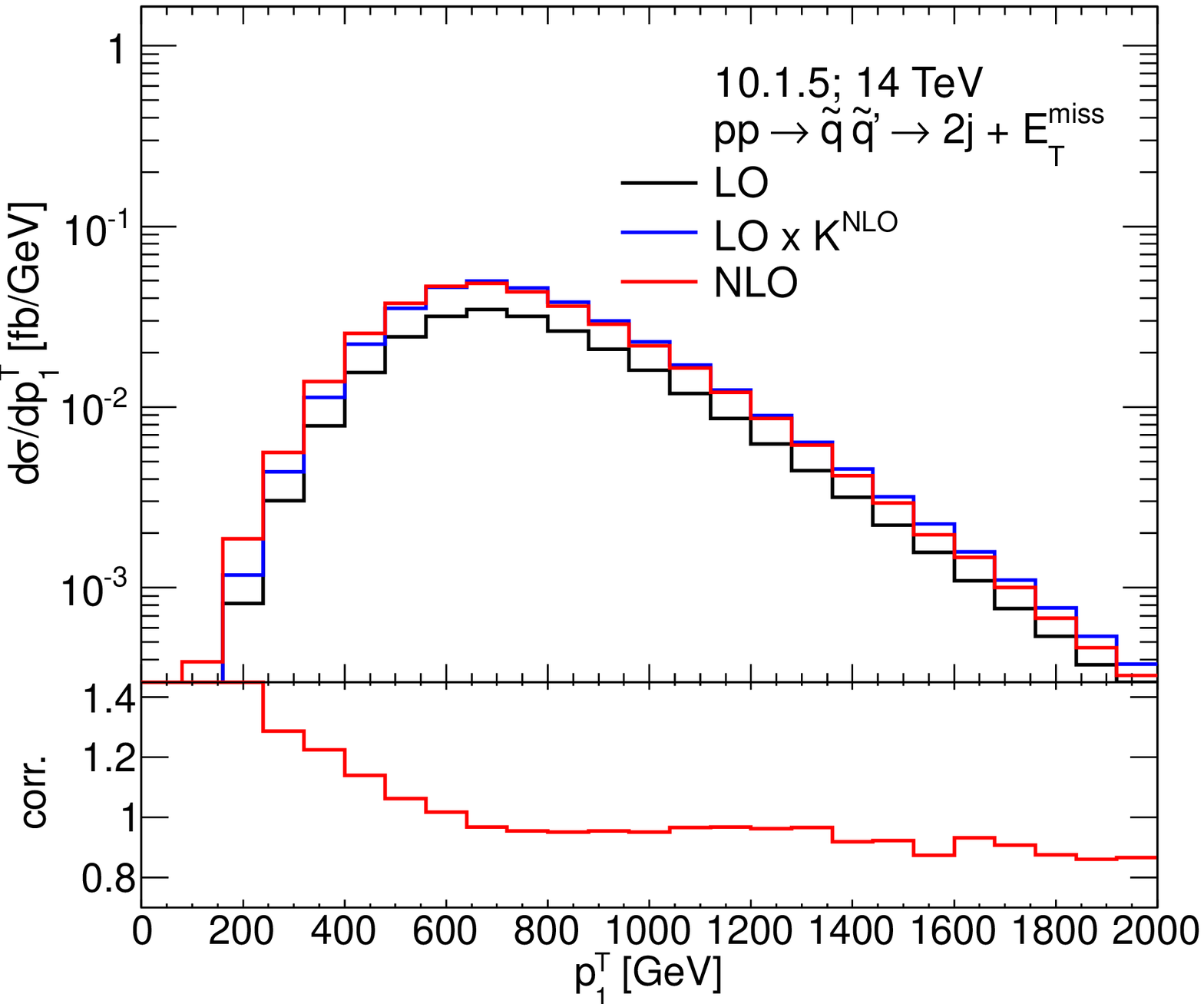}
\includegraphics[width=.49\textwidth]{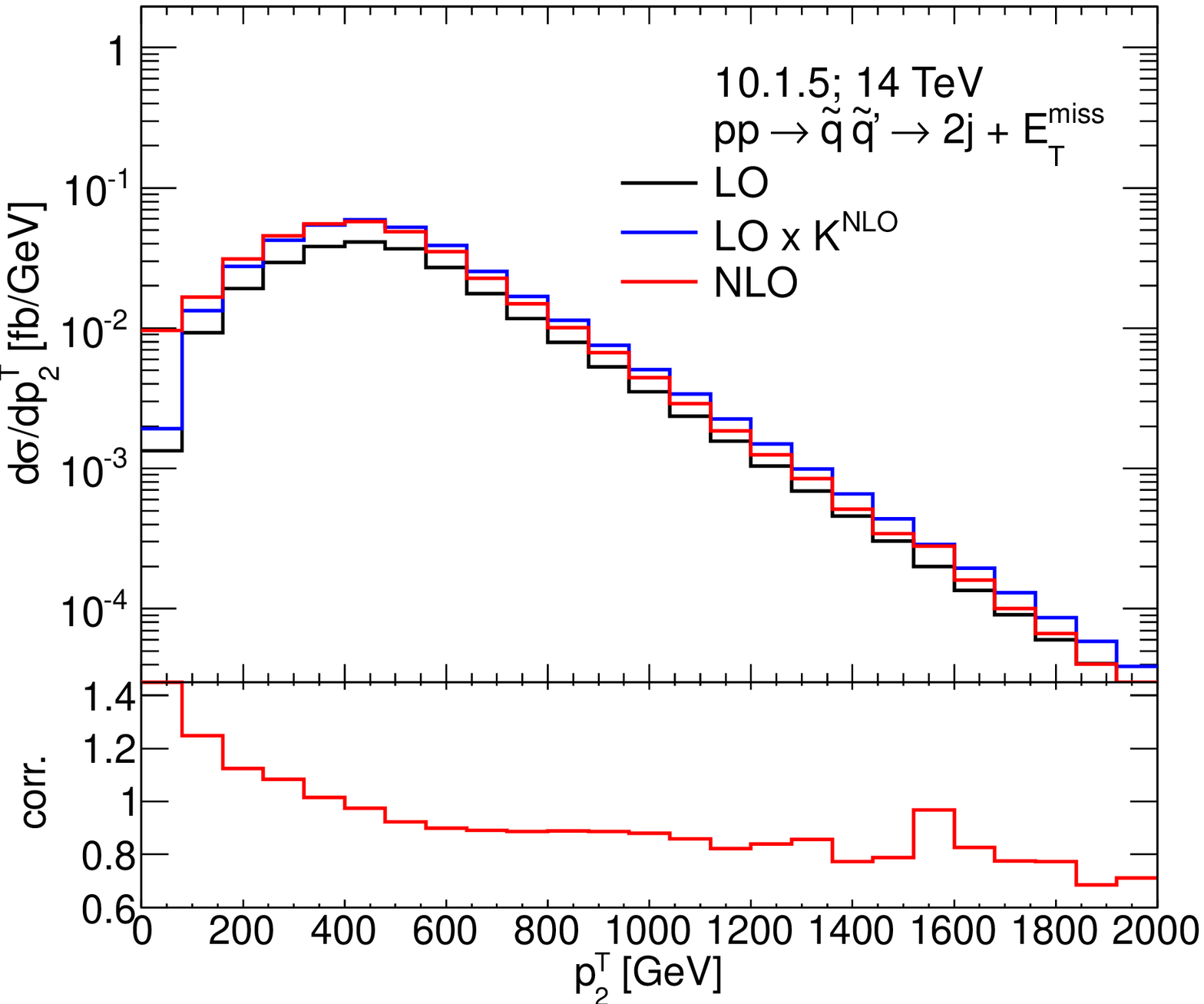}\\
\includegraphics[width=.49\textwidth]{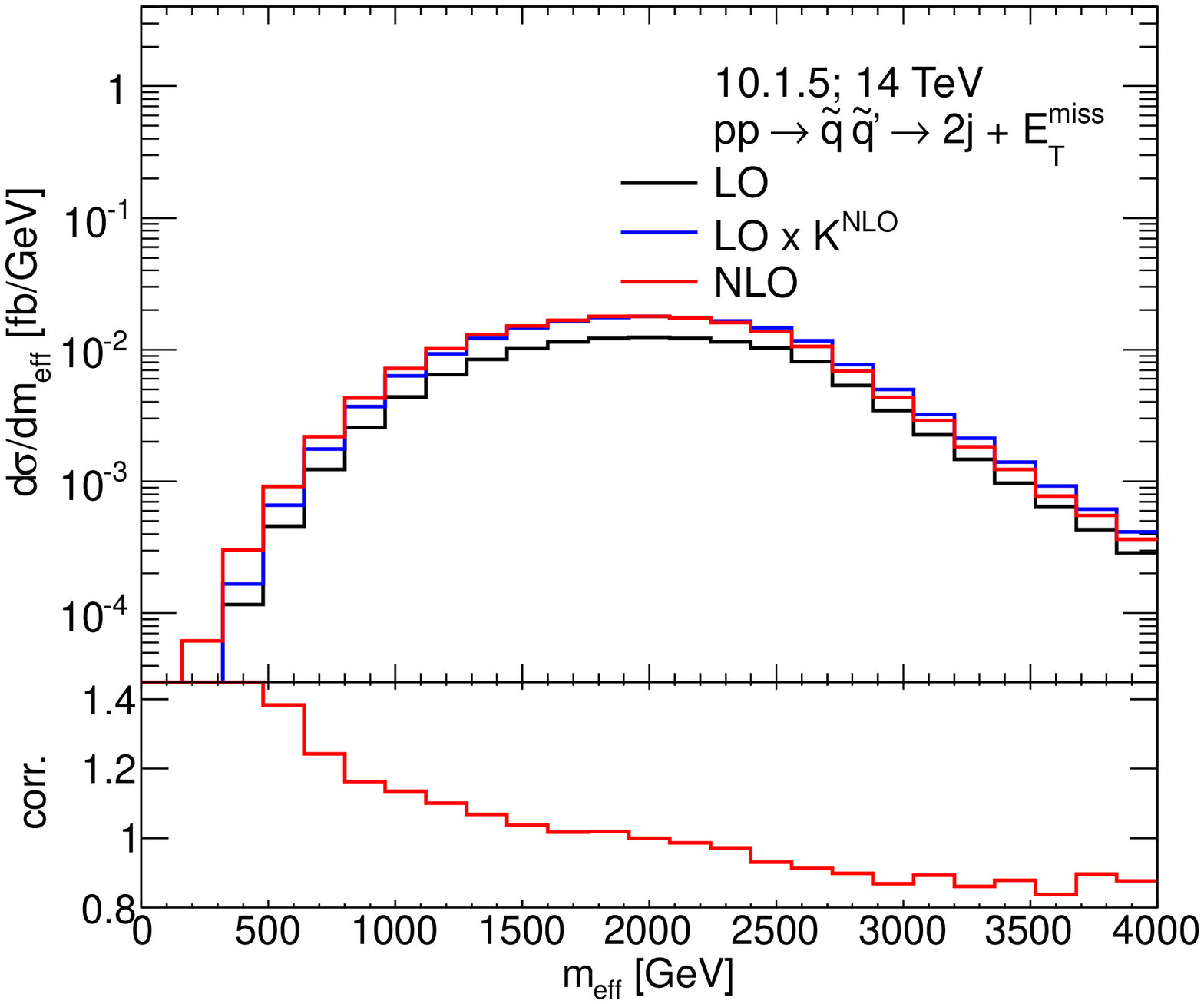}
\includegraphics[width=.49\textwidth]{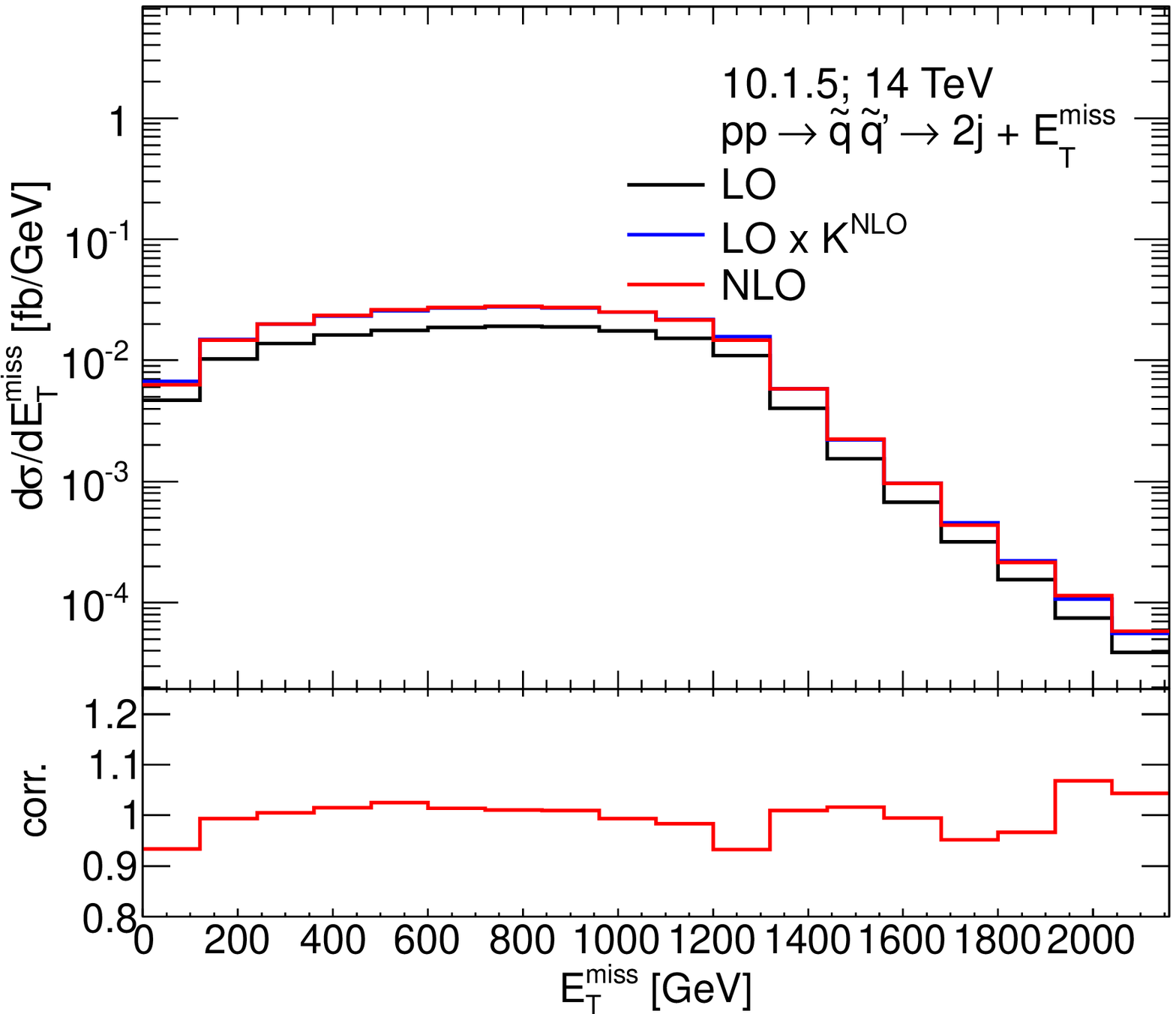}\\
\includegraphics[width=.49\textwidth]{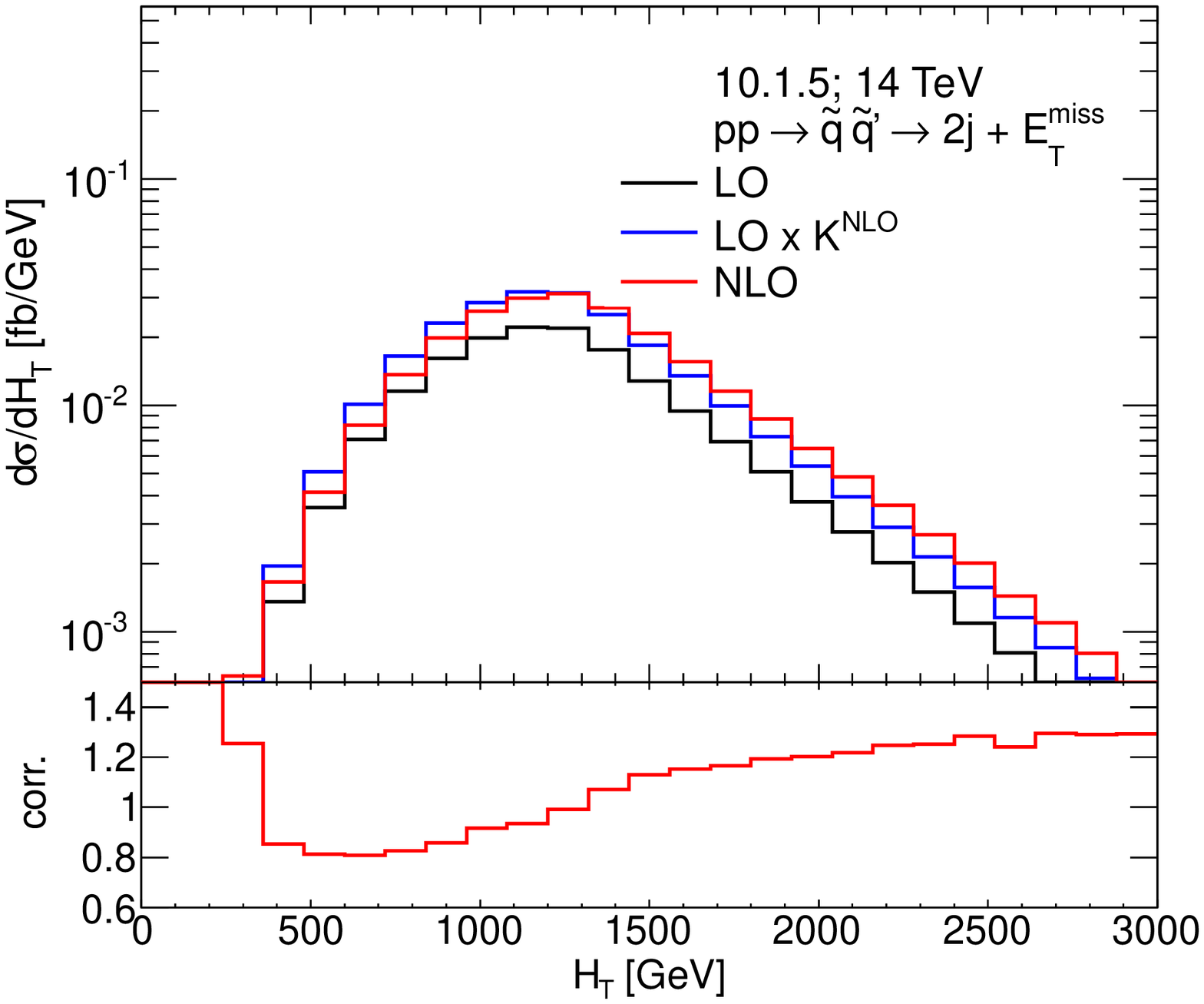}
\includegraphics[width=.49\textwidth]{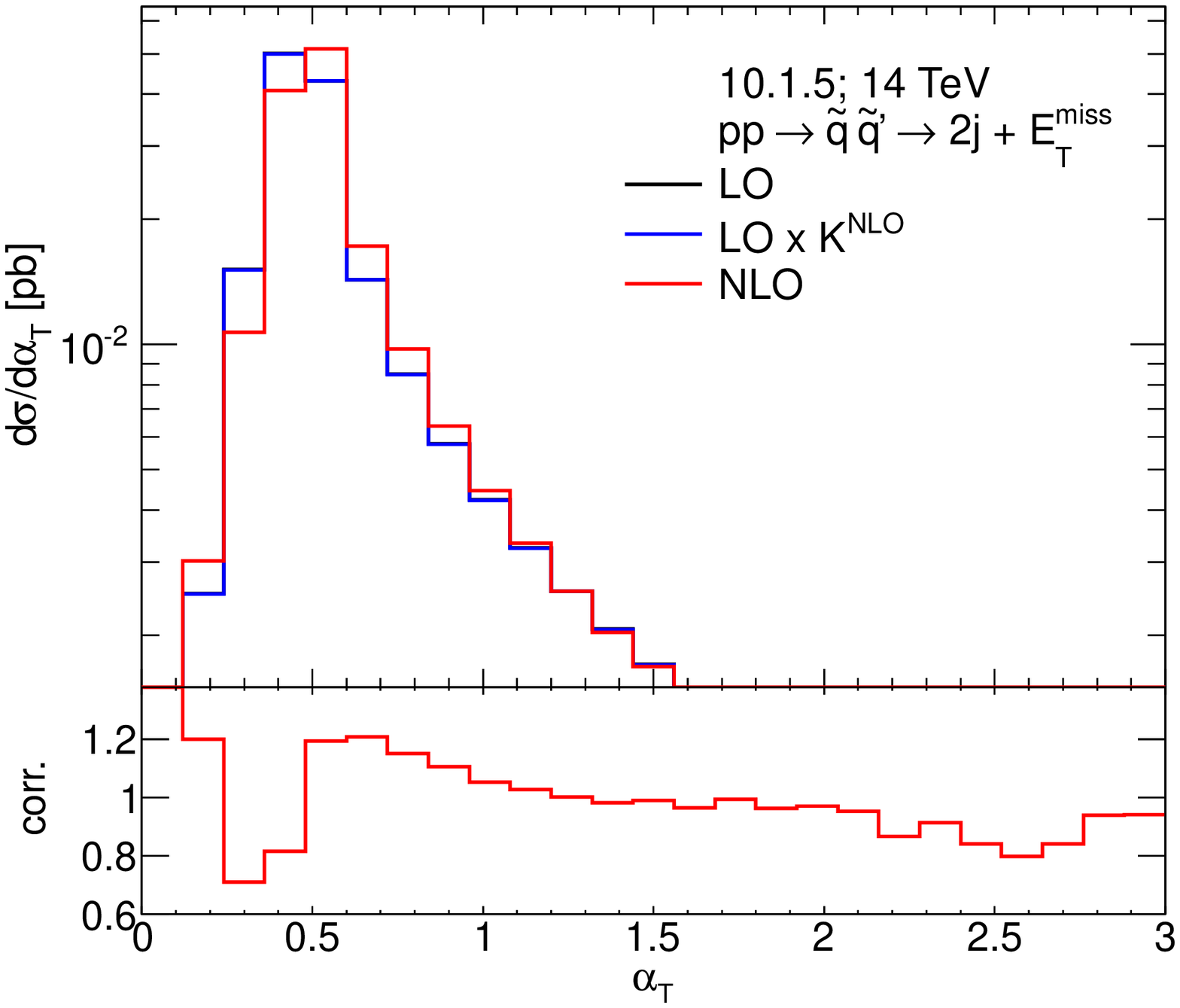}
\caption{Differential distributions of benchmark point 10.1.5 at a center of mass energy
$\SqrtS=14$. In the upper part of the plots we show in black LO, in red NLO  
and in blue LO distributions rescaled by the ratio $K^{\text{NLO}}$
between the integrated NLO and LO results. In the lower part of the plots NLO
corrections in the shapes are shown, defined as the full NLO divided by the rescaled
$\text{LO}\cdot K^{\text{NLO}}$ distribution. 
From top left to bottom right we show differential distributions in $\pT_1$, $\pT_2$,
$\meff$, $\missingET$, $H_{T}$ (all in  fb/\GeV)  and in $\alpha_T$ (in pb).}
\label{fig:plots1015}
}

\FIGURE{
\includegraphics[width=.49\textwidth]{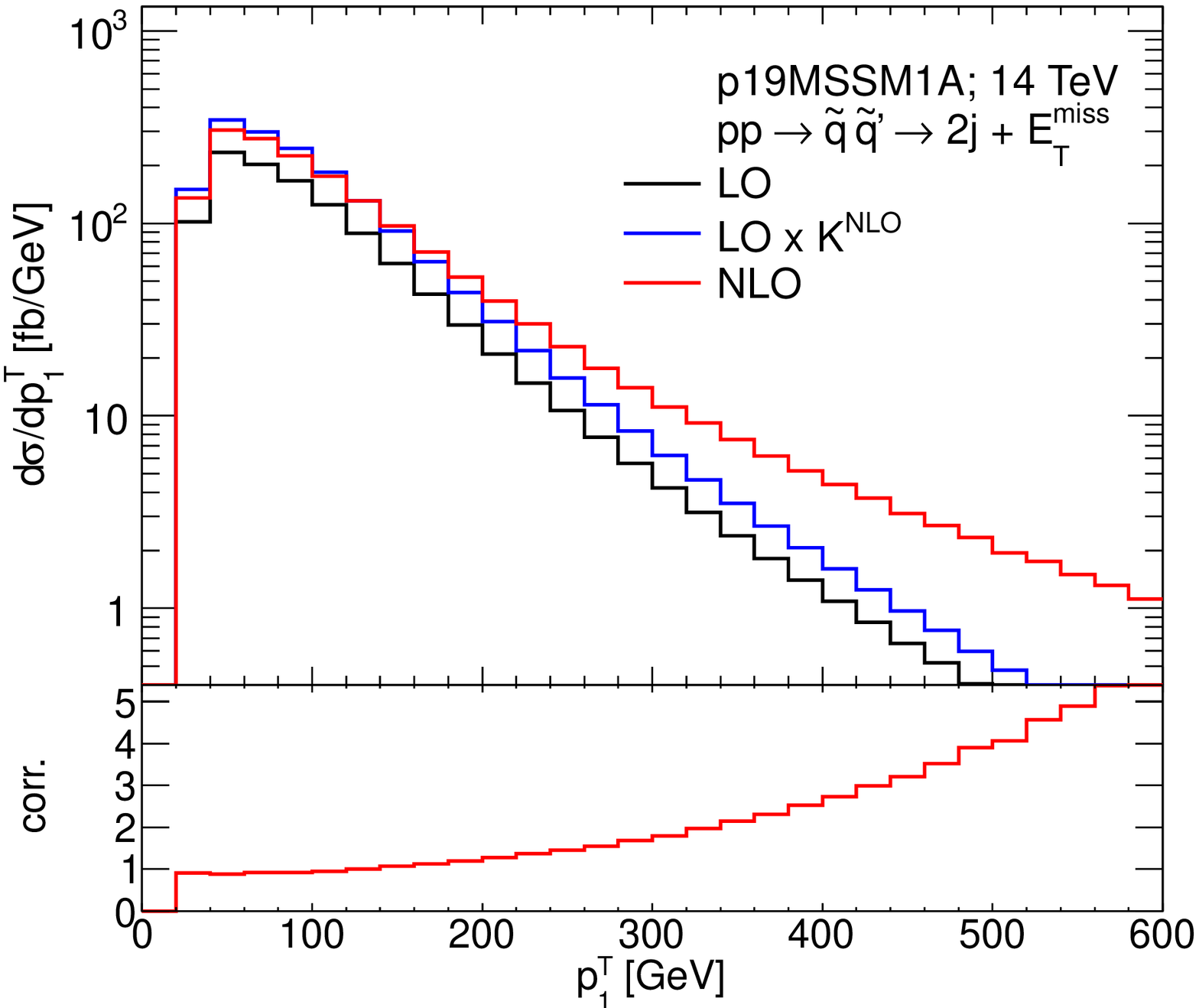}
\includegraphics[width=.49\textwidth]{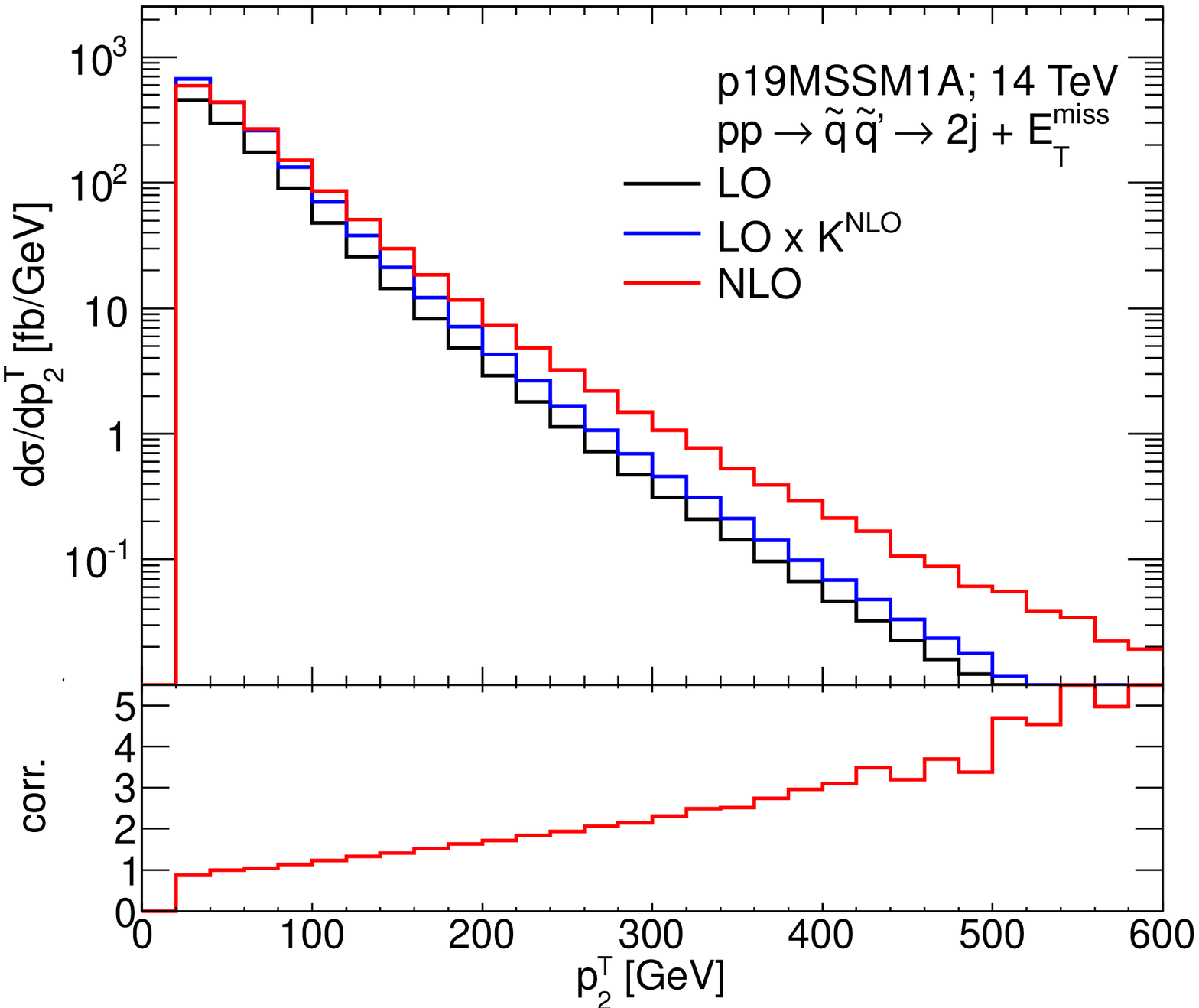}\\
\includegraphics[width=.49\textwidth]{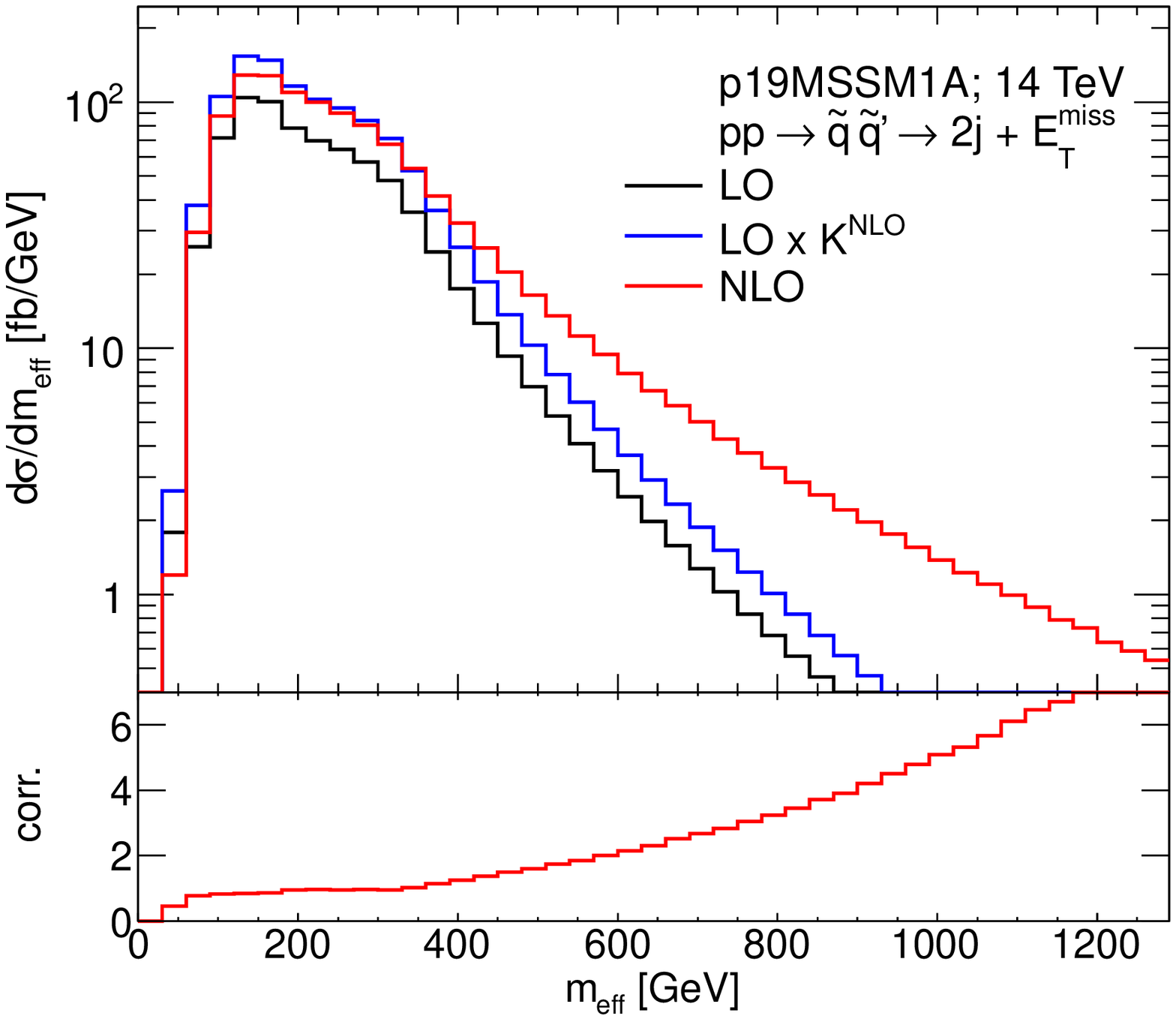}
\includegraphics[width=.49\textwidth]{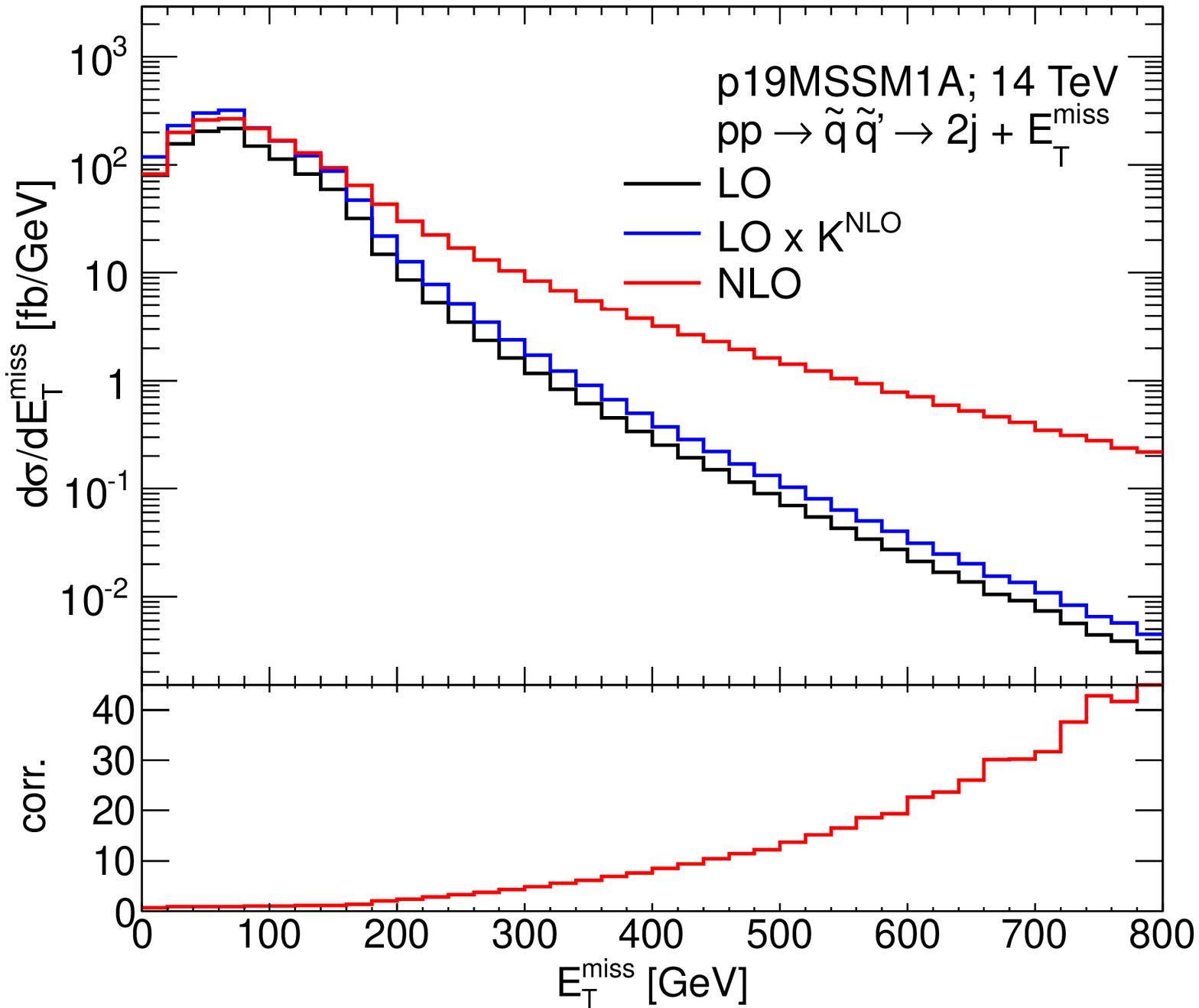}\\
\includegraphics[width=.49\textwidth]{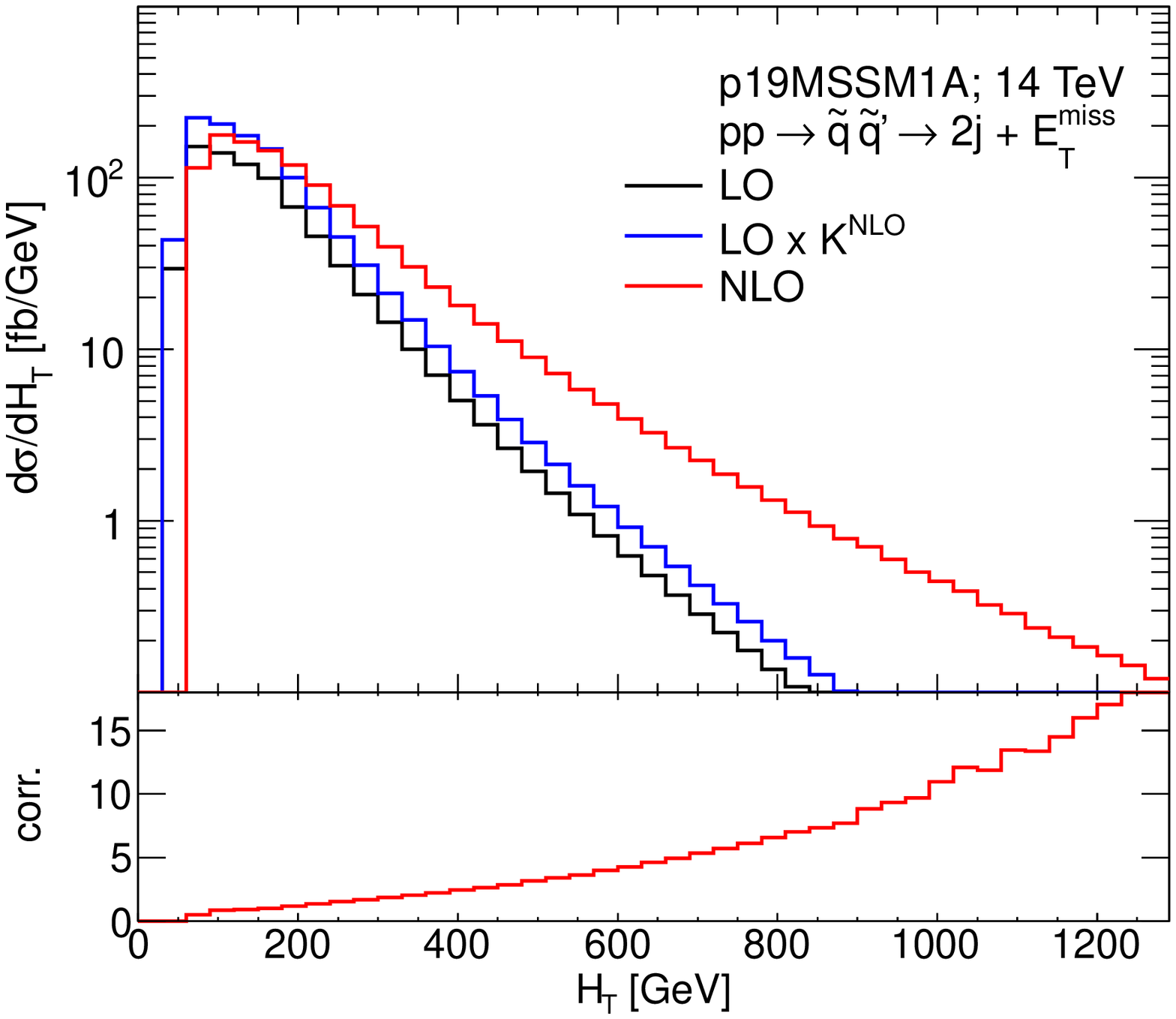}
\includegraphics[width=.49\textwidth]{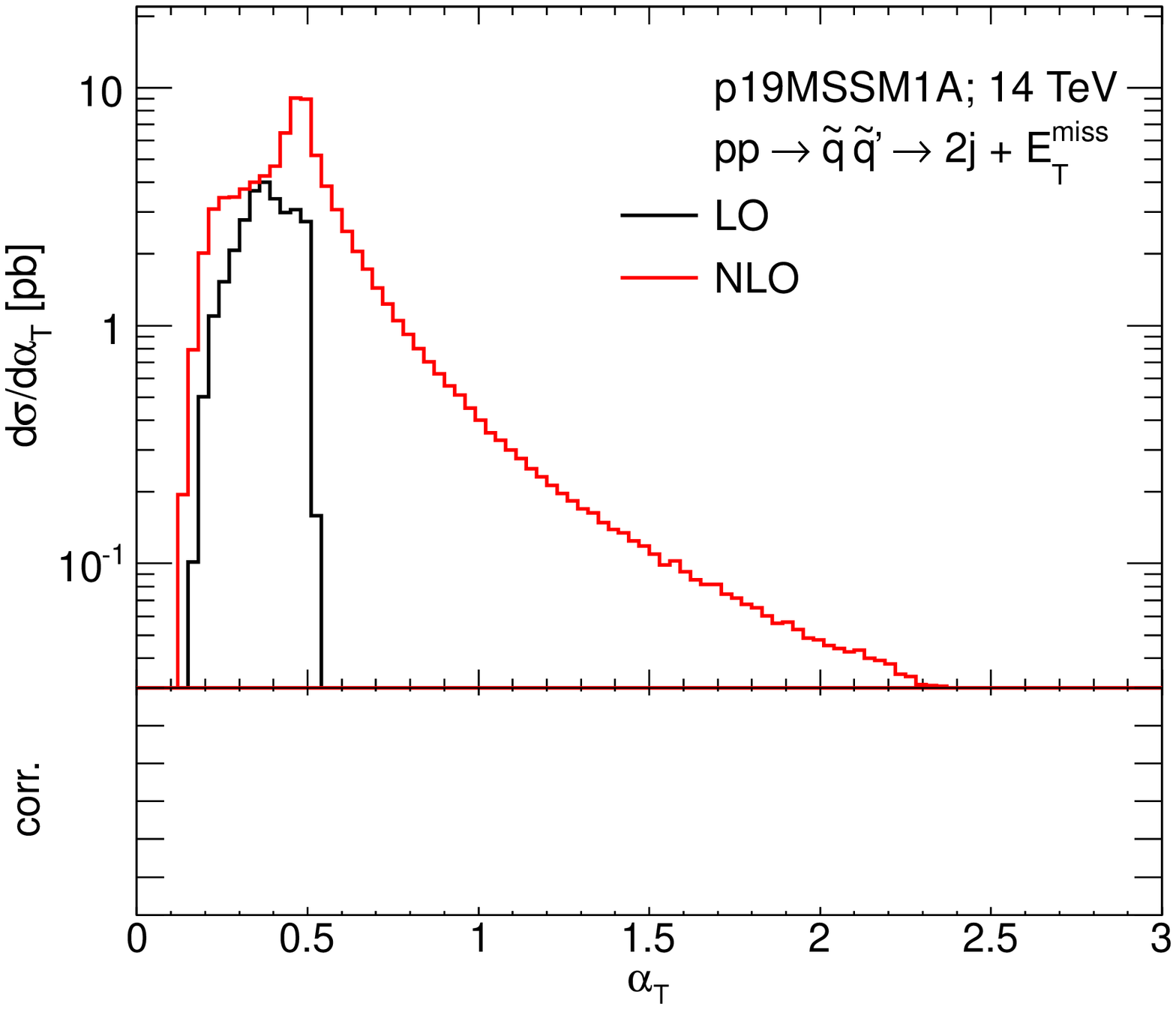}
\caption{Differential distributions of benchmark point p19MSSM1A at a center of mass
energy $\SqrtS=14$. In the upper part of the plots we show in black LO, in red NLO  
and in blue LO distributions rescaled by the ratio $K^{\text{NLO}}$
between the integrated NLO and LO results. In the lower part of the plots NLO
corrections in the shapes are shown, defined as the full NLO divided by the rescaled
$\text{LO}\cdot K^{\text{NLO}}$ distribution. From top left to bottom right we show differential distributions in $\pT_1$,
$\pT_2$, $\meff$, $\missingET$, $H_{T}$ (all in  fb/\GeV)  and in $\alpha_T$ (in pb). }
\label{fig:plotsp19}
}

\FIGURE{
\includegraphics[width=.49\textwidth]{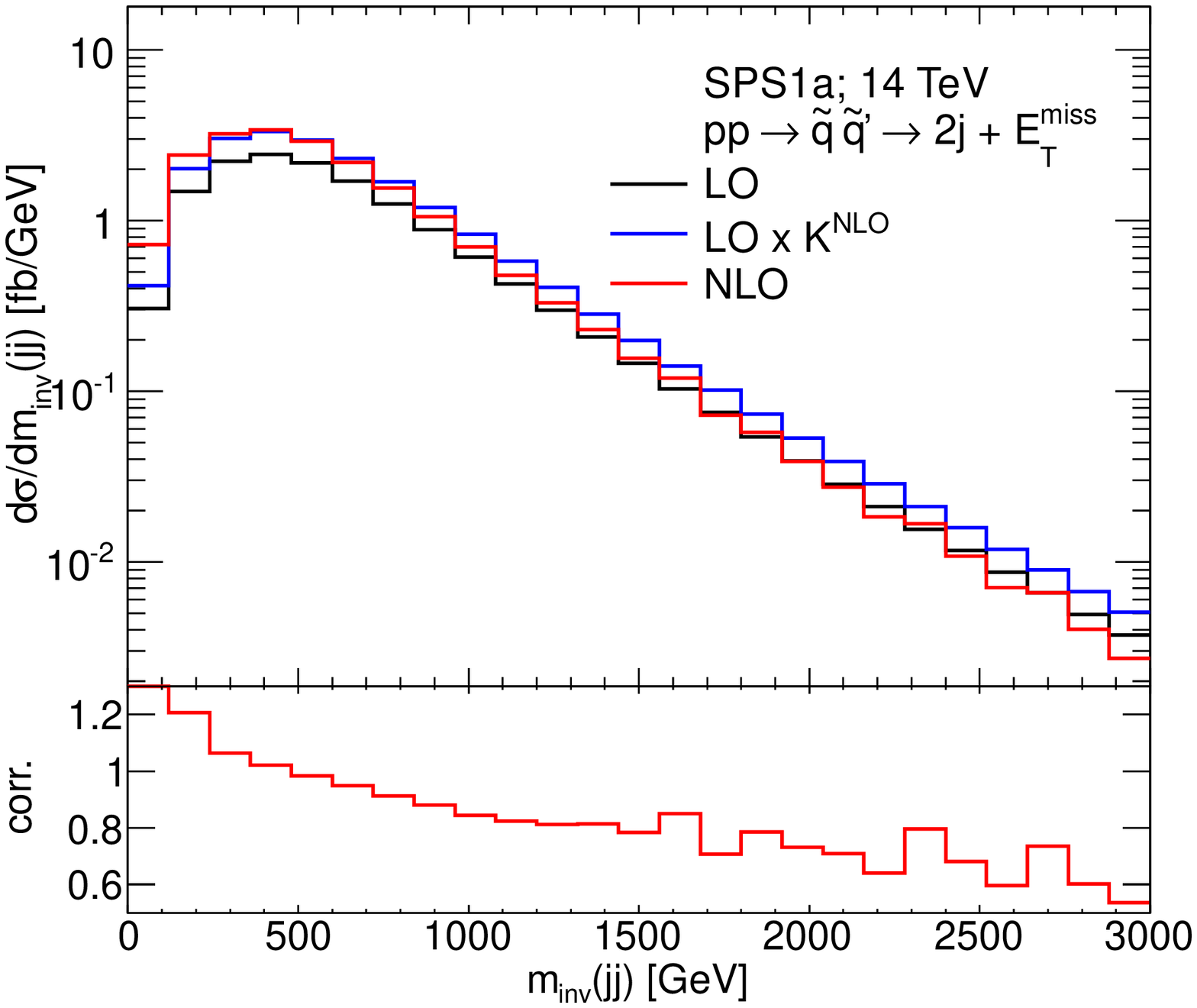}
\includegraphics[width=.49\textwidth]{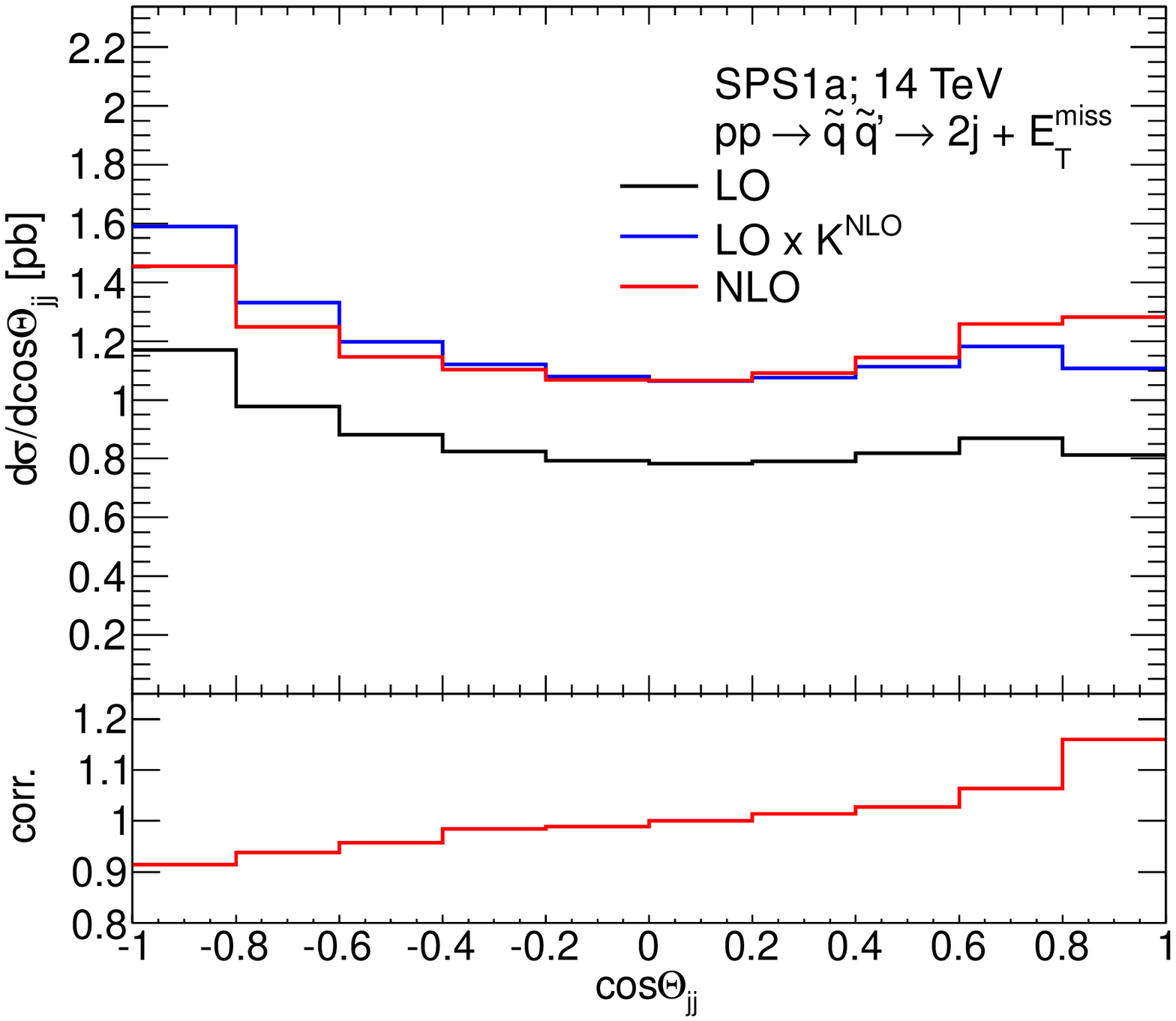}\\
\includegraphics[width=.49\textwidth]{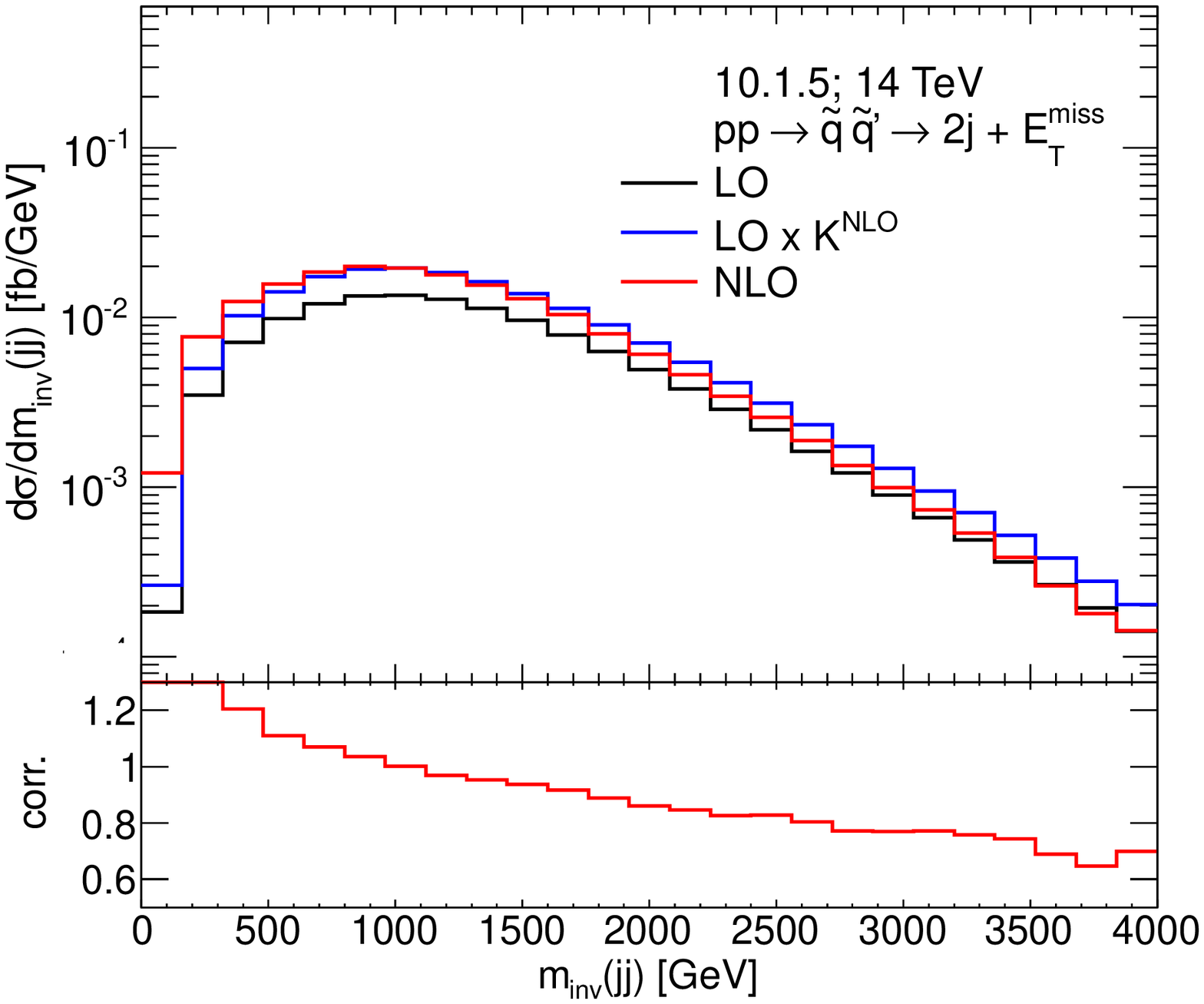}
\includegraphics[width=.49\textwidth]{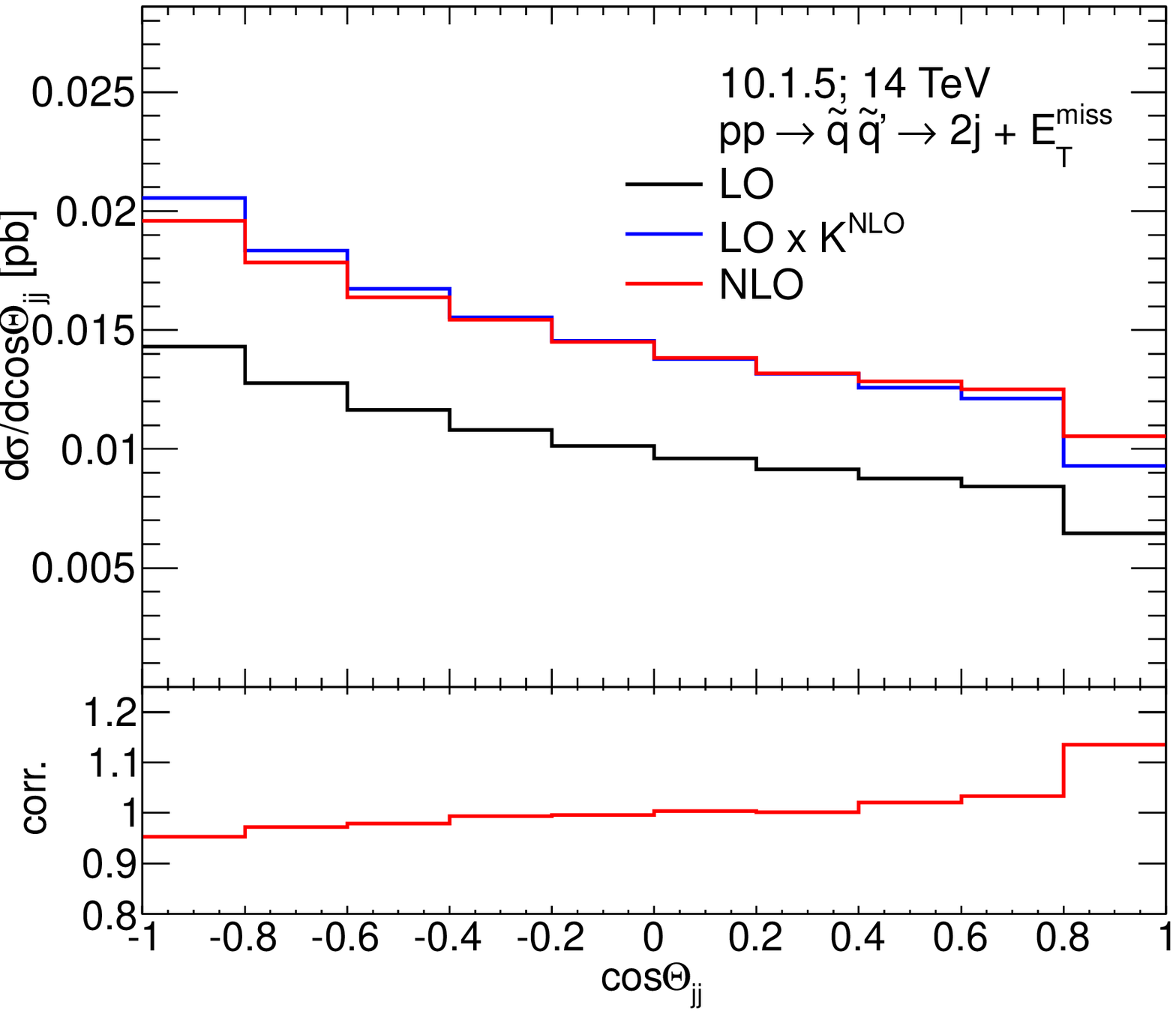}
\includegraphics[width=.49\textwidth]{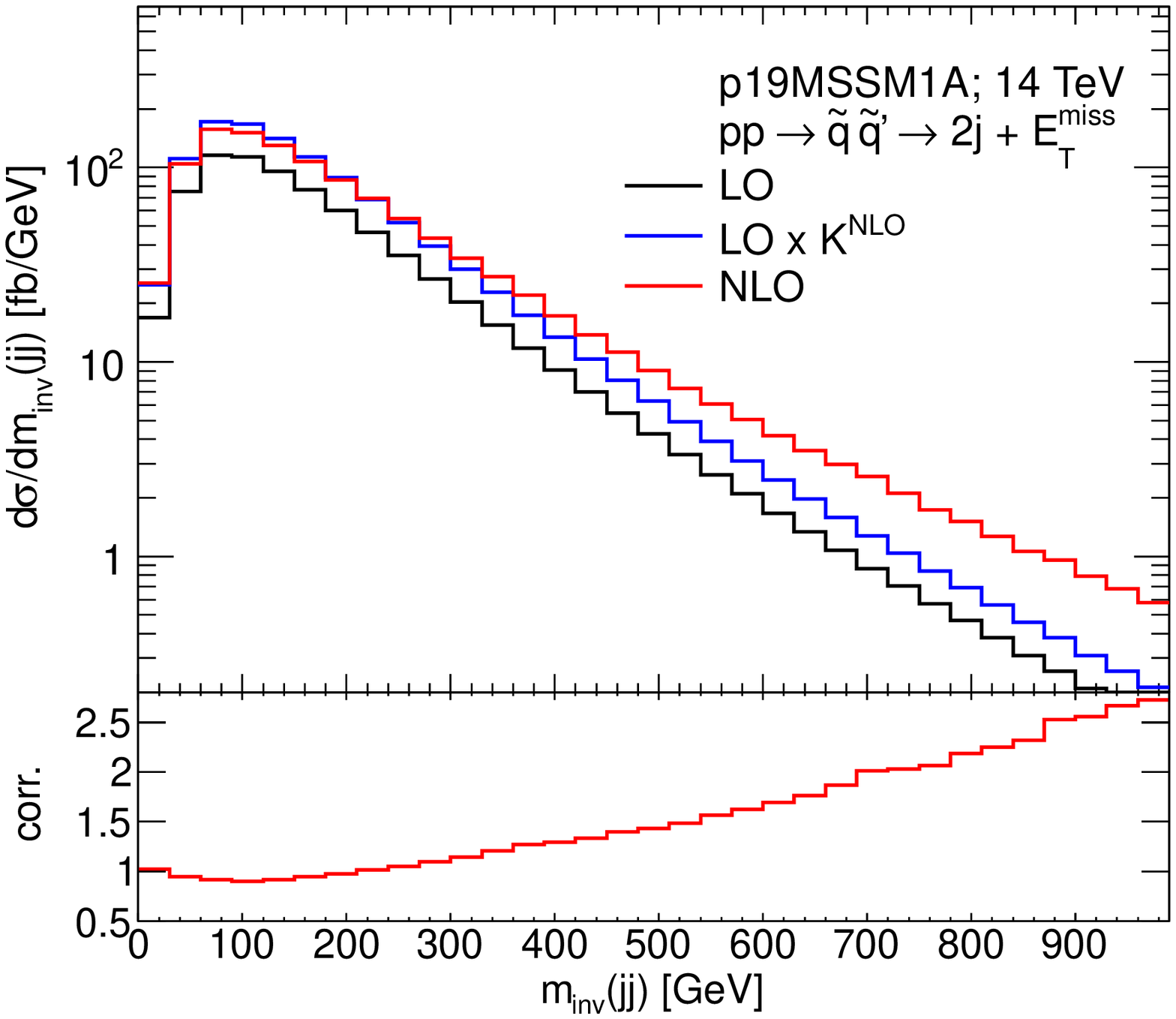}
\includegraphics[width=.49\textwidth]{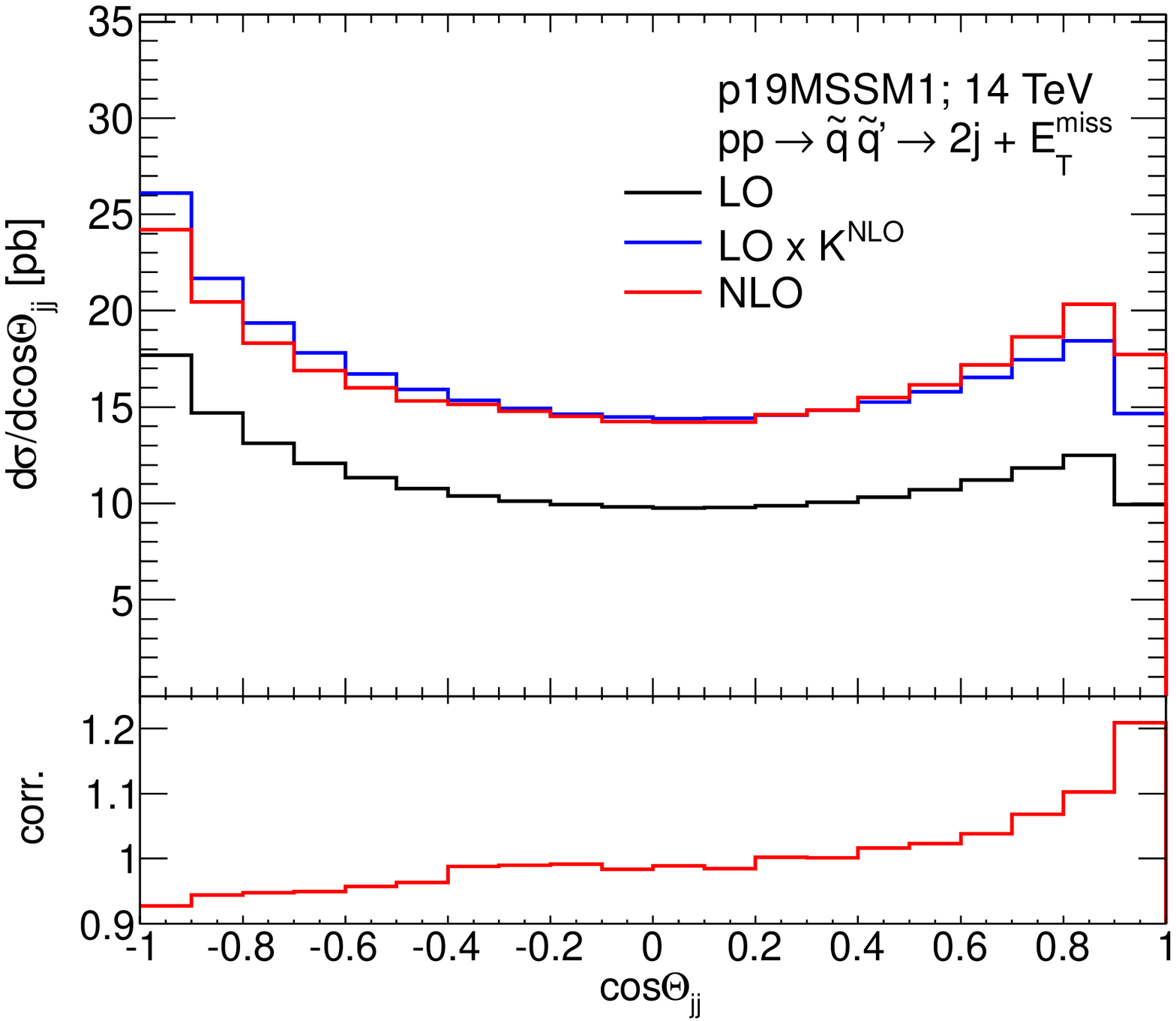}
\caption{Distributions in the invariant mass $m_{\text{inv}}(jj)$ (in fb/GeV) and the
cosine of the angle between the two hardest jets $\cos\Theta_{jj}$ (in pb) for the
benchmark points SPS1a (top), 10.1.5 (middle), p19MSSM1A (bottom) and a center of mass
energy $\SqrtS=14~\TeV$. In the upper part of the plots we show in black LO, in red NLO  
and in blue LO distributions rescaled by the ratio $K^{\text{NLO}}$
between the integrated NLO and LO results. In the lower part of the plots NLO
corrections in the shapes are shown, defined as the full NLO divided by the rescaled
$\text{LO}\cdot K^{\text{NLO}}$ distribution. }
\label{fig:angular}
}

\FIGURE{
\includegraphics[width=.49\textwidth]{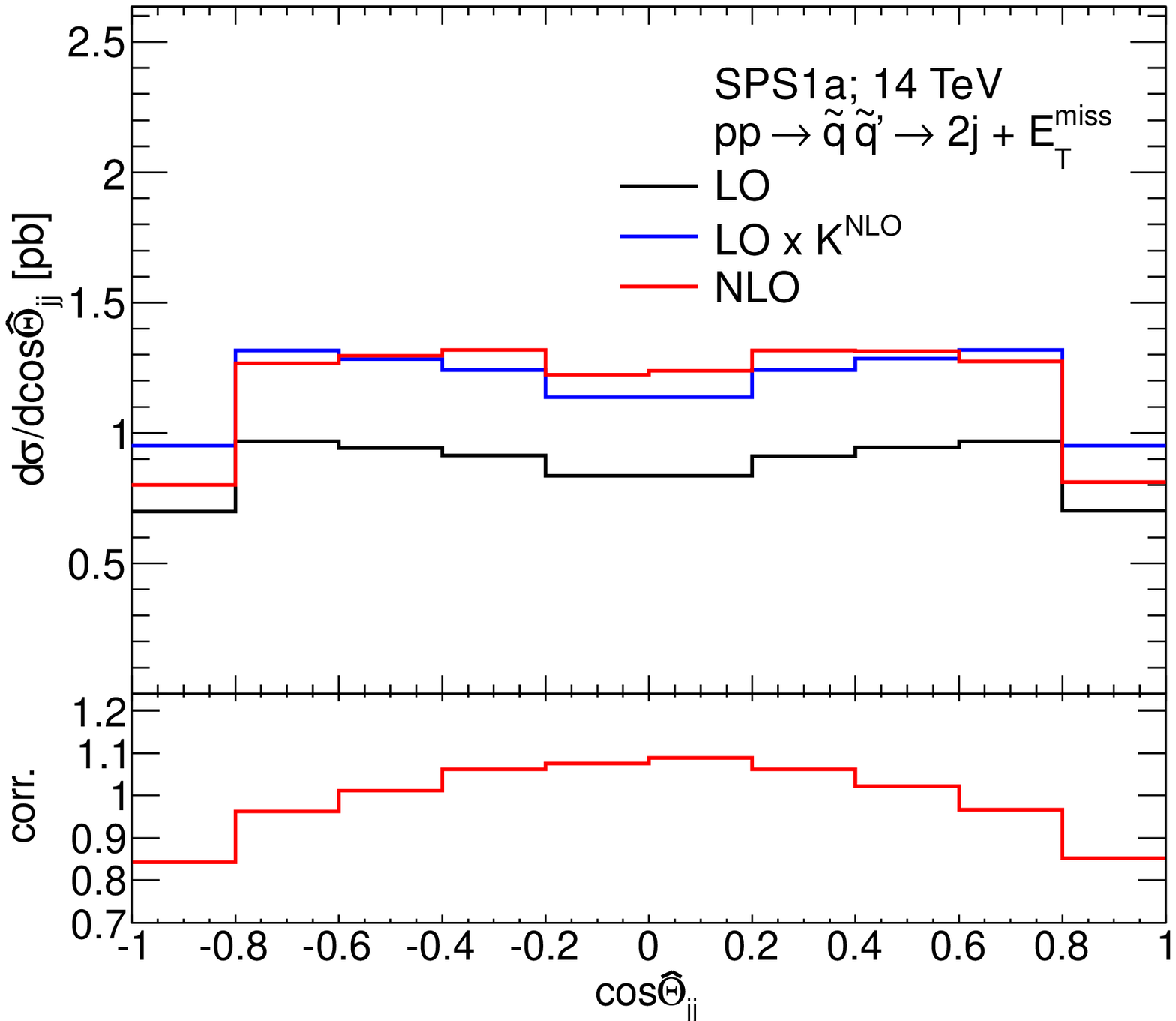}
\includegraphics[width=.49\textwidth]{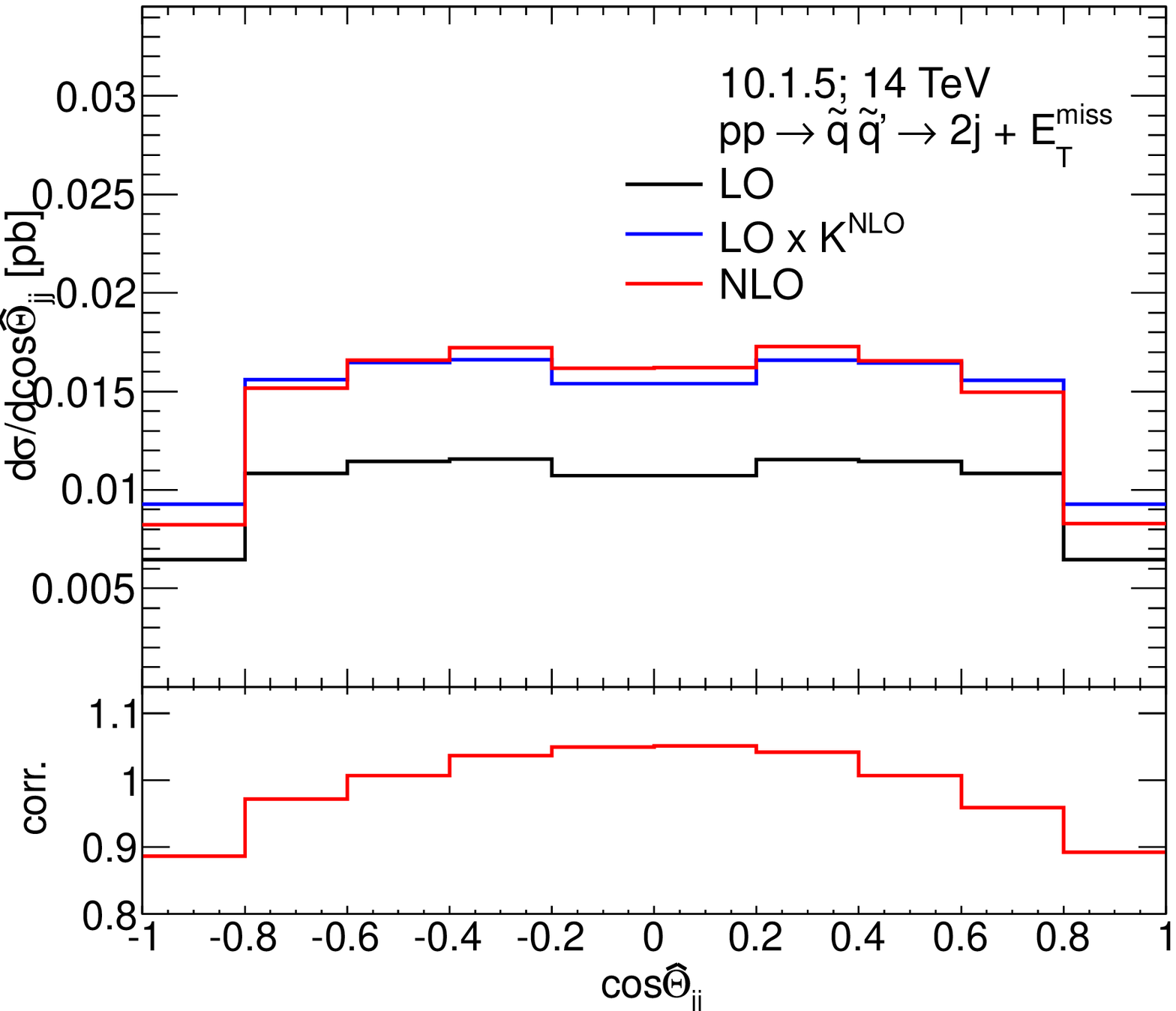}\\
\includegraphics[width=.49\textwidth]{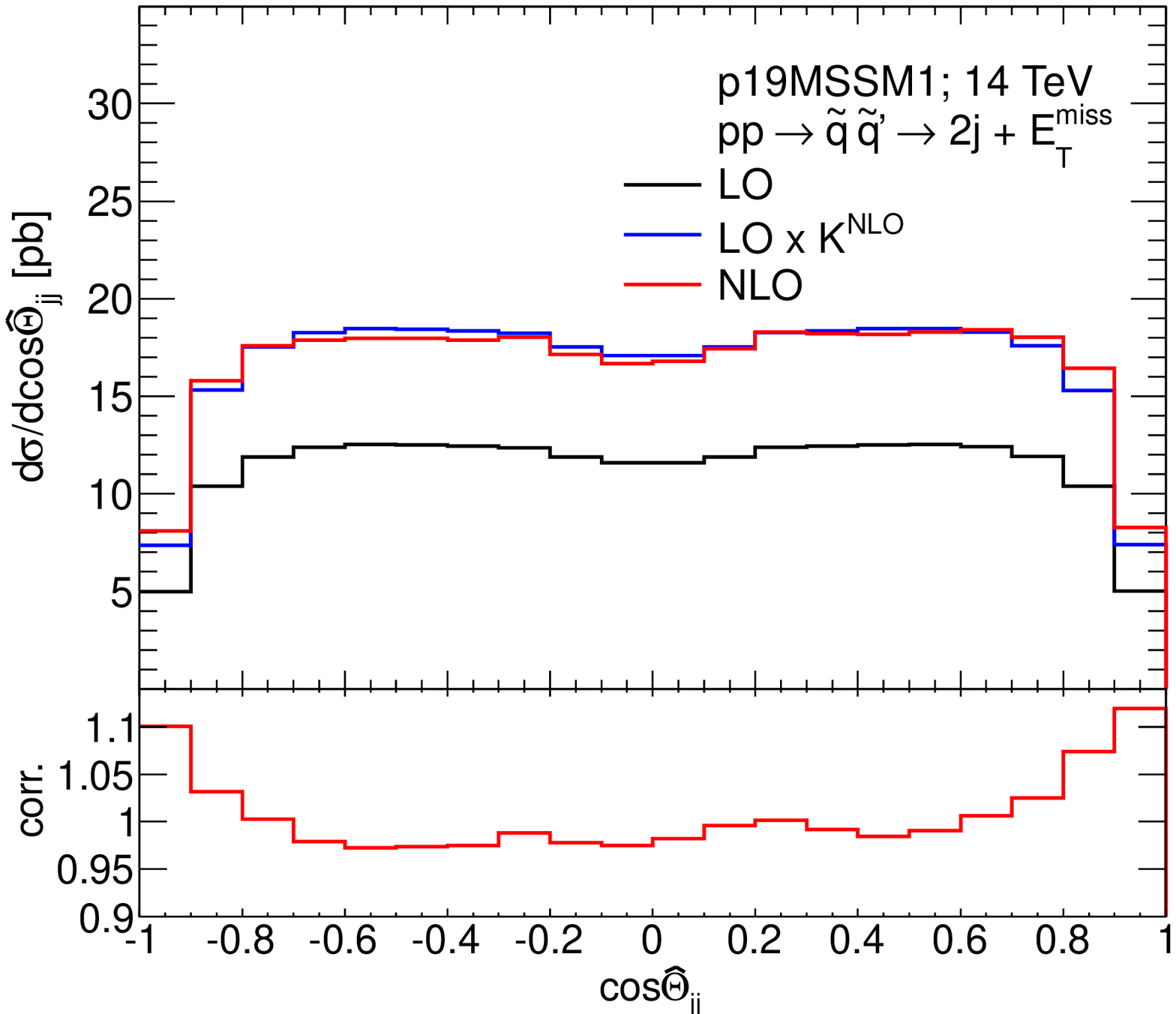}
\caption{Distributions in $\cos{\hat\Theta}$ (in pb) for the benchmark points SPS1a (upper left), 10.1.5 (upper right)
p19MSSM1A (bottom) at a center of mass energy of $\SqrtS=14~\TeV$. In the upper part of the plots we show in black LO, in red NLO  
and in blue LO distributions rescaled by the ratio $K^{\text{NLO}}$
between the integrated NLO and LO results. In the lower part of the plots NLO
corrections in the shapes are shown, defined as the full NLO divided by the rescaled
$\text{LO}\cdot K^{\text{NLO}}$ distribution. }
\label{fig:thetaprime}
}

\clearpage

\subsubsection{Event rates}
\label{sec:rates} 

After investigating inclusive cross sections and differential distributions, we now proceed 
to event rates, \ie, cross sections integrated on signal regions defined to reduce background 
contributions. By this study, we want to quantify a possible impact of our calculation on 
current searches for supersymmetry and future measurements of event rates at the LHC.

In table \ref{tab:atlasrates} we list cross sections after applying cuts of \eqref{atlascuts} 
and in table \ref{tab:cmsrates} cross sections after applying cuts of \eqref{cmscuts}. 
We show LO and NLO cross sections for all three benchmark points and all three energies 
together with resulting K-factors. For comparison we again list inclusive K-factors
of just production, already shown in table \ref{tab:inclusive}. 
From these results, a fully differential description of all squark and gluino channels including
NLO effects in production and decay seems inevitable for a conclusive interpretation of SUSY
searches (or signals) at the LHC. 
Numbers in table \ref{tab:atlasrates} and table \ref{tab:cmsrates} again show that, for compressed
spectra like p19MSSM1A, a pure LO approximation is unreliable for a realistic phenomenological
description of the experimental signatures considered here.
Furthermore, as already suggested in~\cite{Usubov:2010gs} 
and expected from the differential distributions shown in section~\ref{sec:differential}, in particular
interpretations based on $\alpha_T$ seem to be highly affected by higher order corrections.

\TABULAR[h]{c|c||c|c|c||c}{
\hline
   \bf{benchmarkpoint}  & \bf{Energy} [TeV] &
$\boldsymbol{N^{(0)}_{\text{ATLAS}}} $ &$\boldsymbol{N^{(0+1)}_{\text{ATLAS}}} $ & $
\boldsymbol{K_{N_{\text{ATLAS}}}}$ & $\boldsymbol{K_{pp\to \sq
\sq'}}$
\\
\hline
\hline
		&$7$	& $0.066$\pba	& $0.083$\pba	& $1.26$ & $1.37$	\\
SPS1a		&$8$	& $0.097$\pba	& $0.121$\pba	& $1.25$ & $1.35$	\\
		&$14$	& $0.347$\pba	& $0.424$\pba	& $1.22$ & $1.28$	\\
\hline
		&$7$	& $0.313$\fba	& $0.503$\fba	& $1.61$ & $1.57$	\\
10.1.5		&$8$	& $0.861$\fba	& $1.344$\fba	& $1.56$ & $1.52$	\\
	
		&$14$	& $13.82$\fba	& $19.77$\fba	& $1.43$ & $1.40$	\\
	
\hline
		&$7$	& $0.140$\fba	& $20.76$\fba	& $\sim150$ & $1.40$	 \\
p19MSSM1A	&$8$	& $0.339$\fba	& $37.96$ \fba	& $\sim110$ & $1.39$	\\
	
		&$14$	& $0.0044$\pba	& $0.264$ \pba	& $\sim60$ & $1.34$	\\
\hline
}{LO $N^{(0)}_{\text{ATLAS}}$ and NLO $N^{(0+1)}_{\text{ATLAS}}$ cross section
predictions and K-factors $K_{N_{\text{ATLAS}}}$ for the three benchmark scenarios SPS1a,
10.1.5, p19MSSM1A and center of mass energies $\SqrtS=7,8,14~\TeV$ where the cuts of
\eqref{atlascuts} are applied. For comparison we also list is the inclusive NLO production
K-factor $K_{pp\to \sq\sq'}$ already shown in \tabref{tab:inclusive}.  
\label{tab:atlasrates}
}

\TABULAR[h]{c|c||c|c|c||c}{
\hline
   \bf{benchmarkpoint}  & \bf{Energy} [TeV] & $\boldsymbol{N^{(0)}_{\text{CMS}}}
$ &$\boldsymbol{N^{(0+1)}_{\text{CMS}}} $ & $ \boldsymbol{K_{N_{\text{CMS}}}}$ &
$\boldsymbol{K_{pp\to \sq
\sq'}}$
\\
\hline
\hline
		&$7$	& $0.112$\pba	& $0.141$\pba	& $1.26$&$1.37$	\\
SPS1a		&$8$	& $0.157$\pba	& $0.197$\pba	& $1.25$&$1.35$	\\
		&$14$	& $0.488$\pba 	& $0.614$\pba	& $1.26$& $1.28$	\\
\hline
		&$7$	& $0.201$\pba	& $0.261$\pba	& $1.30$& $1.57$	\\
10.1.5		&$8$	& $0.542$\fba	& $0.674$\fba	& $1.24$&$1.52$	\\
	
		&$14$	& $8.129$\fba	& $8.884$\fba	& $1.09$&$1.40$	\\		
\hline
		&$7$	& $10^{-6}$\pba	& $0.095$\pba	& $\ord(10^4)$&$1.40$	\\
p19MSSM1A	&$8$	& $10^{-6}$\pba & $0.151$\pba	& $\ord(10^4)$& $1.39$	\\
	
		&$14$	& $2\cdot 10^{-5}$\pba & $0.687$\pba  & $\ord(10^4)$&$1.34$	\\
\hline}{LO $N^{(0)}_{\text{CMS}}$ and NLO $N^{(0+1)}_{\text{CMS}}$ cross section
predictions and K-factors $K_{N_{\text{CMS}}}$ for the three benchmark scenarios SPS1a,
10.1.5, p19MSSM1A and center of mass energies $\SqrtS=7,8,14~\TeV$ where the cuts of
\eqref{cmscuts} are applied. For comparison we also list is the inclusive NLO production
K-factor $K_{pp\to \sq\sq'}$ already shown in \tabref{tab:inclusive}.  
\label{tab:cmsrates}
}



\section{Conclusions}
\label{sec:conclusion}

In this paper we presented a study of squark--squark production and the subsequent 
squark decay into the lightest neutralino at the LHC, providing for the first time
fully differential predictions for the experimental signature
$2j+\missingET(+X)$ at NLO in QCD. 
For the calculation, each flavour and chirality configuration of the squarks
has been treated individually, allowing the combination of production cross sections
and decay distributions in a consistent way.      

We studied inclusive cross sections, differential distributions for jet observables,
and experimental signatures with cuts, illustrating the effect of the NLO contributions.
In general, the NLO corrections are important.
In particular, NLO effects going beyond a rescaling with a global $K$-factor can in general not 
be neglected for setting precise limits on the sparticle masses and model parameters;
they become specially important in model classes with compressed spectra.
At the same time, the theoretical uncertainty on the level of differential distributions is reduced by our calculation. 

Although the present study is dedicated to the simplest squark decay mode,
the fully differential description facilitates also the 
study of more complex final states including electroweak decay chains
at the same level of the QCD perturbative order.
Since the calculational framework is fully exclusive also with respect to flavour 
and chiralities, it can easily be merged with the electroweak contributions of LO and NLO.


\acknowledgments
This work was supported in part by the Research Executive Agency 
of the European Union under the 
Grant Agreement PITN-GA-2010-264564 (LHCPhenonet).
We are grateful to M.~Flowerdew, P.~Graf, T.~Hahn, E.~Mirabella, 
F.D.~Steffen and particularly A.~Landwehr for valuable discussions. 

\clearpage

\appendix
\section{Diagrams of NLO corrections}
\label{app:diagrams}
Here, for completeness, we display all relevant diagrams used in our NLO calculation of squark-squark production.
The contribution of some of them vanish under the assumption $m_{q}=0$.  For
example, this is the case for the $5^{\rm th}$ diagram on the $1^{\rm st}$ line when 
$a\neq b$; any helicity state of the quark in the propagator can interact either with
$\sq_{ia}$ or with $\sq_{jb}$ but not with both of them.\\

\vspace{1 cm}

\FIGURE{
\centering
\includegraphics{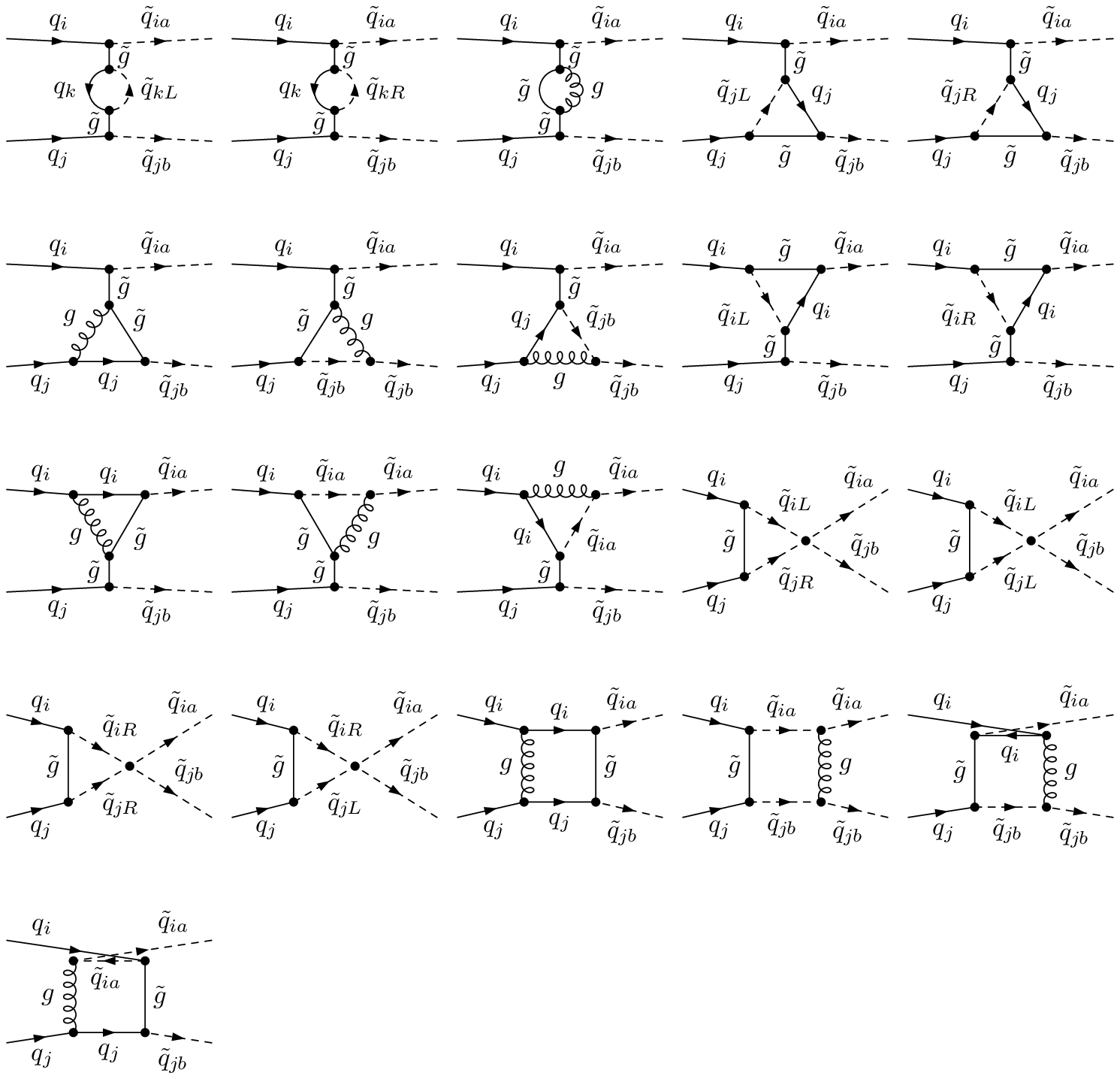}
\caption{Loop diagrams contributing to all flavour and chirality structures of squark--squark production.}
\label{fig:loop-t}
}

\FIGURE{
\centering
\includegraphics{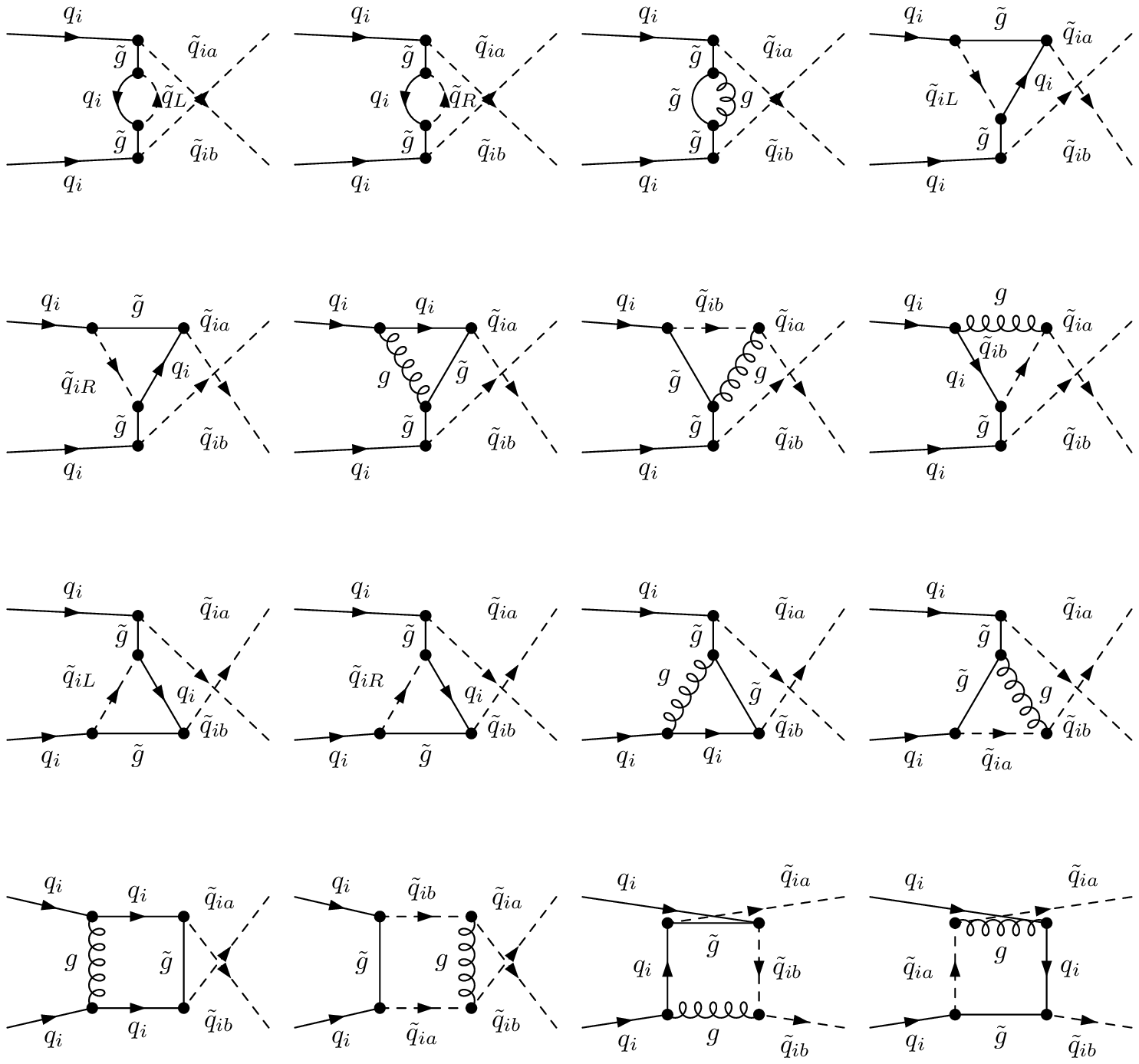}
\caption{Loop diagrams contributing only for squarks with equal flavour.}
\label{fig:loop-u}
}

\FIGURE{
\centering
\includegraphics{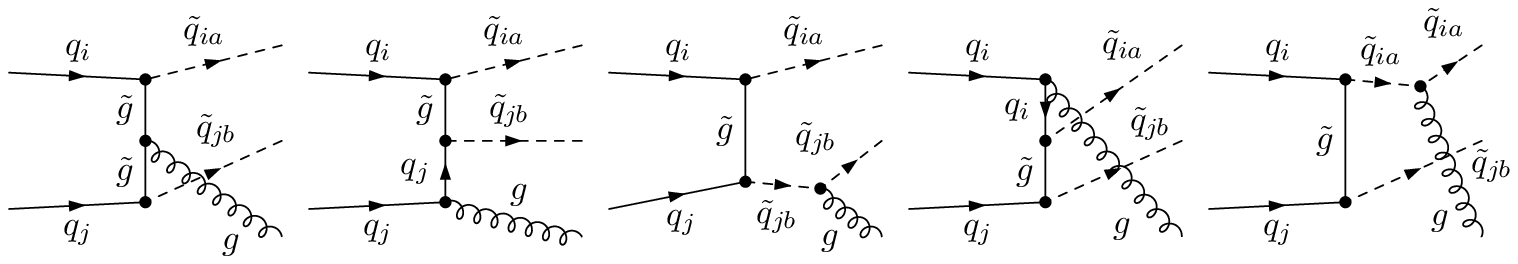}
\caption{Real gluon radiation diagrams contributing to all the flavour and chirality structures. Additional 
diagrams contributing only for equal flavour squarks are obtained from the above ones by a simple crossing
of the initial state quarks.}
\label{fig:real-t}
}

\clearpage
\section{Formulae for soft and collinear radiation}
\label{app:soft-collinear}
\subsection {Soft radiation}
The soft-bremsstrahlung correction factor in \eqref{softintegrals}
involves the kinematical factors $C^{(tt,ut,uu)}_{\sq \sq'}$ from (\ref{ttuuut})
and phase space integrals \cite{Denner:1991kt,Germer:2010vn}. 
Keeping a finite
quark mass $m_q$ only in the mass-singular terms, yields the
following expressions $\mathcal{I}_{ij}$ for the production process
$ q q' \to \tilde{q} \tilde{q}'$, where the labels $i,j$ correspond to the
assignment $q\to1$, $q'\to2$, $\sq\to3$, and $\sq'\to4$,
%
\begin{align}
\begin{split}
  \mathcal{I}_{ii} &=\frac{4}{3} \left[ \ln\left( \frac{4(\Delta E)^2}{\lambda^2} \right) +
  \ln\left(\frac{m_i^2}{ s_{12}}\right)\right]\qquad\qquad\qquad\qquad\qquad\qquad\qquad~~ \text{for }i=\{1,2\},
\\
  \mathcal{I}_{ii} &=\frac{4}{3}\left[ \ln\left( \frac{4(\Delta E)^2}{\lambda^2} \right) +
  \frac{1}{\beta_i} \ln\left(\frac{1-\beta_i}{1+\beta_i}\right)\right] \qquad\qquad\qquad\qquad\qquad\qquad\,
  \text{for } i=\{3,4\},
\\
  \mathcal{I}_{12} &=\left(-\frac{1}{3}C^{(tt)}_{\sq \sq'}-\frac{1}{3}C^{(uu)}_{\sq \sq'}-\frac{5}{3}C^{(ut)}_{\sq \sq'}\right) \sum_{i=1,2}\left[ \ln\left(\frac{ s_{12}}{m_i^2}\right)
  \ln\left(\frac{4(\Delta E)^2}{\lambda^2}\right) - \frac{1}{2}
  \ln^2\left(\frac{ s_{12}}{m_i^2}\right) -\frac{\pi^2}{3} \right],
\\
  \mathcal{I}_{34} &=\left(-\frac{1}{3}C^{(tt)}_{\sq \sq'}-\frac{1}{3}C^{(uu)}_{\sq \sq'}-\frac{5}{3}C^{(ut)}_{\sq \sq'}\right)  \frac{1}{v_{34}} \sum_{i=3,4} \left[
    \ln\left(\frac{1+\beta_i}{1-\beta_i}\right) \ln\left(\frac{4(\Delta
      E)^2}{\lambda^2}\right) -2\Li{\frac{2\beta_i}{1+\beta_i}}
    -\frac{1}{2}\ln^2\left(\frac{1-\beta_i}{1+\beta_i}\right)
    \right],
\\
  \mathcal{I}_{ij} &=\left(\frac{7}{6}C^{(tt)}_{\sq \sq'}-\frac{1}{6}C^{(uu)}_{\sq \sq'}-\frac{1}{6}C^{(ut)}_{\sq \sq'}\right)
\Big[  \ln\left(\frac{ s_{ij}^2}{m_i^2m_j^2}\right)
  \ln\left(\frac{4(\Delta E)^2}{\lambda^2}\right) -\frac{1}{2}
  \ln^2\left(\frac{ s_{12}}{m_i^2}\right)
  -\frac{1}{2}\ln^2\left(\frac{1-\beta_j}{1+\beta_j}\right) -\frac{\pi^2}{3}
\\
  & -2\Li{1-\frac{2p_i^0p_j^0}{ s_{ij}}(1+\beta_j)}
  -2\Li{1-\frac{2p_i^0p_j^0}{ s_{ij}}(1-\beta_j)}\Big]
\qquad
  \text{for } i+j=5\, ,
\\
  \mathcal{I}_{ij} &= \left(-\frac{1}{6}C^{(tt)}_{\sq \sq'}+\frac{7}{6}C^{(uu)}_{\sq \sq'}-\frac{1}{6}C^{(ut)}_{\sq \sq'}\right)
 \Big[ \ln\left(\frac{ s_{ij}^2}{m_i^2m_j^2}\right)
  \ln\left(\frac{4(\Delta E)^2}{\lambda^2}\right) -\frac{1}{2}
  \ln^2\left(\frac{ s_{12}}{m_i^2}\right)
  -\frac{1}{2}\ln^2\left(\frac{1-\beta_j}{1+\beta_j}\right) -\frac{\pi^2}{3}
\\
  & -2\Li{1-\frac{2p_i^0p_j^0}{ s_{ij}}(1+\beta_j)}
  -2\Li{1-\frac{2p_i^0p_j^0}{ s_{ij}}(1-\beta_j)}\Big]
\qquad
 \text{for } i+j=4\quad \text{or}\quad  i+j=6\, ,
\end{split}
\label{Iij}
\end{align}
with $\quad s_{ij}=2\, p_i\!\cdot\!p_j$, $\quad \beta_i=|\vec{p}_i|/p_i^0$, 
$\quad v_{ij}=\sqrt{1-4m_{i}^{2}m^{2}_{j}/ s_{ij}^2}\, .$
\\

For the decay process $\sq\rightarrow q\,\neu$, 
the corresponding expressions for $\mathcal{I}_{ij}$ in the 
decay width \eqref{dGammasoft} 
read as follows, with the the label assignment
$\sq \to1$, $q\to2$ (and $\neu\to3$), 
\begin{align}
\begin{split}
  \mathcal{I}_{11} &= \ln\left( \frac{4(\Delta E)^2}{\lambda^2} \right) -2 \, ,
  \\
  \mathcal{I}_{12} &=
  \ln\left(\frac{4(p^{0}_{2})^2}{m_{q}^{2}}\right)
  \ln\left(\frac{4(\Delta E)^2}{\lambda^2}\right) -\frac{1}{2}
  \ln^2\left(\frac{4(p^{0}_{2})^2}{m_{q}^{2}}\right)
  -\frac{\pi^2}{3}\, ,
  \\
  \mathcal{I}_{22} &=  \ln\left( \frac{4(\Delta E)^2}{\lambda^2} \right) +
  \ln\left(\frac{m_{q}^{2}}{4(p_{2}^{0})^{2}}\right)\, .
\end{split}
\label{Iij2}
\end{align}

\subsection{Collinear radiation}
This appendix collects the entries in eqs.~(\ref{collcross}) and (\ref{subpdf-cross}) 
for hard collinear radiation. They are taken
from \cite{Baur:1998kt} and \cite{Diener:2005me} with the replacement $\alpha Q_{q}^2\to(4/3)\alphas$. 

\smallskip
The (unintegrated) parton luminosity entering both the 
collinear gluon emission~(\ref{collcross})
and the subtraction term~(\ref{subpdf-cross})
is given by
\begin{align}\label{lumcoll}
\mathcal{L}_{qq'}(\tau,x,z)^{\text{coll}} \, =\,
\frac{1}{1+\delta_{qq'}}  
\left[f_{q}\!\left(\frac{x}{z},\mu_{F}\right)\, f_{q'}\!\left(\frac{\tau}{x},\mu_{F}\right)
    + f_{q}(\frac{\tau}{x},\mu_{F})\, f_{q'}\!\left(\frac{x}{z},\mu_{F}\right) \right] \, .
\end{align}
The partonic cross section for the collinear emission of a gluon into the cones with opening angle $\Delta\theta$ around the two quarks in the initial state can be written as follows,
\begin{align}
\label{partoniccollinear}
d\hat{\sigma}^{\text{coll-cone}}_{qq'\rightarrow\sq\sq'}(\tau,z) =
d\hat{\sigma}^{(0)}_{qq'\rightarrow\sq\sq'}(\tau) \; \frac{4}{3\pi} \alphas
\left[\frac{1+z^2}{1-z}\log\left(\frac{s\, \delta_{\theta}}{2m_{q}^{2}z}\right)-\frac{2z}{1-z}\right]\, ,
\end{align}
with $\delta_{\theta}=1-\cos\left(\Delta\theta\right)\simeq\Delta\theta^{2}/2$. 
The variable $z$ is the ratio between the momenta of the emitter parton after and before the emission. 

\smallskip
At the parton level, the $z$-dependent part of the subtraction term for one quark is given by
\begin{align}
\label{subpdf1}
d\hat{\sigma}^{\text{sub1}}_{qq'\rightarrow\sq\sq'}(\tau,z) =
d\hat{\sigma}^{(0)}_{qq'\rightarrow\sq\sq'}(\tau) \; \frac{2}{3\pi} \alphas
\left[\frac{1+z^2}{1-z}\log\left(\frac{\mu^{2}_{F}}{(1-z)^{2}m_{q}^2}\right)-\frac{1+z^2}{1-z}\right]\, ,
\end{align}
involving the factorization scale $\mu_F$, whereas the $z$-independent part 
\begin{align}\label{subpdf2}
d\hat{\sigma}^{\text{sub2}}_{qq'\rightarrow\sq\sq'}(\tau) =
d\hat{\sigma}^{(0)}_{qq'\rightarrow\sq\sq'}(\tau) \; \frac{4}{3\pi} \alphas
\left[1-\log(\delta_{s})-\log^{2}(\delta_{s})
    +\left(\log(\delta_{s})+\frac{3}{4}\right)\log\left(\frac{\mu_{F}^2}{m_{q}^2}\right)\right]  
\end{align}
contains also the soft-gluon phase space cut $\delta_{s}=2\Delta E/\sqrt{s}$.

\smallskip
In an analogous way, the (unintegrated) parton luminosity for 
collinear quark radiation in \eqref{qrad-hadr} is given by
\begin{align}\label{lumcollqrad}
\mathcal{L}_{i}(\tau,x,z)^{\text{coll-quark}} = \,
2~ g\left(\frac{x}{z},\mu_{F}\right) f_{i}\!\left(\frac{\tau}{x},\mu_{F}\right) \, ,
\end{align}
involving also the gluon distribution function $g(x,\mu_F)$.

\smallskip
Collecting in one formula the collinear cone emission of a quark 
and the subtraction term  (last term in the brackets of \eqref{qrad-part}),
one finds
\begin{align}
\label{qrad-part}
d\hat{\sigma}^{\text{coll-quark}}_{q_{i/j}g\rightarrow\sq_{ia}\sq_{jb}\bar{q}_{j/i}}(\tau,z) =& \,
d\hat{\sigma}^{(0)}_{q_{i}q_{j}\rightarrow\sq_{ia}\sq_{jb}} \, \cdot \nonumber \\ 
&  \cdot \, \frac{\alphas}{2\pi}\, P_{qg}(z) \,      
   \Big[\log\Big(\frac{s(1-z)^2\,\delta_{\theta}}{2m_{q}^2 z}\Big)+2z(1-z)   
       -\log\Big(\frac{\mu_{F}^2}{m_{q}^2 }\Big)\Big]\, ,
\end{align}
with the gluon--quark splitting function $P_{qg}(z)= [z^2+(1-z)^2]/2$. 
\\
Finally, the parton luminosity for non-collinear quark radiation is 
\begin{align}
\label{lumnoncollqrad}
\mathcal{L}_{i}(\tau)^{\text{noncoll-quark}} =
2 \int_{\tau}^{1}\frac{dx}{x} \;
g(x,\mu_{F})f_{i}\!\left(\frac{\tau}{x},\mu_{F}\right)\, .
\end{align}

\clearpage

\section{NLO corrections to the squark decay width}
\label{app:decay-an}
In section~\ref{sec:decayTOT} we need the NLO corrections 
to the decay width $\Gamma(\sq\to q\, \tilde{\chi})$, 
where $\tilde{\chi}$ is either a neutralino or a chargino, and $q$ a light quark.  
Here we perform the derivation of the correction factor, following 
the steps of the former calculation~\cite{Djouadi:1996wt}, 
but keeping explicitly the dependence on the masses $m_{q}$ 
for the collinear singularities and $\lambda$ for the IR singularities.
The NLO corrections in \eqref{eq:FQCD}
can be expressed in terms of a form factor $F^{\text{QCD}}$, 
which receives four contributions,
\begin{align}\label{dwnlo}
F^{\text{QCD}}=F_{g}+F_{\tilde{g}}+F_{ct}+F_{r} \, ,
\end{align}
namely loop corrections involving gluons $(F_{g})$ and gluinos $(F_{\tilde{g}})$, 
the counterterm contribution $(F_{ct})$, and the contribution from 
real gluon emission $(F_{r})$ . 

Keeping $m_{q}$  and $\lambda$ as independent mass parameters for the singular terms,
in dimensional reduction $F_{g}$ and $F_{ct}$ can be written as follows,
\begin{align}
F_{g}\, = &
\, \frac{\Delta}{2} - \frac{1}{2} \log\left(\frac{m_{\sq}^2}{\mu^2}\right)
+1-\log\left(\frac{m_{q}^2}{m_{\sq}^2}\right) +
\frac{1}{4}\log^2\left(\frac{m_{q}^2}{m_{\sq}^2}\right)-
\frac{1}{2}\log\left(\frac{\lambda^2}{m_{\sq}^2}\right)\log\left(\frac{m_{q}^2}{m_{\sq}^2}\right) 
     \nonumber \\
&  
   +\log\left(\frac{\lambda^2}{m_{\sq}^2}\right)\log(1-\kappa)-\log^2(1-\kappa)+\log(1-\kappa)-\Li{\kappa}\, ,  \\
F_{\rm ct}\,  = & \, - \frac{\Delta}{2} 
 + \frac{1}{2} \log{\left(\frac{ m_{\sq}^2}{\mu^2}\right)}
 - \log{\left(\frac{ \lambda^2}{ m_{\sq}^2 }\right)}
 + \frac{3}{4}\log{\left(\frac{ m_{q}^2}{ m_{\sq}^2 }\right) } 
 + \frac{\gamma}{4\,(1-\gamma)} -  \frac{\gamma}{2} - \frac{15}{8}   \nonumber \\  
& \,   - \frac{1}{2}(\gamma^2-1)\log{\left(\frac{\gamma -1}{\gamma}\right)}
 + \frac{1}{4}\left[ \frac{2\,\gamma-1}{(1-\gamma)^2}+3  \right] \log(\gamma)
\, , 
\label{Fg}
\end{align}
where $\kappa= m_{\tilde{\chi}}^{2}/m_{\sq}^{2}$, $\gamma=m_{\sq}^{2}/m_{\tilde g}^{2}, $ 
and $\Delta$ denotes the UV divergence, cf.~\eqref{dzgmsbar}.

$F_{\tilde{g}}$ is free of soft, collinear, and UV singularities, hence it
is not affected by the  choice of regulators; it is  
identical to the result in \cite{Djouadi:1996wt} and we do not repeat it here.

The part from real gluon emission, integrated over the full phase space, 
can be expressed with the help of the 
bremsstrahlung integrals given in~\cite{Denner:1991kt}, evaluated 
in the limit $m_{q}=0$ except for the mass-singular terms. 
The fully integrated decay width for $\tilde{q} \to q\tilde{\chi} g$
can be written as follows, 
\begin{eqnarray}
\label{gammareal}
& & \Gamma_{\tilde{q} \to q \tilde{\chi} g} = 
\Gamma^{(0)}_{\tilde{q} \to q \tilde{\chi}} \cdot
\frac{4}{3} \frac{\alpha_s}{\pi} \, F_r \,  , \\
& & F_r = \frac{2}{m_{\tilde{q}}^2 - m_{\tilde{\chi}}^2} 
  \big[ 2 ( m_{\neuc}^2 - m_{\sq}^2 )
                      \, ( m_{\sq}^2 \, I_{00} + m_{q}^2 \, I_{11} 
                          + I_0 + I_1 )
       - 2  ( m_{\neuc}^2 - m_{\sq}^2 )^2  \, I_{01}
       - I
       - I_1^0 \big]  \nonumber
\end{eqnarray}
with $\Gamma^{(0)}$ from \eqref{LOdwn} and (\ref{LOchar}). The
phase space integrals $ I \equiv I(m_{\sq},m_{q},m_{\neuc})$
are given by 
\begin{align}\label{Is}
I_{00}  = &
\frac{1}{4\,m_{\sq}^4}\bigg[m_{\neuc}^2-m_{\sq}^2+m_{\sq}^2 \log\Big[\frac{m_{\sq}^2-m_{\neuc}^2}{\lambda m_{\sq}}\Big]
+ m_{\neuc}^2\log\Big[\frac{\lambda m_{\neuc}^2}{m_{\sq}^3-m_{\sq}m_{\neuc}^2}\Big] \bigg] \nonumber  \, ,\\ 
I_{11}  = & \frac{1}{4\,m_{q}^2\,m_{\sq}^2}\, (m_{\sq}^2-m_{\neuc}^2)\left[\log\left(\frac{m_{q}}{\lambda}\right)-1\right]\, ,\nonumber \\
I_{01}  = & \frac{1}{4\,m_{\sq}^2}\left[\frac{\pi^2}{2}+\log^2\left(\frac{m_q}{\lambda}\right)
-\log^2\left(\frac{m_{\sq}^2-m_{\neuc}^{2}}{\lambda m_{\sq}}\right)
          -\Li{1-\frac{m_{\neuc}^2}{m_{\sq}^2}}\right] \, ,   \nonumber \\
I  = & \frac{1}{8\,m_{\sq}^2} \bigg[m_{\sq}^4-m_{\neuc}^4+4
m_{\neuc}^2 m_{\sq}^2\log\left(\frac{m_{\neuc }} {m_{\sq}}\right)\bigg]\, ,  \nonumber
\end{align}
\begin{align}
I_{0} =&  \frac{1}{4\,m_{\sq}^2}\left[m_{\neuc}^2-m_{\sq}^2-2 m_{\neuc}^2 \,\log\left(\frac{m_{\neuc}} {m_{\sq}}\right)\right]\, , \nonumber \\
I_1  = & \frac{1}{4\,m_{\sq}^2}\bigg[ m_{\neuc} ^2-m_{\sq}^2 +2 m_{\neuc}^2 \,\log\left(\frac{m_{q} m_{\neuc}} {m_{\sq}^2-m_{\neuc}^2}\right)
 -2 m_{\sq}^2 \,\log\left(\frac{m_{q}m_{\sq}}{m_{\sq}^2-m_{\neu_{j}}^2}\right)\bigg]\, , \nonumber \\
I_1^0 = & \frac{1}{16\,m_{\sq}^2}
\bigg[ 5 m_{\neuc}^4-8 m_{\neuc}^2 m_{\sq}^2+3 m_{\sq}^4
        +4 (m_{\neuc}^4-2 m_{\neuc}^2 m_{\sq}^2) \,\log\left(\frac{m_{q} m_{\neuc}} {m_{\sq}^2-m_{\neuc}^2} \right) \nonumber \\
 & +\,4 m_{\sq}^4\, \log\left(\frac{m_{q} m_{\sq}}{m_{\sq}^2-m_{\neuc}^2} \right)\bigg] \ . 
\end{align}
\\
With these expressions, \eqref{gammareal} yields $F_r$ for the real gluon part of $F^{\text{QCD}}$
in \eqref{dwnlo},
\begin{eqnarray}\label{Fr}
F_{r}=& & \frac{- 5 + 8 \kappa - 3 \kappa^2 - 8 \kappa \log(\kappa) + 
   6 \kappa^2 \log(\kappa)}{8 (1 - \kappa)^2}  \nonumber \\ 
 & & +\, 4 - \frac{\pi^2}{2} - \frac{5}{2} \log(1 - \kappa) +  \log(1 - \kappa)^2 - 
     \log^2\left(\frac{m_{q}}{\lambda}\right) + 2 \log\left(\frac{m_{\sq}}{\lambda}\right)  \nonumber\\
     & & +\, \frac{1}{2} \log\left(\frac{m_{q}}{m_{\sq}}\right) + 
    2 \log(1 - \kappa) \log\left(\frac{m_{\sq}}{\lambda}\right) +
    \log^2\left(\frac{m_{\sq}}{\lambda}\right) 
  + \Li{1 - \kappa}\, .
\end{eqnarray}
Combining all four contributions in \eqref{dwnlo}
we obtain a compact analytical expression for the form factor $F^{\text{QCD}}$, 
which agrees with the result in \cite{Djouadi:1996wt},
\bee
F^{\rm QCD} & = & 
               - \frac{1}{8}\left( \frac{4\,\gamma^2-27\,\gamma+25}{\gamma-1}
                                    + \frac{3\,\kappa-5}{\kappa-1} \right)
  - \frac{\pi^2}{3} - 2\, {\rm Li_2}(\kappa)
  - \frac{1}{2}\,(\gamma^2-1)\,\log \left(\frac{\gamma-1}{\gamma}\right) \nonumber \\
   &  & 
  + \, \frac{3\,\gamma^2-4\,\gamma+2}{4\,(1-\gamma)^2}\,\log \left(\gamma \right)
  - \frac{3}{2}\,\log (1- \kappa) 
  + \frac{1}{4} \!\cdot\! \frac{3\,\kappa^2-4\,\kappa}{(\kappa-1)^2}\,\log \left(\kappa \right) 
  - \log\left( \kappa \right) \,\log (1-\kappa) \nonumber \\ 
   &  & +\, \sqrt{ \kappa \gamma} \left[ \frac{1}{ \kappa} \log 
(1-\kappa)+ \frac{1}{1-\kappa} \left[ \gamma \log (\gamma) -(\gamma-1)
\log (\gamma-1) \right] + \frac{ \kappa +\gamma -2}{(1-\kappa)^2} \, I \, 
\right] \, ,
\eee
where for $\kappa \gamma <1$ the function $I$ is given by 
\bee
I= {\rm Li_{2}} \left( \frac{\gamma-1}{\gamma \kappa-1} \right)
 - {\rm Li_{2}} \left( \kappa \frac{\gamma-1}{\gamma \kappa-1} \right)
 - {\rm Li_{2}} \left( \frac{\gamma+\kappa-2}{\gamma \kappa-1} \right)
 + {\rm Li_{2}} \left( \kappa \frac{\gamma+\kappa-2}{\gamma \kappa-1} \right) \, , \nonumber
\eee
and for $\kappa \gamma > 1$ one has
\bee
I & = &-{\rm Li_{2}}\left( \frac{\gamma \kappa-1}{\gamma-1}  \right)
       +{\rm Li_{2}}\left( \frac{\gamma \kappa-1}{\gamma+\kappa-2}  \right)
       +{\rm Li_{2}}\left( \frac{\gamma \kappa-1}{\kappa(\gamma-1)}  \right)
       -{\rm Li_{2}}\left( \frac{\gamma \kappa-1}{\kappa(\gamma+\kappa-2)}  
\right)          \nonumber \\
  &   & -\log (\kappa)\,\log \frac{\gamma+\kappa-2}{\gamma-1} \ . \nonumber
\eee

\clearpage
\section{Comparison between local and global diagram subtraction schemes}
\label{app:DS}
Here, expanding the discussion in section \ref{sec:quarkrad}, we want to elucidate the differences between the implementations of the DS scheme in the global approach, as used in this paper, and in the local approach, as used e.g. in \cite{Beenakker:1996ch,Frixione:2008yi,Binoth:2011xi}. 
In the following discussion we consider the contribution from the resonant diagrams of \figref{fig:radq-u}(a). In a notation similar to the one of appendix B.1 of \cite{Beenakker:1996ch}, the contribution to the partonic cross section emerging from these diagrams can be written as
\begin{equation}\label{xsec-prospino}
\hat{\sigma}=\int_{m^{2}_{\sq_{ia}}}^{q^2_{\text{max}}}dq^{2}\frac{f(q^2)}{(q^2-m_{\tilde{g}}^{2})^2+m_{\tilde{g}}^2 \Gamma_{\tilde{g}}^2}\, ,
\end{equation}
where $f(q^{2})$ is the differential cross section in $q^2$ (the squared invariant mass of $\sq_{ia}$ and  $q_{i}$) without the squared gluino propagator.
Given the squared total energy in the partonic center of mass $s$, the maximum allowed value for $q^2$ is $q^2_{\text{max}}=(\sqrt{s}-m_{\sq_{jb}})^2$.

In the global approach the contribution from on-shell $\sq_{jb}\tilde{g}$ production is subtracted by substituting $\hat{\sigma}$ with
\begin{equation}\label{ds-global}
\Delta\hat{\sigma}_{\text{Global}}=\hat{\sigma}-\hat{\sigma}_{\sq_{jb}\tilde{g}}\frac{\Gamma_{\tilde{g}\to\sq_{ia}}}{\Gamma_{\tilde{g}}}\, ,
\end{equation}
i.e. subtracting exactly the total cross section for on-shell production of $\sq_{jb}\tilde{g}$  multiplied by the branching ratio of $\tilde{g}\to\sq_{ia}q_{i}$.

In the local approach, before phase-space integration, $f(q^2)$ evaluated in the on-shell gluino configuration
\begin{equation}\label{fdig}
f(m^{2}_{\tilde{g}})=\hat{\sigma}_{\sq_{jb}\tilde{g}}\frac{m_{\tilde{g}}\Gamma_{\tilde{g}}}{\pi}\frac{\Gamma_{\tilde{g}\to\sq_{ia}}}{\Gamma_{\tilde{g}}}\
\end{equation}
is subtracted in the numerator of the integrand of \eqref{xsec-prospino},
\begin{equation}
\Delta\hat{\sigma}_{\text{Local}}=\int_{m^{2}_{\sq_{ia}}}^{q^2_{\text{max}}}dq^{2}\frac{f(q^2)-f(m^2_{\tilde{g}})}{(q^2-m_{\tilde{g}}^{2})^2+m_{\tilde{g}}^2 \Gamma_{\tilde{g}}^2}:=\hat{\sigma}-I\, , \quad \text{with} \quad
I=\int_{m^{2}_{\sq_{ia}}}^{q^2_{\text{max}}}dq^{2}\frac{f(m^2_{\tilde{g}})}{(q^2-m_{\tilde{g}}^{2})^2+m_{\tilde{g}}^2 \Gamma_{\tilde{g}}^2}\, .
\end{equation}
In this parton level example the integral $I$ can be analytically calculated. For $m_{\tilde{g}}>m_{\sq_{ia}}$ and $\sqrt{s}>m_{\tilde{g}}+m_{\sq_{jb}}$, i.e. in the region where subtraction is required, it yields
\begin{eqnarray}
I&=&\frac{f(m^{2}_{\tilde{g}})}{\Gamma_{\tilde{g}}m_{\tilde{g}}}\left[\arctan\left(\frac{q^{2}_{\text{max}}-m^{2}_{\tilde{g}}}{\Gamma_{\tilde{g}}m_{\tilde{g}}}\right)-\arctan\left(\frac{m^{2}_{\sq_{ia}}-m^{2}_{\tilde{g}}}{\Gamma_{\tilde{g}}m_{\tilde{g}}}\right)\right] \\\nonumber
&=&\hat{\sigma}_{\sq_{jb}\tilde{g}}\frac{\Gamma_{\tilde{g}\to\sq_{ia}}}{\Gamma_{\tilde{g}}}
\left[1-\frac{\Gamma_{\tilde{g}}m_{\tilde{g}}}{\pi}\left(\frac{q^{2}_{\text{max}}-m^{2}_{\sq_{ia}}}{(q^{2}_{\text{max}}-m^{2}_{\tilde{g}})(m^{2}_{\tilde{g}}-m^{2}_{\sq_{ia}})}\right)\right] + \ord(\Gamma_{\tilde g})\, .
%
\end{eqnarray}
Comparing in this way the global and local approach for the DS subtraction we find
\begin{equation}\label{comparisonds}
\Delta\hat{\sigma}_{\text{Global}}-\Delta\hat{\sigma}_{\text{Local}}=-\hat{\sigma}_{\sq_{jb}\tilde{g}} \Gamma_{\tilde{g}\to\sq_{ia}}\frac{m_{\tilde{g}}}{\pi}
\left(\frac{q^{2}_{\text{max}}-m^{2}_{\sq_{ia}}}{(q^{2}_{\text{max}}-m^{2}_{\tilde{g}})(m^{2}_{\tilde{g}}-m^{2}_{\sq_{ia}})}\right)  + \ord(\Gamma_{\tilde g})\, .
\end{equation}
Hence, even in the limit $\Gamma_{\tilde{g}}\to0$, the two approaches differ by a finite term depending on the physical phase space boundaries, see also \cite{Berdine:2007uv}. This can be understood from the fact that the approximation of the Breit-Wigner distribution $m_{\tilde{g}}\Gamma_{\tilde{g}}/[(q^2-m_{\tilde{g}}^{2})^2+m_{\tilde{g}}^2 \Gamma_{\tilde{g}}^2] \rightarrow \pi\delta(q^2-m^2_{\tilde{g}})$ in the integrand of $I$ is strictly valid only for an integration over the entire real axis. Moreover, the result in \eqref{comparisonds} can be altered if the mapping $q^{2}\to m_{\tilde{g}}^{2}$ in the local subtraction is performed before the integration of the other phase space variables. This mapping is not uniquely defined and can lead to further differences. 
At the hadronic level the numerical differences between the two
approaches can be of the order of a few per mill of the on-shell
$\sq_{jb}\tilde{g}$ production. Thus, depending on the parameter
region, few per-cent differences can appear for the $\sq_{ia}\sq_{jb}$ NLO relative corrections.
For example, for SPS1a and $\SqrtS=14~\TeV$, corrections for the 
$\tilde d_R \tilde d_R$ cross section arising from \eqref{comparisonds} amount to 
$0.08\%$ of $\sigma^{(0)}_{\tilde d_{R}\tilde g}$ and to $1.9\%$ of $\sigma^{(0)}_{\tilde d_{R}\tilde d_{R}}$, 
since $\sigma^{(0)}_{\tilde d_{R}\tilde g}/\sigma^{(0)}_{\tilde d_{R}\tilde d_{R}}\approx23$.
For different flavour and chirality configurations these corrections vary, they are, however, of the same order.
Finally, we want to note that both the local and the global approach can be extended to a fully differential level.



\bibliographystyle{JHEP}
\providecommand{\href}[2]{#2}\begingroup\raggedright
\endgroup


\end{document}